\documentclass[aip,jcp,preprint]{revtex4-1}

\usepackage[mathlines]{lineno} 
\usepackage{graphicx} 
\usepackage{amsmath} 
\usepackage{geometry}                		
\usepackage{graphicx}				
\usepackage{amssymb,color}
\usepackage{setspace}

\usepackage{bm}
\usepackage{subfigure}
\usepackage{amsmath}
\newcommand{\dd}{\mathrm d}

\makeatletter
\def\@email#1#2{%
 \endgroup
 \patchcmd{\titleblock@produce}
  {\frontmatter@RRAPformat}
  {\frontmatter@RRAPformat{\produce@RRAP{*#1\href{mailto:#2}{#2}}}\frontmatter@RRAPformat}
  {}{}
}%
\makeatother

\begin{document}


\preprint{AIP/123-QED}

\title{Anisotropic self-assembly of soft particles mediated by elliptically polarized AC fields}
\author{Carlos Eduardo Estanislau}
 \affiliation{Departamento de Física, Instituto de Ciências Exatas e Biológicas, Universidade Federal de Ouro Preto, 35400-000, Ouro Preto, MG, Brazil}
\author{Thiago Colla}%
\affiliation{Departamento de Física, Instituto de Ciências Exatas e Biológicas, Universidade Federal de Ouro Preto, 35400-000, Ouro Preto, MG, Brazil}
\author{Christos N. Likos}
\affiliation{Faculty of Physics, University of Vienna, Boltzmanngasse 5, 1090 Vienna, Austria}

\begin{abstract}
    Attractive dipole interactions can be induced between equally charged soft nanoparticles under the influence of AC electric fields. The combination of charge repulsion and dipole attraction, along with different screening responses from an underlying electrolyte, lead to complex aggregations ranging from chain-like formation for linear polarizations to isotropic planar structures in the case of circular polarizations. In this work, we analyze the role of varying field anisotropies in these self-assembled structures. To this end, the formalism previously developed for the coarse-grained interactions of soft particles in the presence of linear 
    [T.~Colla {\it et al.}, ACS Nano {\bf 12}, 4321-4337 (2018)] and circular [M.~Reich {\it et al.}, Soft Matter {\bf 21}, 1516-1528 (2025)] field polarizations is naturally extended to incorporate elliptical polarizations of arbitrary asymmetries. A rich variety of self-assembly formations is found at intermediate field anisotropies, thus bridging the gap between linear and circular field-induced self-assembly scenarios. 
\end{abstract}

\maketitle

\section{Introduction}\label{sec:introduction}

The promotion of unusual structural changes in systems of nanoparticles under the influence of suitable external stimuli is a common mechanism underlying many natural phenomena in physics, chemistry, and biology.~\cite{Mas09,Swe20,Lom20,Chag22,Huang24,Chen23,Chen24} For this reason, the dynamic and structural features of self-assembly under various external conditions have been the topic of intensive investigations, laying a common ground of active research in many distinct areas.~\cite{Kat08,Bha12,Xu13,Tal20,Ama22} Understanding the main mechanisms that trigger different particle aggregations -- as well as their sensitivity to controlled external conditions -- has also become of paramount relevance for the engineering of nano-scaled materials.~\cite{Grze10,Tal20,Li21_2,Ama22} Conformation changes at a nano-scale is a key mechanism in the designing of sensor devices,~\cite{Taka15,Kim23,Pur24,Aki24} capable of detecting small changes in external conditions such as temperature~\cite{Rosa15,Rosa16}, chemical composition~\cite{Rick96,Yang97,Cho18} and radiation.~\cite{Kim23,Vec25,Ale25} The tendency of certain components to either aggregate or separate under the influence of certain external fields is also of fundamental relevance in filtering processes and dispersant agents.~\cite{Avra02,Sud07,Whi11,Col13,Sha23} Moreover, the ability of certain particles to aggregate into close-shaped structures under the controlled influence of external fields is an important mechanism for the engineering of encapsulation devices,~\cite{Ani08,Vel09,Ha10,Veli12} as well as in the designing of drug-delivery systems.~\cite{Lu16,Bai18,Qia22,Aja24} Another relevant application of driven self-assembly is the formation of percolating structures in nano-composites,~\cite{And08,Qure24,Sha25} which provides a convenient way of controlling transport properties and signal emissions. 

Despite its great relevance in the understanding of various natural processes and the design of novel materials, the investigation of molecular aggregation properties under the influence of given external conditions poses many challenges~\cite{Kat08,Koe24}. The level of complexity in particle interactions that induce self-assembly quickly becomes prohibitively large as the number of different components -- along with their constituent sub-elements -- increases in size and scale.~\cite{Kaka12,Sca21,Ama22,Ado25,Agu25,Luca24,Koe24} Strong asymmetries in size and composition, as well as different time scales of particle mobilities require a high level of resolution to track particle motions and interactions dictating the formation of large aggregates.~\cite{Lik01,Car16,Sch23,Bor23} Different shapes and anisotropic interactions add a great deal of complexity at the theoretical description of particle interactions and the resulting self-assembly structures.~\cite{Zha04,Cho17,Rom20,Nic22,Koe24} In the case of soft-matter systems, complex conformation changes usually take place at the level of individual particles and interfaces.~\cite{Lik06} Such changes in volume and shapes are in many cases sensitive to external conditions,~\cite{Kaka12,Nic22,Aza25} such as the presence of external fields,~\cite{Avra02,Duv06,Veli12,Kum17} the affinity with an underlying solvent~\cite{Nar13,Don17,Ham22,Qia23,Yu25}, elastic soft interactions between different particles and interfaces~\cite{Mar04,Rey07,Sar11,Uri16,Foy24} and degree of confinement.~\cite{Bia13,Chu21,Krott24} For this reason, soft particles are known for their ability to self-assembly into a much wider variety of topological structures when compared to hard nano-particles of similar types and sizes~\cite{Lik06,Colla18_2,Jos21}. Moreover, their high sensitivity to surrounding conditions makes these particles ideal for the confection of highly susceptible materials, capable of capturing small fluctuations in physical properties at their close vicinity.~\cite{Uri16,Agra18,Alz23} Particle deformations under compression, along with the possibility to adsorb/expel solvent particles into/out of their inner structures, can result in the formation of very compact aggregates, even able to exceed the close-packing density limit of hard-spheres.~\cite{Gott05,Riest12,Uri16,Alz23,Mis24} In addition, soft particles are the cornerstone of many biological and natural systems, such that materials building-up from these particles are usually bio-compatible and environmentally friendly~\cite{Pur24}. Many soft particles of relevance are endowed with fuzzy layers of grafted polymers, a key property to stabilize these particles in solution due to a soft repulsion that prevents their flocculation.~\cite{Lik01,Lik06} All these properties render soft particles suitable building blocks for the designing of responsive materials~\cite{Lik06}. 

A large variety of soft particles are composed of functionalized polymer chains, which can be further entangled to form complex networks of amorphous-like structures~\cite{Gennes80,Fred05,Sch23}. When embedded in an aqueous solvent, the functional groups dissociate their counterions in solution, thus acquiring a residual net charge of opposite sign.~\cite{Lev02,Ven17} Further addition of electrolytes is a commonly used strategy for further controlling the effective repulsion between these particles. This is accomplished through well known ionic-induced mechanisms such as enhanced electrostatic screening, counterion adsorption onto the polymer backbones, along with complex double layer formations.~\cite{Lev02,Pod06,Khan19} The resulting isotropic interactions (in case of spherical particles) can be strongly modified by the presence of external electric fields, which break down rotation symmetry, also leading to complex electro-osmotic flows.~\cite{Ros91,Moh16,Tell21,Bom21} In the case of AC fields, a stationary ion motion is established in which the ion clouds around the nanoparticles become polarized along the field direction~\cite{Ohs95_2,Dho11}.

For hard colloidal particles, ionic polarization is also accompanied by a field-induced polarization due to a frequency-dependent dielectric mismatch between the nanoparticles and the surrounding polar media, resulting in additional image-charge interactions with mobile counterions~\cite{Lev02,San15,Ngu19}. This is usually not the case for systems of soft particles, as a significant solvent adsorption into the particles precludes large dielectric contrasts~\cite{Colla18,Moh16}. In these cases, however, a different polarization mechanism is present, whereby complex electro-osmotic flow into the soft particle can induce an internal net polarization that couples to the driven frequency~\cite{Osh95,Ohs09,Moh16}. In contrast to the case of dielectric particles --  where the internal polarization arises from bound charge re-orientations~`\cite{Jackson} -- the building-up of polarization charges at the soft cores is driven by a reduced charge mobility that leads to local, frequency-dependent charge separations in static regimes~\cite{Osh95,Ohs09,Ohs12}. Such a decreased mobility is a consequence of both charge correlations and enhanced friction forces on the polymeric network that comprises the inner, soft structure~\cite{Osh95,Ohs95_2,Osh00}. The polarized nanoparticles then induce their own dipole fields, resulting in effective dipole interactions depending on the field frequency, strength, and polarization~\cite{Colla18}. Apart from breaking down rotation symmetry from pure monopole interactions, the field-induced polarization also leads to competing contributions, as the induced dipoles on equally charged particles will attract one another along the field direction~\cite{Colla18,reich:sm:2025}. In addition, the range of the induced dipole interactions is dictated by a rearrangement of the isotropic ionic clouds surrounding the nanoparticles, resulting in anisotropic ion distributions that partially screen the polarized charges~\cite{Colla18}. Since the building-up of such static neutralizing clouds depends on a coupling between ion mobility and frequency modes of oscillating dipoles~\cite{Moh16}, the screening of induced dipole-dipole interactions is generally less effective than that of the monopole interactions~\cite{Colla18}. As a result, the relative strength and range between monopole and dipole interactions can be easily controlled by the external field. The possibility of changing the strength, range, and degree of anisotropy of attractive interactions --  combined with intrinsic isotropic repulsions from monopole charges and overlapping soft layers -- provides a promising strategy for exploring a wide variety of self-assembly topologies of soft particles in the presence of controlling fields~\cite{Noij13,Cra14,Cras17,Aza17,Noij19}. For this reason, the aggregation of soft-particle driven by external fields has been a topic of intensive investigations in the last decades, both experimentally and via computer simulations.~\cite{yethiraj:nature:2003,jusufi:rmp:2009,Riest12,yakovlev:scirep:2017,Colla18,reich:sm:2025} Anisotropic interactions
can also be brought about through the influence of tilted external
magnetic fields on colloidal monolayers of (super-)paramagnetic colloids.~\cite{froltsov:pre:2003,hoffmann:prl:2006,hoffmann:jpcm:2006,hoffmann:molphys:2007,osterman:prl:2009,snezhko:prl:2009,chremos:jpcb:2009,dobnkar:sm:2013,mueller:langmuir:2014}

A series of experiments has been recently performed in which ionic microgels are shown to be arranged into chains of different sizes and separations when subjected to AC fields of different frequencies and strengths~\cite{Moh12,Noij13,Moha15,Moh16,Colla18,Kir19,Noij19}. The fields were linearly polarized,~\cite{Colla18} (LP) and the chains were formed along the field direction. Depending on the field frequency and the microgel packing fractions, the chains also display regular, crystal-like arrangements over the lateral, in-plane directions. Dielectric spectroscopy studies have shown a rich scenario in which different polarization mechanisms come into play, depending on the coupling between the driven field frequency and the various natural oscillating modes~\cite{Moh16}. At higher frequencies, the major polarization contribution comes from the polarization of the polymer backbones that make up the internal microgel structure. A theoretical model was then proposed,~\cite{Colla18} in which the chain formation was assigned to attractive dipole-dipole interactions resulting from uniform microgel polarizations, induced along the field direction. The theory was able to qualitatively capture the onset of chain formation and crystal in-plane ordering predicted by the experiments. Recently, this model has been modified to account for the situation of soft-particles in the presence of circularly polarized~\cite{reich:sm:2025} (CP) fields. Changing the field polarization from LP to CP has drastic effects on the underlying aggregation properties. Instead of linear chains along the field direction, CP fields lead to the formation of aggregates arranged across the polarization plane, regularly spaced along the perpendicular, out-of-plane direction. Such marked difference in topological structures results from a changing in dipole anisotropy, which is switched from in-plane attraction and out-of-plane repulsion in LP case to the opposite situation in the case of CP fields~\cite{reich:sm:2025}. 

These findings suggest that the degree of anisotropy in particle interactions is a key factor in determining the formation of different self-assembly structures. One therefore expects that continuously changing the light polarization from LP to CP should give rise to a number of anisotropic, intermediate structures ranging from linear chains to paralleled arranged planes. This can be clearly achieved by considering the case of elliptically polarized (EP) fields, in which case the eccentricity of elliptical light plays the role of a continuous parameter that spans all the degrees of anisotropy lying in-between the LP and CP cases. In this work, we provide such an extension, thus exploring how the increased degree of anisotropy influences the aggregation morphologies of soft particles induced by elliptically polarized fields of different strengths and frequencies. To this end, a general framework is first developed in Section \ref{sec:theory} for computing coarse-grained dipole interactions in the case of spherical nanoparticles featuring polarizing charges which oscillate in phase with elliptically polarized fields. This situation is then further specialized to the particular case of soft particles developing a constant polarization charge. Since this system has been worked out for both linear and circular polarizations in previous works~\cite{Colla18,reich:sm:2025}, the effective dipole interactions can be straightforwardly extended to incorporate elliptical polarizations by a simple matching of dipole coefficients, as shown in Section \ref{sec:ac}. After presenting the general formalism, we then proceed to provide an overview of the role that anisotropy plays in shaping different percolating aggregates by analyzing results from computer simulations,
which is done in Section \ref{sec:results}. Finally, in Section \ref{sec:conclusions}, we summarize the obtained results, alongside with our main conclusions and perspectives for future investigations.

\section{Theoretical Background}\label{sec:theory}

We now proceed with a general analysis of the coarse-grained interactions between spherical nanoparticles (of either soft or hard nature), induced by an elliptically polarized, alternating electric field. We consider that the nanoparticles are embedded on a $1:1$ electrolyte of concentration $c_s$. It is further assumed that the applied field is in a frequency range where its dominant effect is to induce polarizing charges at the nanoparticles, while ion mobility is not directly influenced by the field. Instead, ion flow is coupled to the external field in a rather indirect fashion through the ionic response to field-induced, oscillating dipoles on the nanoparticles. More specifically, the electric external field $\bm{E}(t)\sim \bm{E}_0e^{{\mathrm i}\omega t}$ induces a time-varying polarization field $\bm{P}(\bm{r},t)$ in the interior of the nanoparticles, which in a stationary state oscillates with the same driven frequency $\omega$ (up to a possible phase shift). We shall hereby restrict our attention to linear and
isotropic media, in which case $\bm{P}(\bm{r},t)$ and $\bm{E}(t)$ always oscillate parallel to each other. The oscillating polarization gives rise to a charge density of magnitude $\rho_p(\bm{r},t)=-\nabla\cdot\bm{P}(\bm{r},t)$ inside the particle domains. In the case of an elliptical polarized light oscillating with frequency $\omega$ along the $xy$ plane, the polarizing field can be written as
\begin{equation}
   \bm{E}(t)=E_0[\cos\alpha\cos\omega t\hat{\bm{e}}_x+\sin\alpha\sin\omega t\hat{\bm{e}}_y]\equiv E_0\bm{n}(t),\label{E1}
\end{equation}
where $E_0$ is twice the root-mean-squared (rms) field amplitude, $\omega$ is the driving frequency, $\alpha$ is an angle related to the phase shift between the $x$ and $y$ field components, and ${\bm{n}}(t)=(\cos\omega t\cos\alpha)\hat{\bm{e}}_x+(\sin\omega t\sin\alpha)\hat{\bm{e}}_y$ is a vector pointing at the instantaneous field direction. Note that this vector is not normalized; it is rather chosen such that the rms field over a cycle is $E^2_{\mathrm{rms}}=E^2_0/2$. The parametrization in Eq. (\ref{E1}) is such that the principal axes 
of the ellipse traced by the tip of the 
${\bm E}$-vector over a period lie along the $x$ and $y$ directions, with lengths $E_{0x}=E_0\cos\alpha$ and $E_{0y}=E_0\sin\alpha$, respectively.~\footnote{A perhaps more practical parametrization would be to set a phase shift between the oscillating $E_x$ and $E_y$ components. However, such a parametrization field leads to elliptical polarizations whose major axes are rotated with respect to the $\hat{\bm{e}}_x$ and $\hat{\bm{e}}_y$ axes, and is for this reason avoided in view of unnecessary complications in the numerical implementations that follows.} Values of $\alpha$ lying in the range $0\le \alpha<\pi/4$ correspond to major axes parallel to $\hat{\bm{e}}_x$, and continuously span the whole set of planar polarizations, from linear $(\alpha=0)$ to circular light ($\alpha=\pi/4$). Similarly, the range $\pi/4<\alpha\le\pi/2$ describes light with major axes along the $y$ direction, and all corresponding degrees of eccentricity. 
In a stationary situation, the polarizations induced on the nanoparticles display a similar oscillation pattern as that of Eq. (\ref{E1}). The induced polarization charges give raise to effective, field-induced interaction between nanoparticles, partially mediated by a time-averaged redistribution of mobile ions, in response to the oscillating fields. 

\begin{figure}[h!]
    \centering
    \includegraphics[width = 13cm, height= 8cm]{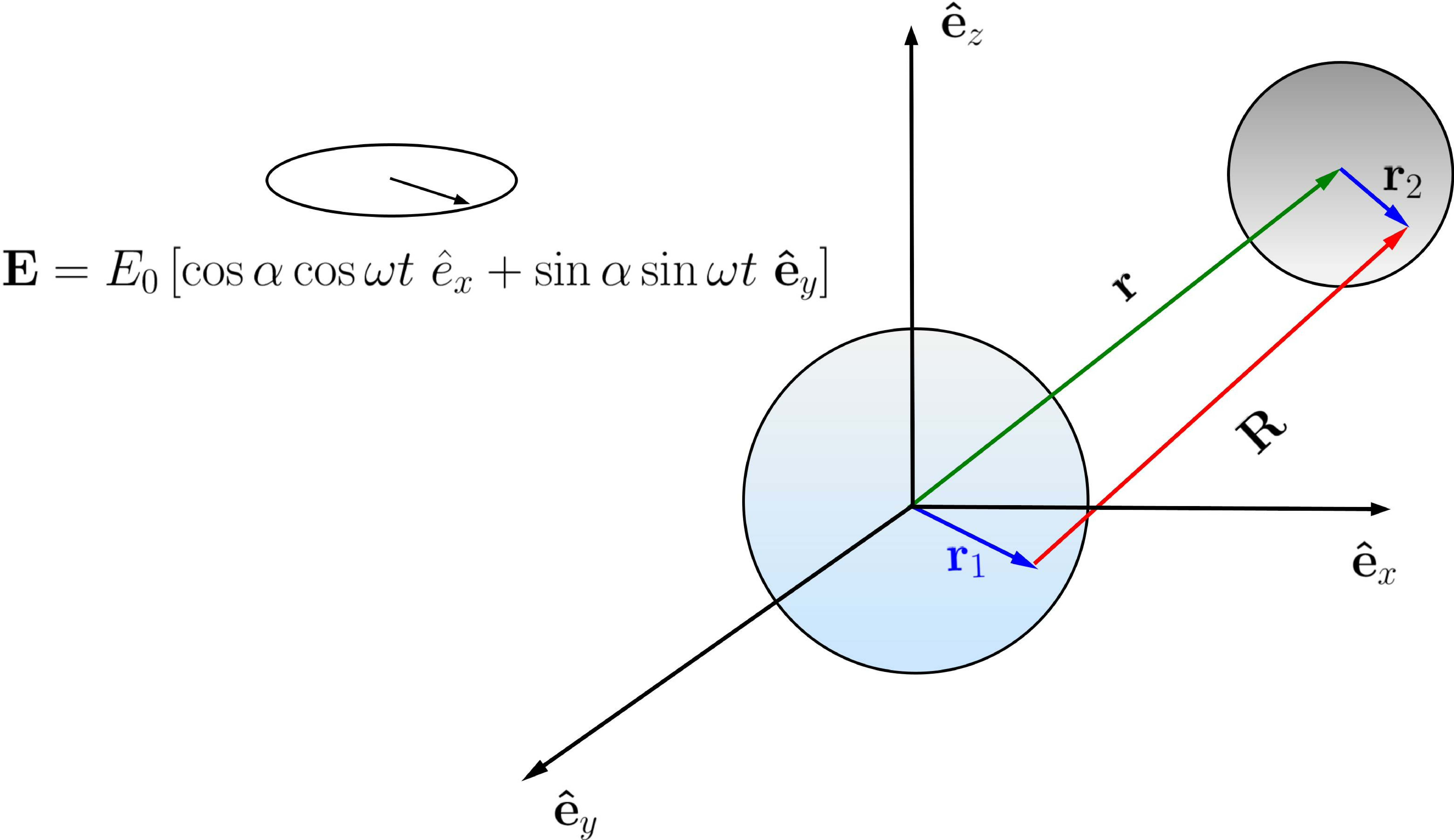}
    \caption{Sketch of our model system. Two spherical particles of center-to-center separation $\bm{r}$ are polarized by an elliptically polarized field, lying along the $x$-$y$ plane. The origin of the coordinate system
    is placed at the center of colloid 1 and colloid
    2 has its center at position ${\bm r}$.
    The induced dipole interactions are obtained by considering the interactions between bound charges at points located at positions $\bm{r}_1$ and $\bm{r}_2$ from the particle centers $1$ and $2$, respectively, and separated by a distance $\bm{R}=\bm{r}-(\bm{r}_1 - {\bm r}_2)$. The 
    electrostatic interaction is mediated by a response function $G(R)$ which incorporates the implicit effects of a underlying electrolyte. }
    \label{fig:fig1}
\end{figure}

In order to investigate some general features of such induced dipole interactions, let us consider the two spherical particles depicted in Fig.~\ref{fig:fig1}, whose center-to-center separation is $\bm{r}$,  both polarized by the external field in a similar way. At a given time $t$, the particles have parallel, instantaneous polarizations $\bm{P}_1(\bm{r}_1,t)=P_1(\bm{r}_1){\bm{n}}(t)$ and $\bm{P}_2(\bm{r}_2)=P_2(\bm{r}_2){\bm{n}}(t)$, where $\bm{r}_1$ and $\bm{r}_2$ denote position vectors relative
to the colloids' centers of mass, as shown 
in Fig.~\ref{fig:fig1}. The instantaneous, field-induced electrostatic interaction between the dipole charges can be formally written as
\begin{equation}
    u_{dd}(\bm{r},t)=\int\psi(\bm{r}+\bm{r}_2,t)\rho_{p2}(\bm{r}_2,t)\dd^3r_2,
    \label{udd1}
\end{equation}
where $\psi({\bm r} + {\bm r}_2,t)$ is the instantaneous potential induced by polarizing charges on particle $1$, and $\rho_{p2}(\bm{r}_2,t)=-\nabla_2\cdot\bm{P}_2({\bm r}_2,t)$ is the polarizing charge density at particle $2$. The potential $\psi(\bm{r}',t)$ at an arbitrary point $\bm{r}'$ is assumed to implicitly incorporate ionic-mediated effects in response to an external charge source. In cases in which such sources act on an open system, the potential due to polarizing charges at particle $1$ can be formally written as
\begin{equation}
    \psi(\bm{r}',t)=\int\rho_{p1}(\bm{r}_1,t)G(|\bm{r}'-\bm{r}_1|)\dd^3{r}_1,
    \label{psi1}
\end{equation}
where $G(|\bm{r}'-\bm{r}_1|)$ is the Green function representing the response of a point charge at $\bm{r}'$ due to a point, unit source located at $\bm{r}_1$, in the presence of the embedding electrolyte. For the special case of bare electrostatics, this function
satisfies the equation:
\begin{equation}
\nabla_1^2 G_0(|\bm{r}'-\bm{r}_1|) = -\frac{4\pi}{\varepsilon}\delta(\bm{r}'-\bm{r}_1),
\label{eq:laplace}
\end{equation}
where $\varepsilon$ is the dielectric constant of the
electrolyte and the subscript denotes the absence
of counterions that screen the Coulomb 
potential.

Before combining these expressions, it is convenient to rewrite them in a more suitable form. Under the replacement $\rho_{p2}(\bm{r}_2,t)=-\nabla_2\cdot\bm{P}_2(\bm{r}_2,t)$, followed by an integration by parts, Eq. (\ref{udd1}) can be transformed into:
\begin{equation}
    u_{dd}(\bm{r},t)=\int\nabla_2\psi(\bm{r}+\bm{r}_2,t)\cdot\bm{P}_2(\bm{r}_2,t)\dd^3{r}_2,
    \label{udd2}
\end{equation}
where the operator $\nabla_2$ is meant to act only on the coordinates of particle $2$. A similar transformation can be applied in (\ref{psi1}), which is then rewritten as
\begin{equation}
    \psi(\bm{r}',t)=\int \bm{P}_1(\bm{r}_1,t)\cdot \nabla_1G(|\bm{r}'-\bm{r}_1|)\dd^3{r}_1.
    \label{psi2}
\end{equation}
The quantity $\nabla_2\psi(\bm{r}+\bm{r_2},t)$ in (\ref{udd2}) thus reads 
\begin{equation}
    \nabla_2\psi(\bm{r}+\bm{r}_2,t)=\int \nabla_2[\bm{P}_1(\bm{r}_1,t)\cdot \nabla_1G(R)]\dd^3{r}_1,
    \label{psi3}
\end{equation}
where $R\equiv |\bm{r}-\bm{r}_{12}|$ is the distance between the observation point $\bm{r}$ and the relative position between source points $\bm{r}_{12}=\bm{r}_1-\bm{r}_2$. This expression can be further simplified by means of the identity $\nabla(\bm{a}\cdot\bm{b})=\bm{a}\times(\nabla\times\bm{b})+\bm{b}\times(\nabla\times\bm{a})-(\bm{a}\cdot\nabla)\bm{b}-(\bm{b}\cdot\nabla)\bm{a}$, where $\bm{a}$ and $\bm{b}$ are arbitrary vector fields. Noticing that $\nabla_2\bm{P}_1({\bm r}_1,t)=\nabla_2\times\bm{P}_1({\bm r}_1,t)=0$ and $\nabla_1G(R)=-G'(R)\bm{\hat{e}}_R$ (where $\bm{\hat{e}}_R\equiv \bm{R}/R$), is an irrotational field, Eq. (\ref{psi3}) becomes
\begin{equation}
    \nabla_2\psi(\bm{r}+\bm{r}_2,t)=\int \dd^3{r}_1(\bm{P}_1(\bm{r}_1,t)\cdot\nabla_2)G'(R)\bm{\hat{e}}_R.
    \label{psi4}
\end{equation}
Also noting that $\nabla_2G'(R)=G''(R)\nabla_2(R)=G''(R)\hat{\bm{e}}_R$ and $(\bm{P}_1(\bm{r}_1,t)\cdot\nabla_2)\hat{\bm{e}}_R=\bm{P}_1(\bm{r}_1,t)-(\bm{P}_1(\bm{r}_1,t)\cdot\bm{\hat{e}}_R)\bm{\hat{e}}_R$, this expression can be rewritten as
\begin{equation}
\begin{split}    
    \nabla_2\psi(\bm{r}+\bm{r}_2,t)&=\int \dfrac{G'(R)}{R}\left[\bm{P}_1(\bm{r}_1,t)-(\bm{P}_1(\bm{r}_1,t)\cdot\bm{\hat{e}}_R)\bm{\hat{e}}_R\right]\dd^3{r}_1\\
    &+\int G''(R)\left(\bm{P}_1(\bm{r}_1,t)\cdot\bm{\hat{e}}_R\right)\bm{\hat{e}}_R\dd^3{r}_1.
    \label{psi5}
\end{split}    
\end{equation}
Replacement of this relation into Eq. (\ref{udd2}) leads to the following dipole potential:
\begin{equation}
\begin{split}
    u_{dd}(\bm{r},t)&=\int \dd^3{r}_1\dd^3{r}_2(\bm{P}_1(\bm{r}_1,t)\cdot\bm{\hat{e}}_R)(\bm{P}_2(\bm{r}_2,t)\cdot\bm{\hat{e}}_R)\left(G''(R)-\dfrac{G'(R)}{R}\right)\\
    &+\int \dd^3{r}_1\dd^3{r}_2\dfrac{G'(R)}{R}\bm{P}_1(\bm{r}_1,t)\cdot\bm{P}_2(\bm{r}_2,t).
\end{split}
    \label{udd3}
\end{equation}
Since, by assumption, both polarizations $\bm{P}_1$ and $\bm{P}_2$ both point along the field direction $\bm{n}(t)$, and $\bm{n}(t)\cdot\bm{\hat{e}}_R=\dfrac{\bm{n}(t)\cdot{(\bm{r}-\bm{r}_{12})}}{R}$, this relation can be also written as
\begin{equation}
\begin{split}
    u_{dd}(\bm{r},t)&=\int P_1(\bm{r}_1)P_2(\bm{r}_2)\left(G''(R)-\dfrac{G'(R)}{R}\right)\left[\dfrac{(\bm{n}(t)\cdot\bm{r})-(\bm{n}(t)\cdot\bm{r}_{12})}{R} \right]^2\dd^3{r}_1\dd^3{r}_2\\
    &+\int P_1(\bm{r}_1)P_2(\bm{r}_2)\dfrac{G'(R)}{R}|\bm{n}(t)|^2\dd^3{r}_1\dd^3{r}_2.  
    \end{split}
    \label{udd4}
\end{equation}

It is interesting to note that, in the case 
of {\textit{static}} point dipoles of magnitudes $p_1$ and $p_2$, we have $P_1(\bm{r}_1)=p_1\delta(\bm{r}_1)$ and  $P_2(\bm{r}_2)=p_2\delta(\bm{r}_2)$, and the above relation is reduced to
\begin{equation}
    u_{dd}(\bm{r})=p_1p_2\left[G''(r)(\bm{\hat{n}}\cdot\bm{\hat{e}}_r)^2+\dfrac{G'(r)}{r}\left(|\bm{{\hat{n}}}|^2-(\bm{{\hat{n}}}\cdot\bm{\hat{e}}_r)^2\right)\right],
    \label{u_point1}
\end{equation}
where $\bm{\hat{e}}_r=\bm{r}/r$ is the unit vector connecting the two point dipoles and now
${\bm{\hat{n}}}$ is a {\textit {unit}} vector. Considering 
the Green's function of
Eq.~(\ref{eq:laplace}) alongside a static case where both dipoles are aligned in the $z$ direction, $\bm{n}=\bm{\hat{e}}_z$ forming an angle $\theta$ with the vector $\bm{r}$, the above relation can be further simplified as
\begin{equation}
    u_{dd}(\bm{r},t)=\dfrac{p_1p_2}{\varepsilon r^3}\left(3\cos^2\theta-1\right),
    \label{u_point2}
\end{equation}
thus properly recovering the limit of point-like dipoles in the absence of screening. 

The most general case, Eq. (\ref{udd4}), requires further time averaging over the field vectors $\bm{n}(t)$. Notice that these averages only involve non-trivial, quadratic terms in $\bm{n}(t)$, and must be carefully evaluated. We start by averaging Eq. (\ref{udd4}) over one field cycle, thus yielding the following time-averaged potential $\bar{u}_{dd}(\bm{r})\equiv\langle u_{dd}(\bm{r},t)\rangle$: 
\begin{equation}
\begin{split}
    \bar{u}_{dd}(\bm{r})&=\langle(\bm{n}(t)\cdot\bm{r})^2\rangle\int P_1(\bm{r}_1)P_2(\bm{r}_2)\dfrac{f'(R)}{R}\dd^3{r}_1\dd^3{r}_2\\
    &+\langle|\bm{n}(t)|^2\rangle\int P_1(\bm{r}_1)P_2(\bm{r}_2)f(R)\dd^3{r}_1\dd^3{r}_2\\
    &-2\int\langle(\bm{n}(t)\cdot\bm{r})(\bm{n}(t)\cdot\bm{r}_{12})\rangle P_1(\bm{r}_1)P_2(\bm{r}_2)\dfrac{f'(R)}{R}\dd^3{r}_1\dd^3{r}_2\\
    &+\int\langle(\bm{n}(t)\cdot\bm{r}_{12})^2\rangle P_1(\bm{r}_1)P_2(\bm{r}_2)\dfrac{f'(R)}{R}\dd^3{r}_1\dd^3{r}_2,
    \end{split}
    \label{udd5}
\end{equation}
where $f(R)\equiv G'(R)/R$. If the two polarizing particles have same size and polarizations, the third integral on the rhs should vanish, as it is clearly anti-symmetric under particle exchange $1\leftrightarrow 2$. The same holds if the polarizations are radially symmetric, $P(\bm{r})=P(r)$, since the angular integral of $\bm{n}(t)\cdot\bm{r}_{12}$ over the unit sphere should vanish on symmetry considerations. We shall henceforth restrict our attention to this particular situation, in which case the third integral contributes nothing to the dipole-dipole interactions. Note also that, in this case, the first and second integrals on the rhs cannot depend on the orientation of $\bm{r}$, as the coordinate frames can be chosen in such a way that $\bm{r}$ lies along the $z$-axis. 

Let us now consider the explicit form of the time averages in Eq. (\ref{udd5}). The average $\langle|\bm{n}(t)|^2\rangle$ can be readily computed as
\begin{equation}
  \langle|\bm{n}(t)|^2\rangle=\langle\cos^2\omega t\cos^2\alpha+\sin^2\omega t\sin^2\alpha\rangle=\dfrac{1}{2},
  \label{av1}
\end{equation}
where use was made of the relations $\langle \cos^2\omega t\rangle=\langle \sin^2\omega t\rangle=\dfrac{1}{2}$. Similarly, the time average $\langle(\bm{n}(t)\cdot\bm{r})^2\rangle$ can be explicitly evaluate as:
\begin{equation}
\begin{split}  
\langle(\bm{n}(t)\cdot\bm{r})^2\rangle =r^2\sin^2\theta\langle\left[\cos\varphi\cos\omega t\cos\alpha+\sin\varphi\sin\omega t\sin\alpha\right]^2\rangle\\
=r^2\sin^2\theta[\cos^2\alpha\cos^2\varphi\langle\cos^2\omega t\rangle+\sin^2\alpha\sin^2\varphi\langle\sin^2\omega t\rangle\\
 +\dfrac{1}{4}\sin2\alpha\sin2\varphi\langle\sin2\omega t\rangle]\\
 =\dfrac{r^2\sin^2\theta}{4}\left[\cos^2\varphi(1+\cos2\alpha)+\sin^2\varphi(1-\cos2\alpha)\right]\\
 =\dfrac{r^2\sin^2\theta}{4}\left(1+\cos2\alpha\cos2\varphi\right).
\end{split}
\label{av2}
\end{equation}
In these relations, we have used the fact that $\langle\sin2\omega t\rangle=0$, together with usual trigonometric identities to express the time average in the simplest form. Relations (\ref{av1}) and (\ref{av2}) can be combined with Eq. (\ref{udd4}) to provide explicit expressions for the time-averaged dipole interactions induced by elliptically polarized light in molecules featuring radial polarizations, provided they response to the applied field is linear. Before proceeding, it is instructive to note that relation (\ref{av2}) implies that the term proportional to $\langle(\bm{n}(t)\cdot\bm{r})^2\rangle$ in Eq. (\ref{udd4}) can be can be resolved in terms of spherical harmonics $Y_\ell^m(\theta,\varphi)$ as
\begin{equation}
    \langle(\bm{n}(t)\cdot\bm{r})^2\rangle=\dfrac{r^2}{6}\left(1-2\sqrt{\dfrac{\pi}{5}}Y_0^2(\theta,\varphi)\right)+\dfrac{r^2\cos2\alpha}{2}\sqrt{\dfrac{2\pi}{15}}[Y_2^{2}(\theta,\varphi)+Y_2^{-2}(\theta,\varphi)].
    \label{Y1}
\end{equation}
The coefficient multiplying this term (the first integral on the rhs of (\ref{udd5})) depends on the radial coordinate $r$ alone, as the $z$ axis can be chosen as lying on the $\bm{r}$ direction. By the same reasoning, it follows from Eq. (\ref{av1}) that the second term in Eq. (\ref{udd5}) has no dependency on the orientation of $\bm{r}$. In order to explicitly evaluate the angular dependence of Eq. (\ref{udd5}), it is still necessary to consider how the last integral in this relation depends on the orientation of the two dipoles. For this reason, it is convenient to use Eq. (\ref{Y1}) to explicitly write this integral in terms of the relative orientations as
\begin{equation}
\begin{split}
     &\int P_1(r_1)P_2(r_2)\dfrac{f'(R)}{R}\langle(\bm{n}(t)\cdot\bm{r}_{12})^2\rangle \dd^3{r}_1\dd^3{r}_2 =\\
     &\int P_1(r_1)P_2(r_2)\dfrac{f'(R)}{R} \\
     &\times \left[a_0+a_{20}Y_2^{0}(\theta_{12},\varphi_{12})+a_{22}(Y_2^{2}(\theta_{12},\varphi_{12})+Y_2^{-2}(\theta_{12},\varphi_{12}))\right] \dd^3{r}_1\dd^3{r}_2,
    \end{split}
    \label{n12_1}
\end{equation}
where $(\theta_{12},\varphi_{12})$ are angular components of the vector $\bm{r}_{12}=\bm{r}_1-\bm{r}_2$ in the cartesian system depicted in Fig.~\ref{fig:fig1}, and the coefficients $a_0$, $a_{20}$ and $a_{22}$ should be chosen such as to match the corresponding prefactors in Eq. (\ref{Y1}),
i.e.:
\begin{eqnarray}
a_0 & = & \frac{r^2}{6};\label{a0}
\\
a_{20} & = & -\frac{r^2}{3}\sqrt{\frac{\pi}{5}};\label{a20}
\\
a_{22} & = & \frac{r^2\cos 2\alpha}{2}\sqrt{\frac{2\pi}{15}}.\label{a22}
\end{eqnarray}

Since both integrals over the solid angles $\Omega_1$ and $\Omega_2$ in Eq.~(\ref{n12_1}) above are performed over the whole unit sphere, a simple changing of variables $\Omega_2\rightarrow\Omega_{12}$ can be performed in the integration over $\Omega_2$. Relation (\ref{n12_1}) can thus be written as
\begin{equation}
\begin{split}
    & \int P_1(r_1)P_2(r_2)\dfrac{f'(R)}{R}\langle(\bm{n}(t)\cdot\bm{r}_{12})^2\rangle \dd^3{r}_1\dd^3{r}_2 =\\
    &-\int P_1(r_1) r_1^2 \dd r_1\int P_2(r_2)r_2^2\dd r_2\\
    & \times\int\dfrac{f'(R)}{R}\left[a_0+a_{02}Y_2^{0}(\theta_{12},\varphi_{12})+a_{22}(Y_2^{2}(\theta_{12},\varphi_{12})+Y_2^{-2}(\theta_{12},\varphi_{12}))\right] \dd\Omega_1\dd\Omega_{12},
    \end{split}
    \label{n12_2}
\end{equation}
where the minus sing on the rhs comes from the fact that $\dd\Omega_2=-\dd\Omega_{12}$ (since $\bm{r}_{12}=\bm{r}_1-\bm{r}_2)$. Notice that the isotropic integral, proportional to $a_0$, does not depend on the orientation of the connecting vector $\bm{r}$, and can be suppressed in the analysis that follows. We also note that, conceived as a function of $\bm{r}$, the radial distance $R=|\bm{r}-\bm{r}_{12}|$ depends on $r$ and $\cos\gamma=\bm{\hat{e}}_r\cdot{\bm{\hat{e}}}_{12}$, where $\bm{\hat{e}}_r$ and $\bm{\hat{e}}_{12}$ are unit vector pointing along the directions of $\bm{r}$ and $\bm{r}_{12}$, respectively, and $\gamma$ is the angle between them. Therefore, the radial function $f'(R)/R$ in (\ref{n12_2}) can be formally expanded as
\begin{equation}
  \dfrac{f'(R)}{R}=\sum_{\ell=0}^{\infty}b_\ell(R) \mathcal{P}_{\ell}(\cos\gamma),
    \label{f1}
\end{equation}
where $\mathcal{P}_{\ell}(x)$ are Legendre polynomials of order $\ell$, and the coefficients $b_{\ell}(R)$ are functions of $R$, given by
\begin{equation}
b_{\ell}(|{\bm r} - {\bm r}_{12}|)=\dfrac{2\ell + 1}{2}\int_{0}^{\pi}\dfrac{f'(R)}{R}\mathcal{P}_{\ell}(\cos\gamma)\sin\gamma\dd\gamma. 
    \label{inv_leg}
\end{equation}
Now, by invoking the Addition Theorem for the spherical harmonics~\cite{Jackson}, equation (\ref{f1}) can be expressed in terms of the angles $(\theta_{12},\varphi_{12})$ and $(\theta,\varphi)$ between the $z$-axis and the unit vectors $\bm{\hat{e}}_{12}$ and $\bm{\hat{e}}_{r}$, respectively, as
\begin{equation}
  \dfrac{f'(R)}{R}=\dfrac{4\pi}{2\ell +1}\sum_{\ell=0}^{\infty}b_{\ell}(R)\sum_{m=-\ell}^{\ell} Y_{\ell}^{m}(\theta,\varphi)Y^{*m}_{\ell}(\theta_{12},\varphi_{12}).
    \label{f2}
\end{equation}
Performing the angular integrations in Eq. (\ref{n12_2}) thus yields
\begin{equation}
\begin{split}
    &\int\dfrac{f'(R)}{R}\left[a_{20}Y_2^{0}(\theta_{12},\varphi_{12})+a_{22}(Y_2^{2}(\theta_{12},\varphi_{12})+Y_2^{-2}(\theta_{12},\varphi_{12}))\right] \dd\Omega_1\dd\Omega_{12}=\\
   &\sum_{\ell=0}^{\infty}\dfrac{4\pi}{2\ell +1}\sum_{m=-\ell}^{\ell
   }\int\dd\Omega_{12}\left[a_{20}Y_2^{0}(\theta_{12},\varphi_{12})+a_{22}(Y_2^{2}(\theta_{12},\varphi_{12})+Y_2^{-2}(\theta_{12},\varphi_{12}))\right]\\
   &\times Y_{\ell}^{m*}(\theta_{12},\varphi_{12})
    Y_{\ell}^{m}(\theta,\varphi)\left[\int b_{\ell}(R)\dd\Omega_1\right].
   \label{f3}
    \end{split}
\end{equation}
Clearly, integration of $\Omega_{1}$ over the unit sphere leads to a function which depends solely on the radial distances $r$, $r_1$ and $r_2$. It is convenient to introduce a function $g_{\ell}(r,r_1,r_2)$ of these variables, defined as
\begin{equation}
g_{\ell}(r,r_1,r_2)\equiv  \int b_{\ell}(|{\bm r} - {\bm r}_{12}|)\dd\Omega_1 =
\dfrac{2\ell + 1}{2}\int \dd \Omega_1\int_{0}^{\pi}\dfrac{f'(R)}{R}\mathcal{P}_{\ell}(\cos\gamma)\sin\gamma\dd\gamma,
\label{g_l}
\end{equation}
where, in the last equality, use was made of Eq. (\ref{inv_leg}).
Since this function does not depend on particle orientations, the integrations over $\Omega_{12}$ become trivial due to the orthogonality relation between spherical harmonics. All but the $\ell=2$ mode vanish, and Eq. (\ref{f3}) is significantly simplified as
\begin{equation}
\begin{split}
    \int\dfrac{f'(R)}{R}\left[a_{20}Y_2^{0}(\theta_{12},\varphi_{12})+a_{22}(Y_2^{2}(\theta_{12},\varphi_{12})+Y_2^{-2}(\theta_{12},\varphi_{12}))\right] \dd\Omega_1\dd\Omega_{12}=\\
  \dfrac{4\pi}{5}\left[a_{20}Y_2^{0}(\theta,\varphi)+a_{22}(Y_2^{2}(\theta,\varphi)+Y_2^{-2}(\theta,\varphi))\right]g_2(r,r_1,r_2).
   \label{f4}
    \end{split}
\end{equation}

Substitution of the above expression into (\ref{n12_1}) thus provides
\begin{equation}
\begin{split}
    & \int P_1(r_1)P_2(r_2)\dfrac{f'(R)}{R}\langle(\bm{n}(t)\cdot\bm{r}_{12})^2\rangle \dd^3{r}_1\dd^3{r}_2\\
    &=a_0\int\int P_1(r_1)P_2(r_2)\dfrac{f'(R)}{R} \dd^3{r}_1 \dd^3{r}_2\\
    &-\dfrac{4\pi}{5}\left[a_{20}Y_2^{0}(\theta,\varphi)+a_{22}(Y_2^{2}(\theta,\varphi)+Y_2^{-2}(\theta,\varphi))\right]\\
    &\times \int\int P_1(r_1) P_2(r_2)g_2(r,r_1,r_2)r_1^2 r_2^2\dd r_1\dd r_2.
    \end{split}
    \label{n12_3}
\end{equation}

In the last line above, use was made of Eq. (\ref{inv_leg}), together with the fact that the integrals over $\Omega_1$ and $\gamma$ render the integration of particle $2$ isotropic. It now becomes apparent from Eqs. (\ref{Y1}) and (\ref{n12_3}) that the averaged potential, Eq. (\ref{udd5}), can be written as a second-order expansion in Spherical Harmonics as
\begin{equation}
    \bar{u}_{dd}(\bm{r})=u_{00}(r)+u_{20}Y_{2}^0(\theta,\varphi)+u_{22}(r)[Y_{2}^{2}(\theta,\varphi)+Y_{2}^{-2}(\theta,\varphi)].
    \label{udd6}
\end{equation}
The coefficients $u_{00}(r)$, $u_{20}(r)$, and $u_{22}(r)$ can be determined by combining Eqs. (\ref{udd5}), (\ref{av1}), (\ref{Y1}), (\ref{n12_3}), together with the explicit definitions of the coefficients $a_0$, $a_{20}$, and $a_{22}$, given relations (\ref{a0}), (\ref{a20}), and (\ref{a22}), respectively. As a result, the coefficients that define the dipole interactions in Eq. (\ref{udd6}) and can be explicitly written as 
\begin{small}
\begin{eqnarray}
    u_{00}(r) &= & \int\int P_1(r_1)P_2(r_2)\left(\dfrac{r^2f'(R)}{3R}+\dfrac{f(R)}{2}\right)\dd^3{r}_1\dd^3{r}_2;\\
    u_{20}(r) & = &-\dfrac{r^2}{3}\sqrt{\dfrac{\pi}{5}}\int\int \dd^3{r}_1\dd^3{r}_2P_1(r_1)P_2(r_2)\\
    &\times&\left[\dfrac{f'(R)}{R}-\dfrac{1}{2}\int_{-1}^1\dfrac{f'(R)}{R}\mathcal{P}_2(\cos\gamma)\dd\cos\gamma\right];\\
    \nonumber
    u_{22}(r) & = & \dfrac{r^2\cos2\alpha}{2}\sqrt{\dfrac{2\pi}{15}}\int\int \dd^3{r}_1\dd^3{r}_2P_1(r_1)P_2(r_2)\\
    &\times&\left[\dfrac{f'(R)}{R}-\dfrac{1}{2}\int_{-1}^1\dfrac{f'(R)}{R}\mathcal{P}_2(\cos\gamma)\dd\cos\gamma\right].
    \label{coef1}
\end{eqnarray}
\end{small}

Once the polarization responses of a pair of molecules are known, these expressions provide us with a general framework for computing their mutual, dipole-dipole interactions, induced by elliptically polarized light. 
Note that these coefficients are functions of the radial distance $r$ alone, as the $z$-axis can be chosen as lying along the radial direction $\bm{r}$. In practice, the obtained dipole interactions can be written in a much simpler form by noticing that the combination of spherical harmonics in Eq. (\ref{n12_3}) is the same as the one in Eq. (\ref{Y1}). Using (\ref{av2}), Eq. (\ref{udd6}) can be considerably simplified as 
\begin{equation}
\bar{u}_{dd}(\bm{r})= u_0(r)+u_2(r)\sin^2\theta[1+\cos2\alpha\cos2\varphi],
\label{udd7}
\end{equation}
where the coefficients $u_0(r)$ and $u_2(r)$ read
\begin{eqnarray}
 u_{0}(r) &  =  & \dfrac{r^2}{2}\int\int \dd^3{r}_1\dd^3{r}_2P(r_1)P(r_2)f(R);\\
 \nonumber
 u_{2}(r) & = & r^2\int\int \dd^3{r}_1\dd^3{r}_2P_1(r_1)P_2(r_2)\\
 &\times&\left[\dfrac{f'(R)}{R}-\dfrac{1}{2}\int_{-1}^1\dfrac{f'(R)}{R}\mathcal{P}_2(\cos\gamma)\dd\cos\gamma\right].
\label{u2}
\end{eqnarray}

In the case of linear polarized fields, $\alpha=0$, Eq. (\ref{udd7}) reduces to
\begin{equation}
\bar{u}_{dd}(\bm{r})= u_0(r)+2u_2(r)\sin^2\theta\cos^2\varphi=u_0(r)+2u_2(r)\dfrac{x^2}{r^2},
\label{udd8}
\end{equation}
where $x$ is the component of $\bm{r}$ along the field direction. Likewise, setting $\alpha=\pi/2$ results in light polarized along the $y$ axis, $\bar{u}_{dd}=u_0(r)+2u_2(r)y^2/r^2$. The same goes for light polarized along the $z$-axis, in which case Eq. (\ref{udd8}) can be written as
\begin{equation}
\bar{u}_{dd}(\bm{r})= u_0(r)+2u_2(r)\dfrac{z^2}{r^2}=\left(u_0+\dfrac{2u_2}{3}\right)+\dfrac{4u_2}{3}\mathcal{P}_2(\cos\theta)\equiv u_{D0}+u_{D2}\mathcal{P}_2(\cos\theta),
\label{udd9}
\end{equation}
where $\mathcal{P}_2(\cos\theta)=(3\cos^2\theta-1)/2$ denotes the second-order Legendre polynomial. The radial functions $u_{D0}(r)$ and $u_{D2}(r)$ introduced above are the zeroth and second-order coefficients, respectively, of an expansion of $u_{dd}(\bm{r})$ in the case linearly polarized light at the $z$ direction. A direct comparison between these coefficients and those from elliptical light allows one to identify $u_{D0}\leftrightarrow u_0+2u_2/3$ and $u_{D2}\leftrightarrow 4u_2/3$. In terms of these coefficients, Eq. (\ref{udd7}) can be written as 
\begin{equation}
\bar{u}_{dd}(\bm{r})= u_{D0}(r)-\dfrac{u_{D2}(r)}{2}[1+\cos2\alpha\cos2\varphi]\mathcal{P}_2(\cos\theta).
\label{udd10}
\end{equation}
This relation will be used in what follows to directly relate the induced potential for elliptical fields with that of linearly polarized AC field, for which the coefficients $u_{D0}$ and $u{D2}$ have been obtained in previous works. We note in particular that the case of circularly polarized light ($\alpha=\pi/4$) amounts to a re-scale of $u_{D2}$ by a factor $-1/2$, as previously pointed out in a recent investigation of microgels under circularly polarized fields.

\section{Soft particles under AC fields}\label{sec:ac}

We now turn our attention to the specific case
of two penetrable, spherical microgels of 
radius $a$ each. Due to ion dissociation,
these acquire in aqueous
solutions a net charge $Zq$,
where $q$ is the elementary charge,~\cite{Riest12} which is present 
even in the absence of a polarizing electric
field ${\bm E}(t)$.
The dissociation leads to the release of mobile counterions, leaving behind charged residues of opposite sign in the particle backbones. The solution can also contain dissociated ions from monovalent salts. All ions are allowed to penetrate the charged spheres, which can also penetrate one another under compression. The particles are also penetrable to the underlying polar solvent, and in most cases become swollen due to the usual hydrophilic nature of the entangled polymers. As a result, effects from dielectric mismatch are usually negligible, even in the presence of AC fields. This model system can be used as a standard representation of many soft-matter systems composed of dense polymers in solution, such as ionic microgels and spherical polymer brushes. 
Accordingly, we assign to
each microgel a monopole charge density 
$\rho_M({\bm r}) = 3Zq/(4\pi a^3)\Theta(a - r)$, 
such that 
\begin{equation}
\int_{B(a)} \rho_M({\bm r})\dd^3 r = Zq,
\label{eq:monopole}
\end{equation}
where $q$ is the elementary charge and $B(a)$ is the ball of radius $a$.
The charge monopoles interact via screened
Coulomb potentials and contribute to the 
overall microgel-microgel interactions, as 
will be demonstrated below. 

\begin{figure}[h!]
    \centering
    \includegraphics[width=0.6\linewidth]{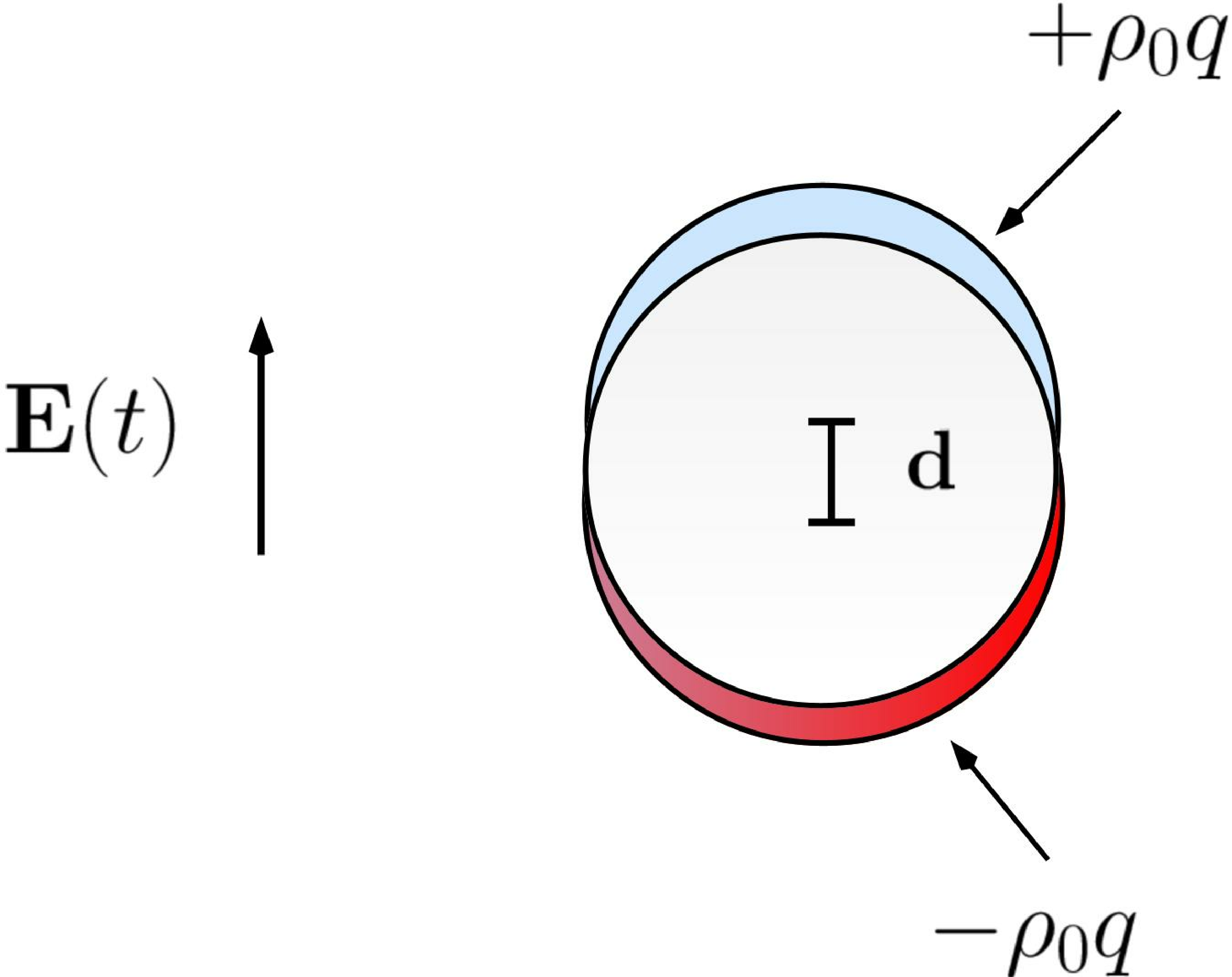}
    \caption{Sketch of the polarizing mechanism of soft particles driven by an alternating field $\bm{E}(t)$, given rise to uniform polarizations. Under the influence of this field, some fixed charges of magnitude $\rho_0$ undergo rigid displacements of amplitude $d$ along the instantaneous field direction. Such a charge displacement leads to the building-up of net charges at the particle boundaries, bearing opposite signs along the field direction.}
    \label{fig:fig2}
\end{figure}

Under the influence of AC fields of sufficient high frequencies, dielectric spectroscopy studies predict the presence of oscillating modes which couples directly to the internal, backbone charges~\cite{Moh16}. The induced polarized charges produce an oscillating dipole field which modifies the static, time averaged ionic distributions. Such anisotropic ionic profiles are characterized by a screening length that depends on the ionic mobility and driven frequencies, and measures the ionic ability of screening the dipole field in the characteristic time scales of oscillating backbone charges.
To account for field-induced polarization, 
we model
each microgel as a superposition of the net
monopole-charged colloid mentioned above and a
polarizable, penetrable, {\textit{neutral}} colloid of radius $a$. The latter has zero charge but
it consists of the superposition of two
oppositely distributions of inner charges, 
modeled by two opposite charge densities
\begin{equation}
\pm q\rho_0({\bm r}) = \pm q\rho_0\Theta(a - r),
\label{eq:plusminus}
\end{equation}
where $\Theta(x)$ represents the unit step function. As shown 
in Fig.~\ref{fig:fig2}, an external field displaces
the centers of charge of these two distributions by a vector ${\bm d}$, giving
rise to a time-dependent polarization $\bm{P}(t)$ inside the spheres, which points at the instantaneous field direction and oscillates in phase with the external field. In this picture, the instantaneous charge density of bound charges reads as
\begin{equation}
  \rho_p(\bm{r},t) = q\rho_0(\bm{r}+d\bm{\hat{n}}(t)) - q\rho_0(\bm{r}) \cong qd\nabla\rho_0(\bm{r})\cdot\bm{\hat{n}}(t),
  \label{rho1}
\end{equation}
where $\bm{\hat{n}}(t)$ is the unit vector pointing at the field direction, and $d\ll a$ is a small charge displacement promoted by the driven field. The situation is depicted in Fig.~\ref{fig:fig2}. It is clear from Eq. (\ref{rho1}) that the inner charge now entails a polarization (bound) charge distribution $\rho_p(\bm{r},t)=d\nabla\rho_0(\bm{r})\cdot\bm{\hat{n}}(t)$, in addition to the static, monopole charges $\rho_M(\bm{r})$. Making now use of Eq.~(\ref{eq:plusminus}), the resulting bound charges of Eq.~(\ref{rho1}) are evaluated as
\begin{equation}
    \rho_p(\bm{r},t)=-\dfrac{\dd}{\dd r}\left[q\rho_0\Theta(a-r)\right](\bm{\hat{e}}_r\cdot\bm{d}(t))=q\rho_0d\delta(r-a)\cos\theta(t),
    \label{rhop1}
\end{equation}
where $\theta(t)$ is the instantaneous angle between the radial and field vectors. Since ${\rho}_p(\bm{r},t)=qd\nabla\rho_0(\bm{r})\cdot\bm{\hat{n}}(t)$, it follows that the induced polarization can be written as
\begin{equation}
  \bm{P}({\bm r},t)=-q\rho_0(\bm{r})\bm{d}(t),
  \label{P1}
\end{equation}
where $\bm{d}(t)=d\bm{\hat{n}}(t)$. The analogy with the ideal case of point dipoles is rather clear: the field induces the formation of point-like dipoles of magnitude $\dd\bm{p}=-(\dd q)\bm{d}$ over the spherical core, attempting to screen the external field. 

The framework of the preceding
Section can now be used to compute the effective dipole interactions between two microgels 
in a solution that contains counterions and salt,
which screen the bare electrostatics. The Green's function $G(|\bm{r}'-\bm{r}_1|)$ now satisfies the equation 
[cf.~Eq.~(\ref{eq:laplace})]:
\begin{equation}
\nabla_1^2 G(|\bm{r}'-\bm{r}_1|) 
- \kappa_d^2 G(|\bm{r}'-\bm{r}_1|) = -\frac{4\pi}{\varepsilon}\delta(\bm{r}'-\bm{r}_1),
\label{eq:debye}
\end{equation}
where $\kappa_d$ is the inverse Debye screening length for the dipole static interactions. The solution which is continuous and bound everywhere reads $G(R)={e^{-\kappa_d R}}/{R}$. Explicit expressions for the coefficients $u_{D0}(r)$ and $u_{D2}(r)$ follow by combining the preceding 
results. In particular,
these coefficients in regions of particle overlaps, $r<2a$, read as:
\begin{eqnarray}
\nonumber
& \beta u_{D0}(r)  =  -\left(\frac{3\tau}{\kappa_d a}\right)^{2}\frac{\lambda_{\mathrm{B}}}{6a} \dfrac{e^{-\kappa_d
  a}}{\kappa_d r}\bigg\{(\kappa_d a+1)\bigg[\bigg(\frac{r^2}{2a^2}-1\bigg)\sinh(\kappa_d a)\\
  \nonumber
&-\sinh(\kappa_d(r-a))
+\frac{1}{\kappa_d a}[\cosh(\kappa_d a)-\cosh(\kappa_d(r-a))]\bigg]\\
&+F(\kappa_d a)\bigg[\frac{r^2}{2a^2}-\frac{1}{\kappa_d a}-1+\bigg(1+\frac{1}{\kappa_d a}\bigg)e^{-\kappa_d r}\bigg]\bigg\}, 
\label{udd11:a}.
\end{eqnarray}
and
\begin{eqnarray}
\nonumber
& \beta u_{D2}(r)  =  -\left(\frac{3\tau}{\kappa_d a}\right)^{2}\frac{\lambda_{\mathrm{B}}}{a}\frac{ e^{-\kappa_d
  a}}{\kappa_d^{2}r^{2}}\bigg\{(\kappa_d a+1)\bigg[\bigg(\frac{r-a}{\kappa_d ra}-\frac{\kappa_d r}{3}\bigg)\\
\nonumber
&\times\sinh(\kappa_d(r-a))+\bigg(1-\frac{1}{\kappa_d^{2}ra}-\frac{r}{3a}\bigg)\cosh(\kappa_d(r-a))\\
&+\bigg(\frac{\kappa_d r}{6}+\frac{\kappa_dr^3}{24a^2}-\frac{1}{\kappa_dr}\bigg)\sinh(\kappa_da)+\bigg(\frac{1}{\kappa_d^2ra}-\frac{r}{6a}\bigg)\cosh(\kappa_d a)\bigg]
\nonumber\\
\nonumber
& +\dfrac{F(\kappa_d a)}{\kappa_da}\bigg[\frac{\kappa_d r}{6}+\frac{\kappa_d^2r^3}{24a}+\frac{\kappa_d^2ra}{6}-
\frac{\kappa_d a+1}{\kappa_d r}\\
&+\bigg(\frac{\kappa_d^{2}ra}{3}+\frac{\kappa_d r}{3}
+\frac{\kappa_d(r+a)+1}{\kappa_d r}+\kappa_d a\bigg)e^{-\kappa_d r}\bigg]\bigg\}
\label{udd11:b}. 
\end{eqnarray}
Here, $\beta=1/k_BT$ is the inverse thermal energy, 
$\lambda_{\mathrm{B}}=\beta q^2/\varepsilon$ denotes the Bjerrum length, $\kappa_d$ is the inverse Debye screening length characterizing the static dipole interactions, and $\tau\equiv P_0/qa$ is a parameter measuring the strength of the induced polarization,  averaged over a full cycle of the external,
periodic field ${\bm E}_0e^{{\mathrm i}\omega t}$.
The function $F(x)$ above is defined as $F(x)\equiv x\cosh x-\sinh x$. 

Beyond particle overlap, $r>2a$, these coefficients take the form:
\begin{equation}
\beta u_{D0}(r)  =  -\left(\frac{3\tau F(\kappa_d a)}{\kappa_d a}\right)^{2}\frac{\lambda_{\mathrm{B}}}{3}\frac{e^{-\kappa_d
    r}}{r}, 
    \label{udd12:a}
\end{equation}    
and    
\begin{equation}    
\beta u_{D2}(r)  =  -\left(\frac{3\tau F(\kappa_d a)}{\kappa_d a}\right)^{2}\frac{2\lambda_{\mathrm{B}}}{3}\frac{e^{-\kappa_d
    r}}{\kappa_d^2 r^3}(\kappa_d r +1)\left(3+\frac{\kappa_d^2r^2}{(\kappa_d
    r+1)}\right).
\label{udd12:b}
\end{equation}
These coefficients can be used in Eq. (\ref{udd10}) to investigate the structural features of these systems in the case of elliptical fields of varying $\alpha$, interpolating between those of linear and circular light, investigated in previous studies. Apart from the induced dipole interactions, the coarse-graining pair potential also comprises an isotropic screened repulsion $u_{M}(r)$ from the monopole charges, in addition to a soft repulsion upon close contact. The soft repulsion arises from elastic deformations from overlapping polymer layers, and can be described via the following Hertzian potential:
\begin{equation}
\beta u_{\mathrm H}(r)=\epsilon_{\mathrm H}\left(1-\frac{r}{2a}\right)^{5/2}\Theta(2a-r),
    \label{Hertz}
\end{equation}
where $\epsilon_{\mathrm H}$ is a parameter measuring the strength of soft repulsions. Note that this interaction vanishes at non-overlapping distances. 

Finally, we add the isotropic, monopole-monopole 
repulsive contribution $u_M(r)$ to the total effective
interaction.
For regions of overlap, $r<2a$, the expression reads
\begin{equation}
\begin{split}
\beta u_{M}(r) &=\frac{9Z^{2}\lambda_B}{2\kappa^6
  a^6 r}\bigg[\gamma(r)-e^{-\kappa a}(\kappa a +1)[\kappa a
  \sinh(\kappa(r-a))\\
  &+\cosh(\kappa(r-a))-e^{-\kappa r}F(\kappa
  a)]\bigg]\hspace{1cm}(r<2a),
\end{split}
\label{umm1}
\end{equation}
where $\kappa$ now stands for the inverse Debye screening length for monopole interactions, and is given by $\kappa=\sqrt{4\pi\lambda_{\mathrm{B}}(c_++c_-)}$, where $c_{\pm}$ are the bulk concentrations of cations and anions. The function $\gamma(r)$ above is defined as:
\begin{eqnarray}
\nonumber
\gamma(r) &=&  \kappa^4
\bigg[\left(\frac{r^2-a^2}{4}\right)(r^2-2ra) + \frac{r}{3}(a^3-(r-a)^3)-\frac{a^4-(r-a)^4}{8}\bigg]
\\ 
&+&\frac{\kappa^2}{2}(r^2-2a^2)+1.
\label{gamma}
\end{eqnarray}
At regions beyond particle overlapping, ($r>2a$), this pair potential takes the form
\begin{equation}
\beta u_{M}(r)=\lambda_B\left(\frac{3Z F(\kappa a)}{\kappa^2 a^2}\right)^{2}\frac{e^{-\kappa r}}{r},\qquad(r>2a).
\label{umm2}
\end{equation} 
Combined with Eq.~(\ref{udd10}), Eqs.~(\ref{udd11:a}), (\ref{udd11:b}), (\ref{udd12:a}), and (\ref{udd12:b}) fully specify the interaction between ionic microgels of polarization $P_0$, induced by elliptically polarized fields whose anisotropy is measured by the parameter $\alpha$ in Eq. (\ref{E1}), with $0\le\alpha\le\pi/4$.  Together with the monopole repulsions of Eqs. (\ref{Hertz}), (\ref{umm1}), and (\ref{umm2}), these interactions can be used to study particle aggregation under different degrees of dipole strengths and field anisotropies.

\section{Results}\label{sec:results}

In order to investigate the structural properties of our soft-particle model under the influence of elliptically polarized fields of different anisotropies, the pair interactions deduced above will now be combined with Molecular Dynamics (MD) simulations. These simulations are performed in the NVT ensemble, employing periodic boundary conditions (with minimal image convention) in a cubic box containing $N=1000$ soft particles. All simulation runs were performed with a total of 
$N_{\mathrm{step}}=5\times 10^6$ time-steps, where configurations were stored  in regular intervals of $N_{\mathrm{int}}=1000$ steps. The chosen length of simulation runs is large enough to ensure the formation of static aggregates that stay far apart from one another. The temperature was kept fixed using the Andersen thermostat. 

The packing fraction $\phi$ is defined as the fraction of the overall box volume
$V = L^3$ occupied by the $N$ microgels of radius $a$:
\begin{equation}
\phi = \frac{4\pi a^3}{3}\frac{N}{L^3},
\label{eq:phi}
\end{equation}
where $L$ is the size of the cubic simulation box, chosen in such a
way that the value of the packing fraction 
 is fixed at $\phi=0.1$. The strength of the Hertzian potential is chosen to be $\epsilon_{\mathrm H}=800$, and we consider particles of size $a=0.53\,{\mathrm{\mu m}}$, while the 
 inverse Debye length $\kappa$ of equilibrium, monopole interactions is set at $\kappa a=2.79$. These parameters are chosen such as to represent typical experimental conditions of ionic microgels in the presence of AC fields, as previously reported in the literature.\cite{Riest12,Colla18} Finally, we are considering an
 aqueous solution at room temperature, resulting into a Bjerrum
 length $\lambda_{\mathrm B} = 7.2\,$\AA.

\subsection{Pair potentials}

\begin{figure}[h!]
    \centering
    \includegraphics[width = 4.9cm, height = 4.2cm]{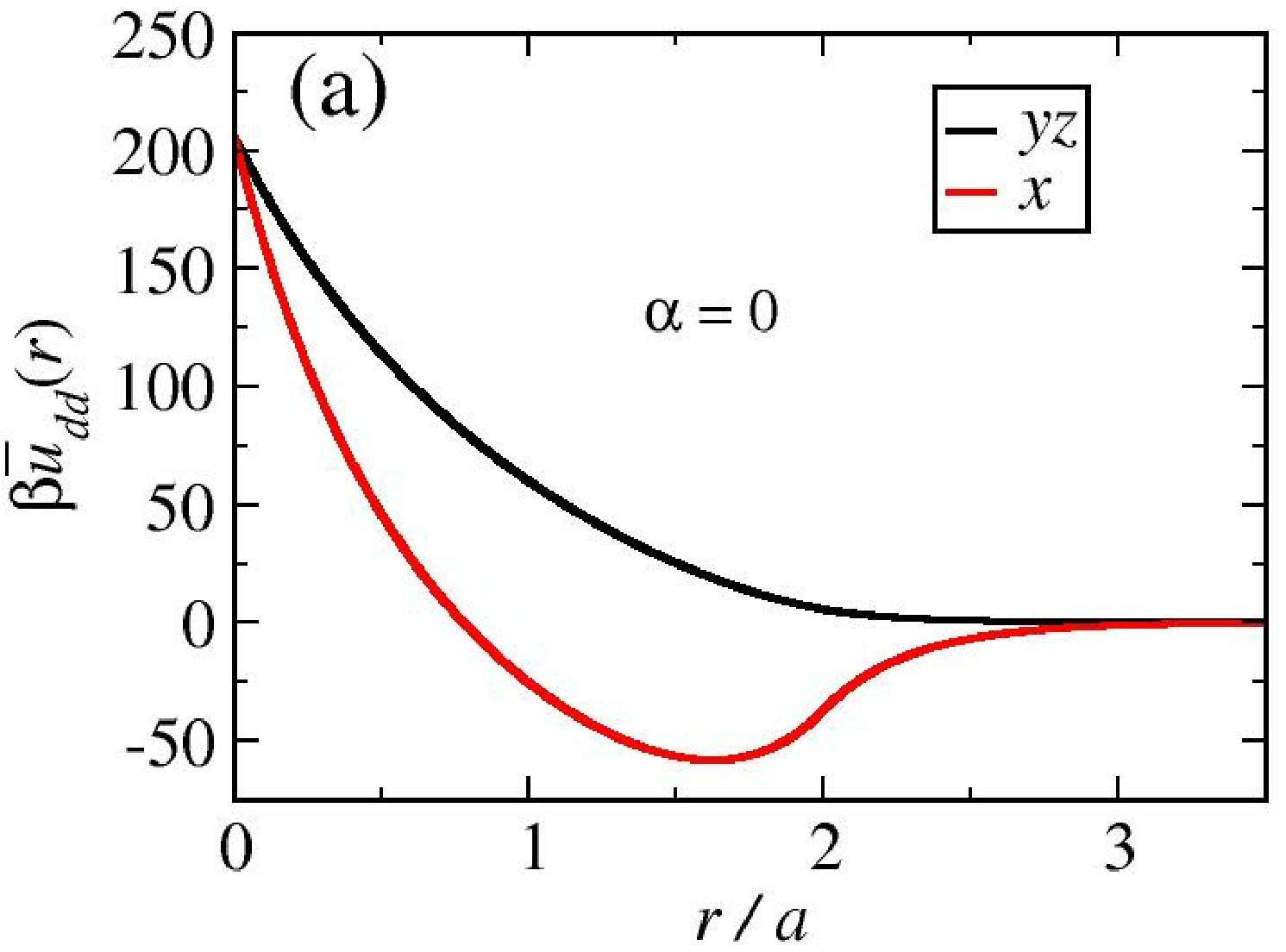}
    \includegraphics[width = 4.9cm, height = 4.2cm]{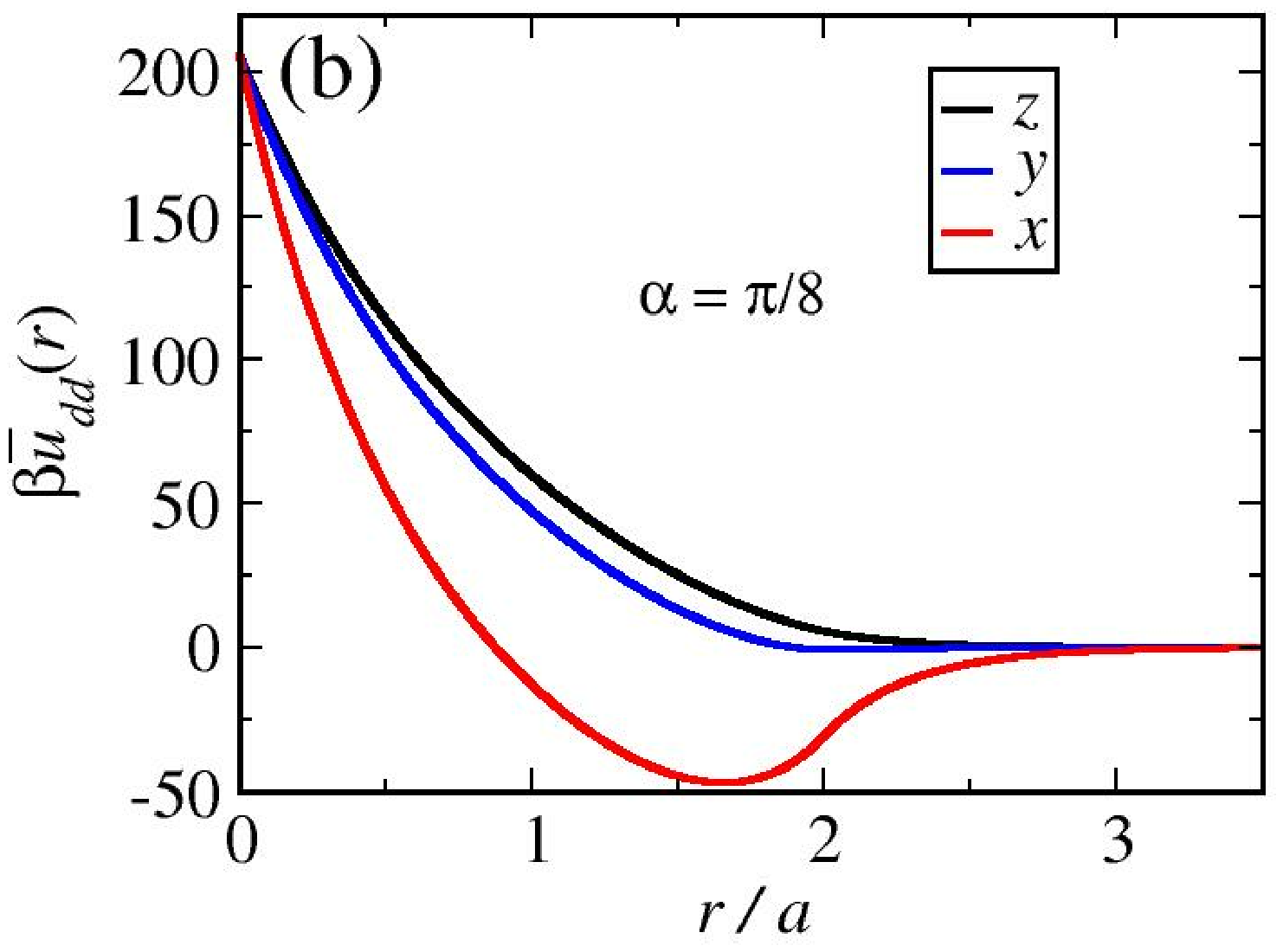}
    \includegraphics[width = 4.9cm, height = 4.2cm]{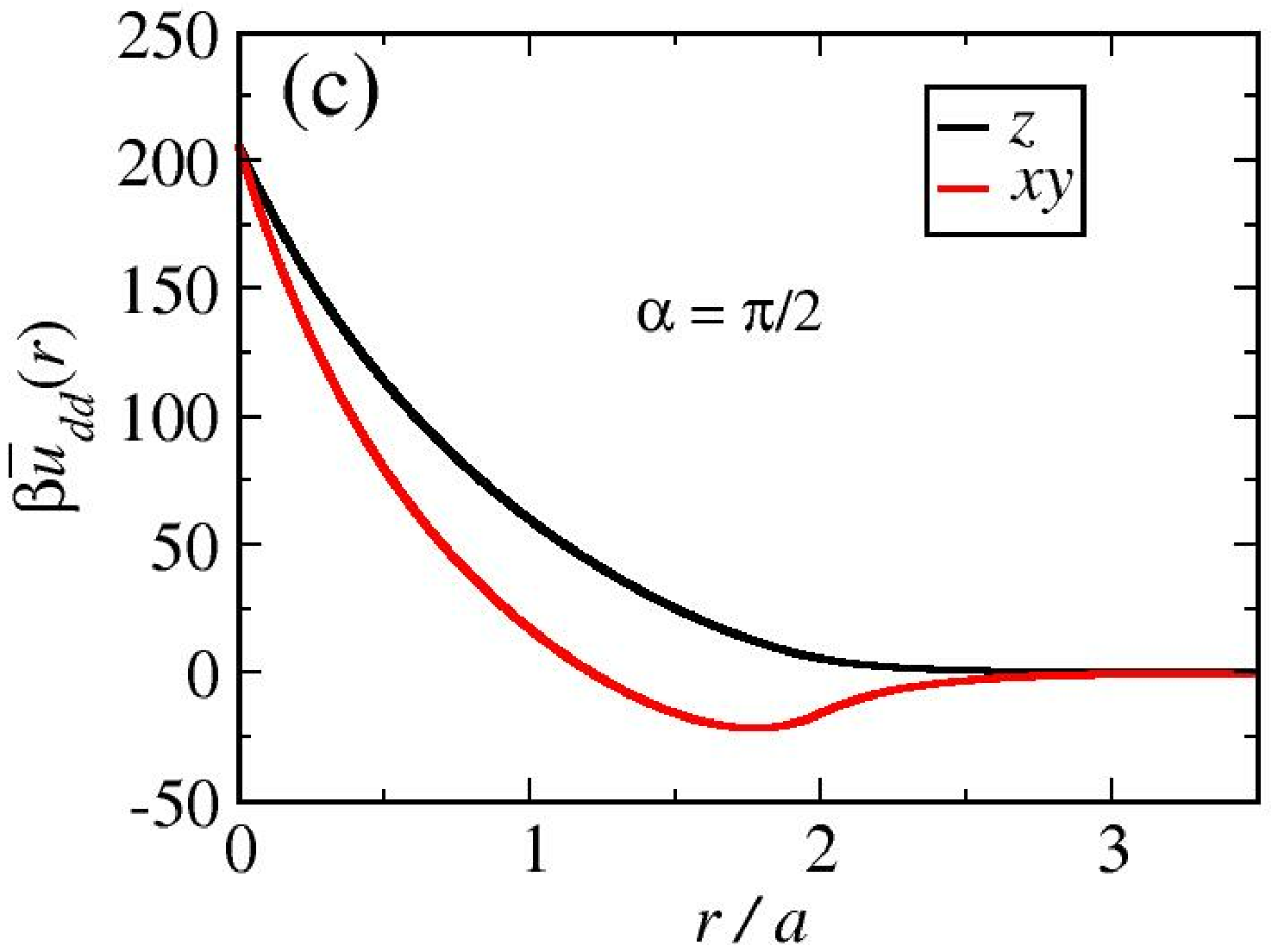}\\
    \caption{The time-averaged dipole-dipole pair potential of ionic microgels, $\bar{u}_{dd}(\bm{r})$, induced by the elliptically polarized field, Eq.\ (\ref{E1}). The reduced polarization is $\tau=P_0/(qa)=400$. In (a), the field is linearly polarized along the $x$ direction ($\alpha=0$), and the potential is isotropic along the $yz$ plane. In (b), the the elliptic polarization angle is $\alpha=\pi/8$, and the interactions are fully anisotropic, whereas in (c) the field is circularly polarized, $\alpha=\pi/4$, and the interactions are isotropic along the $xy$ plane.}
    \label{fig:fig3}
\end{figure}

We start by analyzing the dipole pair interaction at different directions, and how it is influenced by the degree of 
ellipse anisotropy (as measured by the angle $\alpha$), as well as by the dipole strength and range. Figure \ref{fig:fig3} shows the dipole potentials obtained from Eq. (\ref{udd10}), at three representative $\alpha$ values: $\alpha=0$ (left panels), corresponding to a field polarized in the $x$ direction, $\alpha=\pi/8$ (middle panels), where the polarization lies in-between linear ($\alpha = 0$) and circular ($\alpha = \pi/4$) polarizations, whereby the linear polarization
is along the $x$-axis and the circular one on the $xy$-plane. 
The range $0 \leq \alpha \leq \pi/4$ is sufficient to explore all
elliptic polarizations, since values in the domain $\pi/4 < \alpha \leq \pi/2$ can be mapped to the original one via $\alpha \to \pi/2 - \alpha$, accompanied by an 
interchange between $x$ and $y$.

A clear trend is observed in Figure \ref{fig:fig3}, in which the dipole interactions along the $x$ direction feature an attractive, short-range well, whereas the interactions along $z$ direction are purely repulsive. On the other hand, the interactions along the $y$ direction start to be less repulsive as $\alpha$ increases. Eventually, the interactions in the $y$ direction cross over from purely repulsive to partially attractive, such that in the case of circular polarizations ($\alpha=\pi/4$), the interactions in $xy$ plane become isotropic, featuring a short ranged potential well in both directions. At the intermediate polarization $\alpha=\pi/8$, the potential along the $x$ direction still exhibits a deep well of about $\sim -50~k_BT$, and the interactions in the $y$ direction become slightly less repulse in comparison to those along the $z$ direction.  

\begin{figure}[h!]
    \centering
    \includegraphics[width = 7cm, height = 5.3cm]{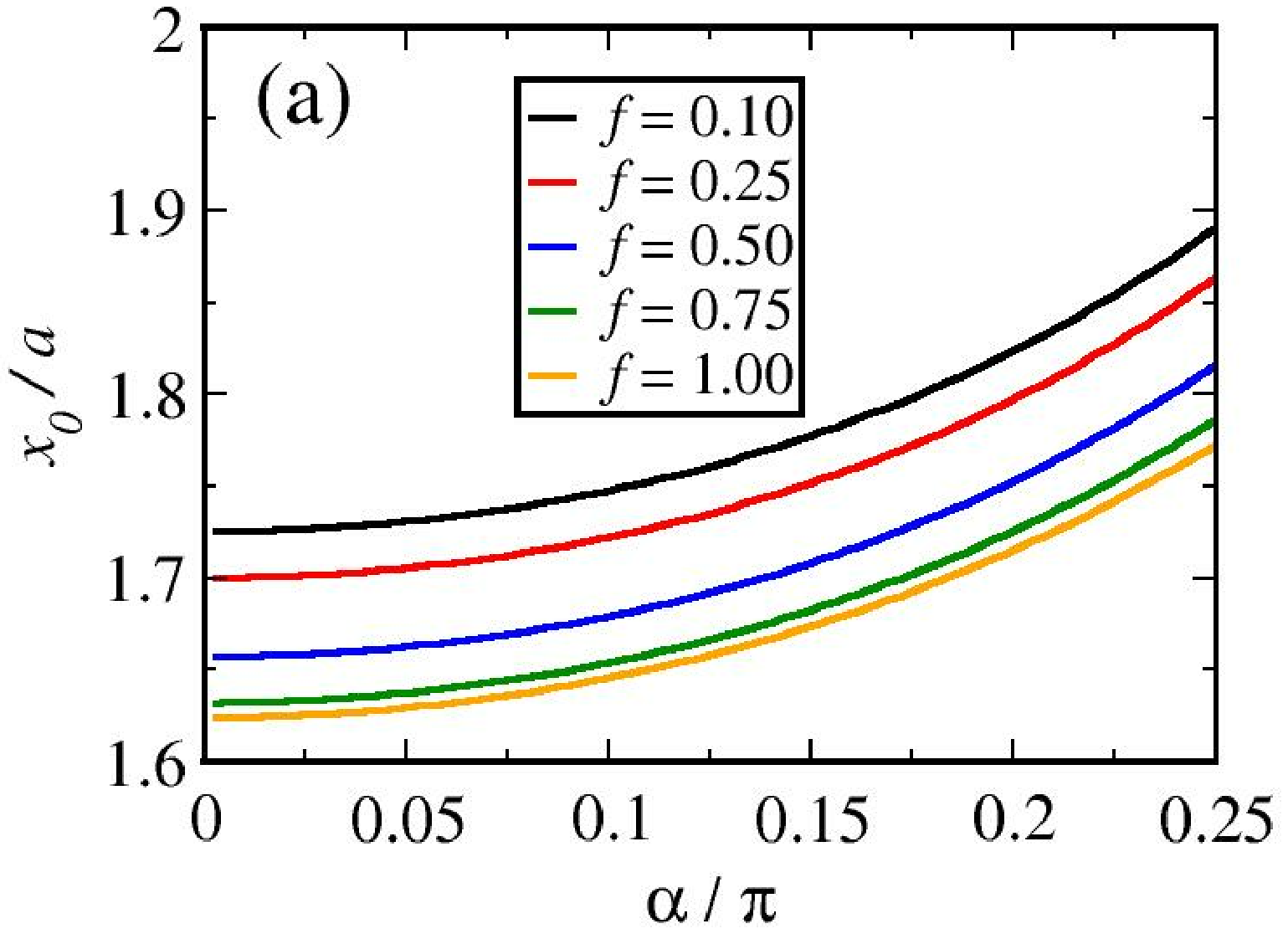}
     \includegraphics[width = 7cm, height = 5.3cm]{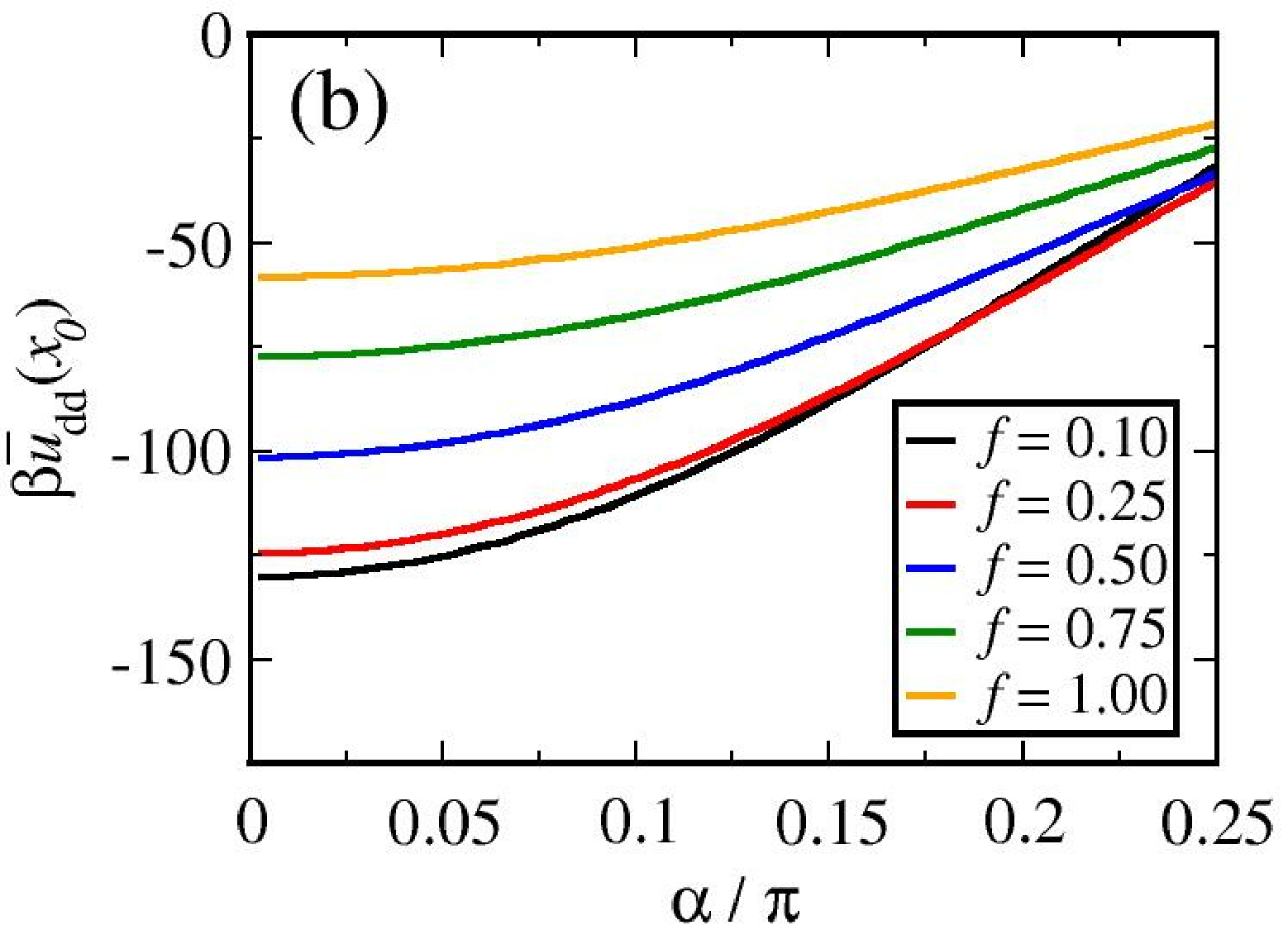}\\
    \includegraphics[width = 7cm, height = 5.3cm]{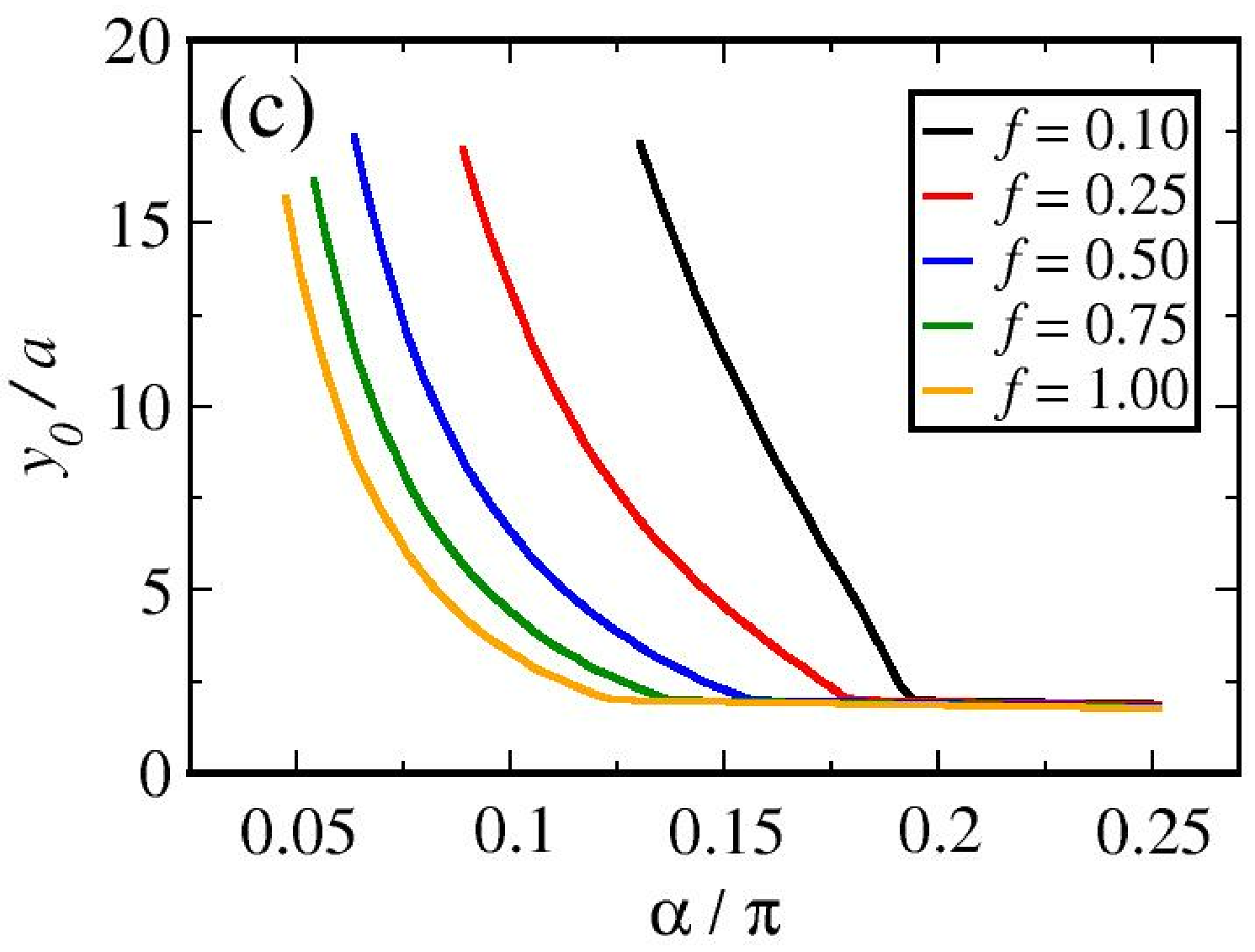}
     \includegraphics[width = 7cm, height = 5.3cm]{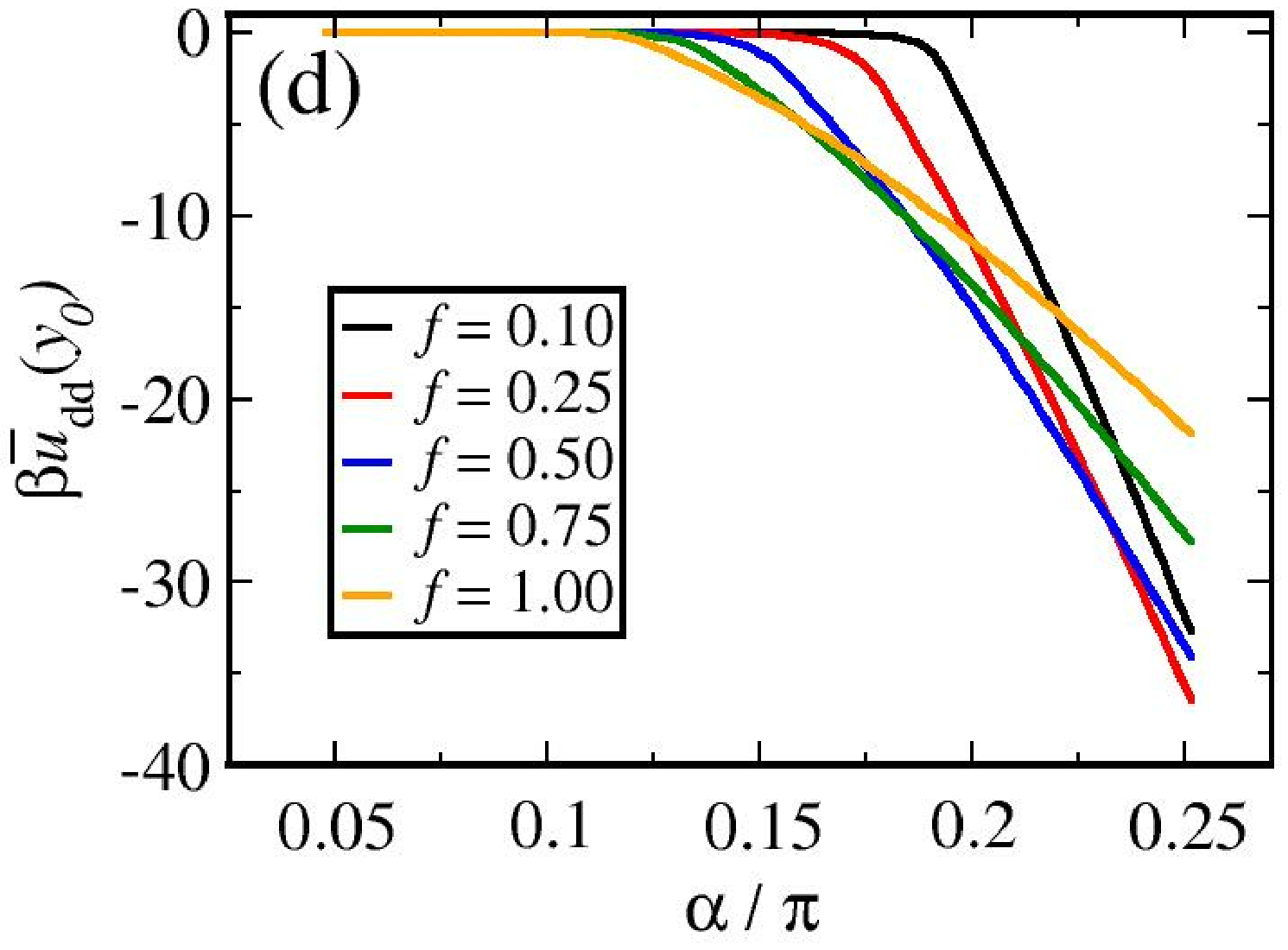}     
    \caption{Equilibrium positions (left panels) and depth (right panels) of dipole pair correlations along the $x$ (top panels) and $y$ (bottom panels) directions, as a function of polarization angles $\alpha$ for different fractions $f$ that determine the ratio between dipole and monopole inverse screening lengths. In all cases, the dimensionless dipole interaction strength is fixed at $\tau=400$.}
    \label{fig:fig4}
\end{figure}

A more detailed overview of the behavior of $\bar{u}_{dd}(\bm{r})$ upon changing the eccentricity of elliptical polarization is provided in Fig.~\ref{fig:fig4}, which shows both the position,
Fig.~\ref{fig:fig4}(a), and depth, Fig.~\ref{fig:fig4}(b), of the potential well across the $x$ and $y$ directions as the polarization angle $\alpha$ is continuously changed from linear to circular polarizations. As pointed out earlier, the dipole inverse screening length $\kappa_d$ is generally different from the 
monopole screening length $\kappa$, as the former is a static quantity depending on ionic mobilities, whereas the latter is an equilibrium property that does not depend on the coupling between driven frequency and ion diffusion. As a result, one can consider that only a fraction of the total amount of available counterions takes part in dynamic dipole screening, such that $\kappa_d$ is always smaller than $\kappa$. We then introduce a parameter $f=\kappa_d/\kappa$ which controls the relative range between monopole and dipole interactions, with $f\le 1$. Figure \ref{fig:fig4} shows how the equilibrium positions and depths of potential minima are influenced by this parameter. As expected, the positions of mechanical equilibrium along the $x$ direction are shifted to smaller values as $f$ increases, while the absolute value of the corresponding potential wells decreases. Increasing the range of dipole interactions leads to stronger particle attractions along the $x$ direction, a trend which is more pronounced at smaller values of $\alpha$, see Fig.~\ref{fig:fig4}(b). This indicates that the range of dipole interactions plays a most prominent role in the case of chain formations,\cite{Colla18} while the planar domains across the $xy$ plane in cases of circular polarizations\cite{reich:sm:2025} should be less sensitive to this quantity. On the other hand, a stronger dipole screening leads to onset of particle attractions along the $y$ direction at smaller values of $\alpha$, as shown in Fig.~\ref{fig:fig4}(c). In all cases, the equilibrium values $y_0$ display a rapid decay towards the distance of closest approach $y_0\approx 2a$ as $\alpha$ increases. However, the polarization angles in which such short range approach along the $y$ direction takes place strongly depends on the fraction $f$. As shown in Fig.~\ref{fig:fig4}(d), the behavior is also non-monotonic with $f$. In particular, in the case of circular polarizations, the potential depths increase at small $f$, but then start to decrease in magnitude as $f$ increases. These results indicate an intriguing, non trivial, coupling between potential range and the degree of anisotropy in pair interactions.   

\begin{figure}[h!]
    \centering
    \includegraphics[width = 4.9cm, height = 4.3cm]{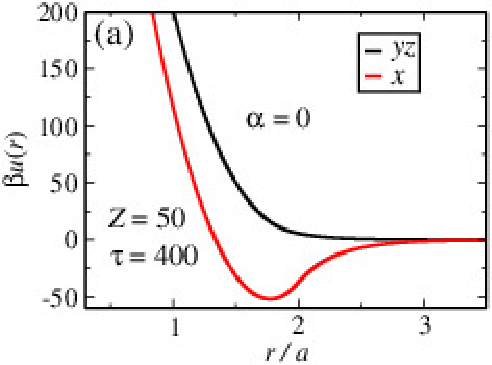}
    \includegraphics[width = 4.9cm, height = 4.3cm]{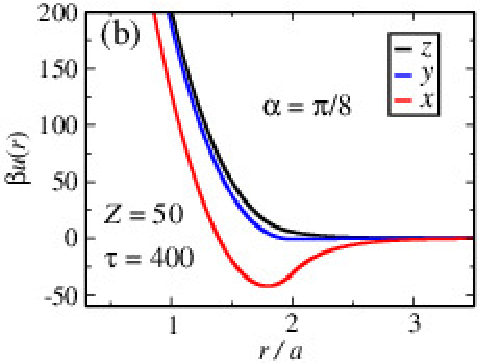}
      \includegraphics[width = 4.9cm, height = 4.3cm]{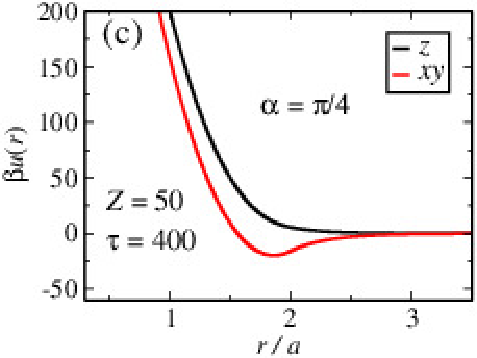}\\
     \includegraphics[width = 4.9cm, height = 4.3cm]{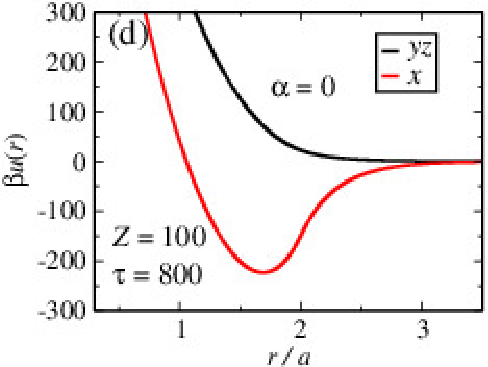}
    \includegraphics[width = 4.9cm, height = 4.3cm]{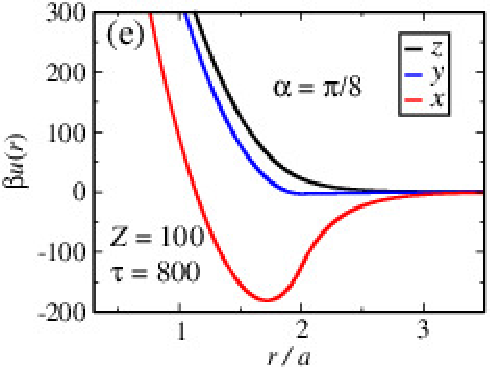}
      \includegraphics[width = 4.9cm, height = 4.3cm]{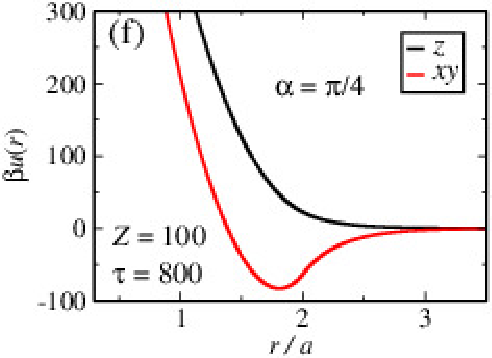}  
    \caption{Total pair potential $u(\bm{r}) = u_M(r) + \bar u_{dd}(\bm{r}) + u_{\mathrm H}(r)$
    for two representative systems. In panels (a), (b), and (c), microgels have charge $Z=50$ and polarizations $\tau=400$, whereas panels (d), (e), and (f) pertain to microgels with larger charges, $Z=100$, and polarizations $\tau=800$. In all cases, the screening lengths for monopole and dipole interactions are the same, $\kappa=\kappa_d=2.79a^{-1}$, i.e., $f = 1$.}
    \label{fig:fig5}
\end{figure}

\begin{figure}[h!]
    \centering
    \includegraphics[width = 6.5cm, height = 5cm]{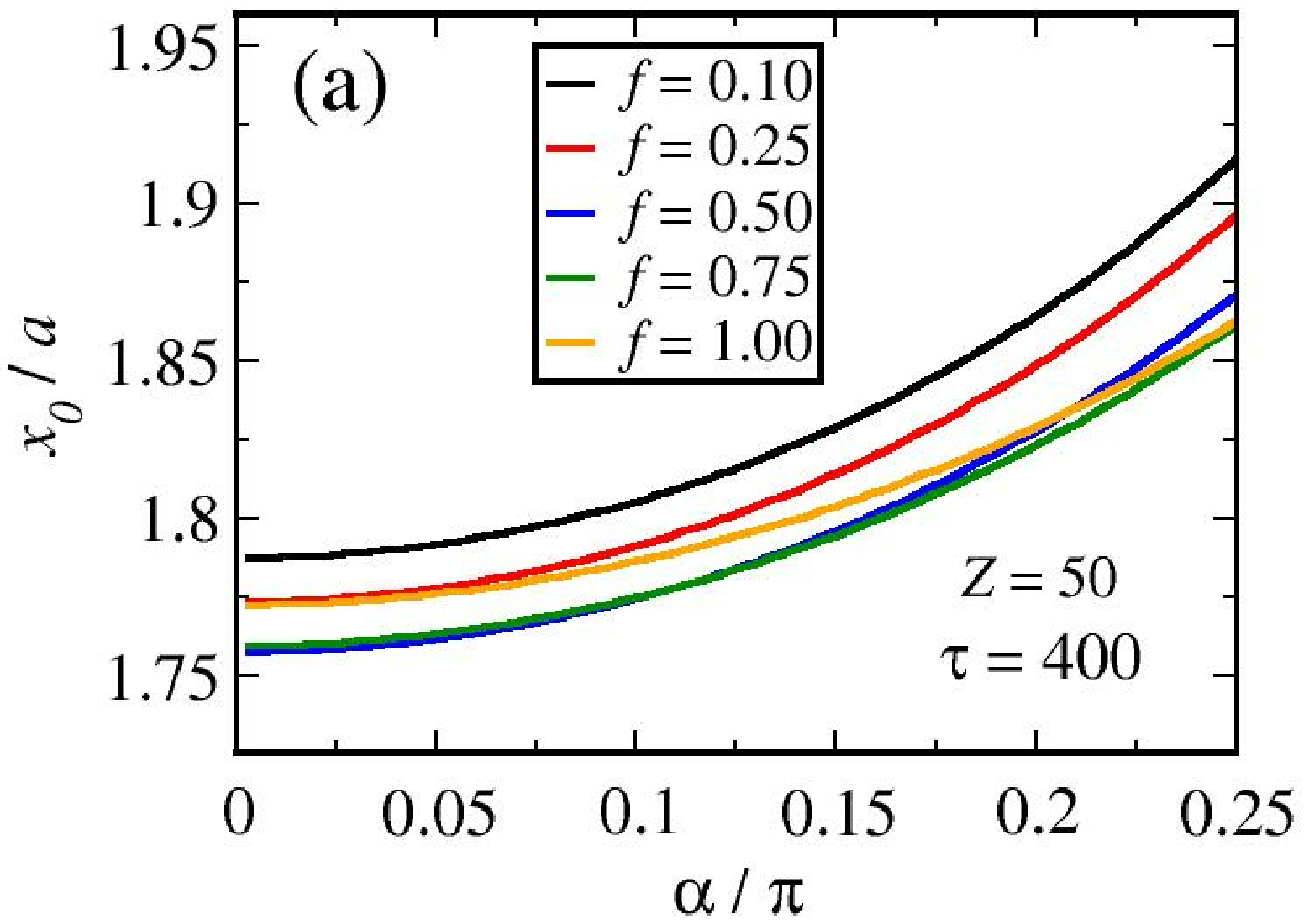}
    \includegraphics[width = 6.5cm, height = 5cm]{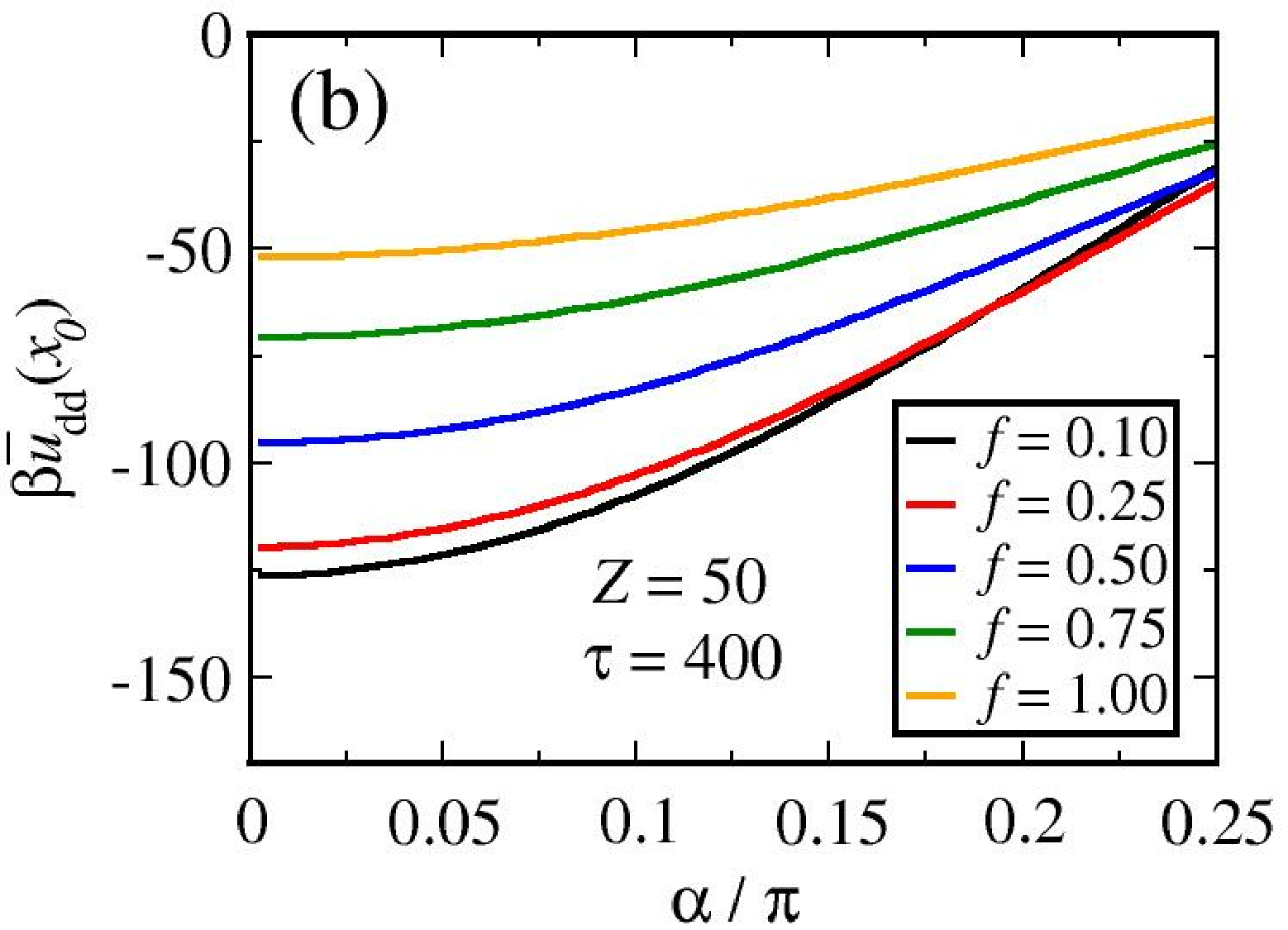}\\
     \includegraphics[width = 6.5cm, height = 5cm]{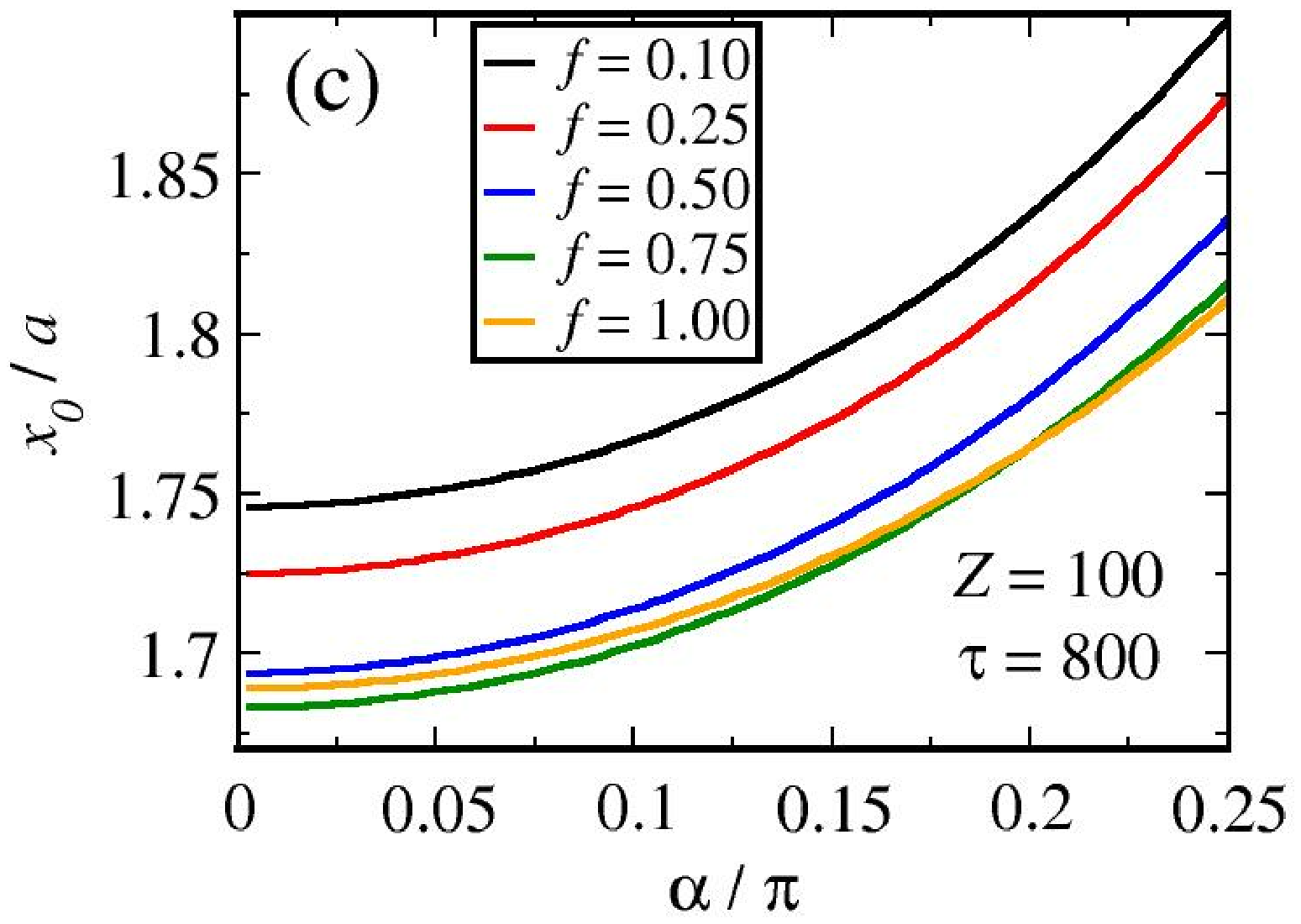}
    \includegraphics[width = 6.5cm, height = 5cm]{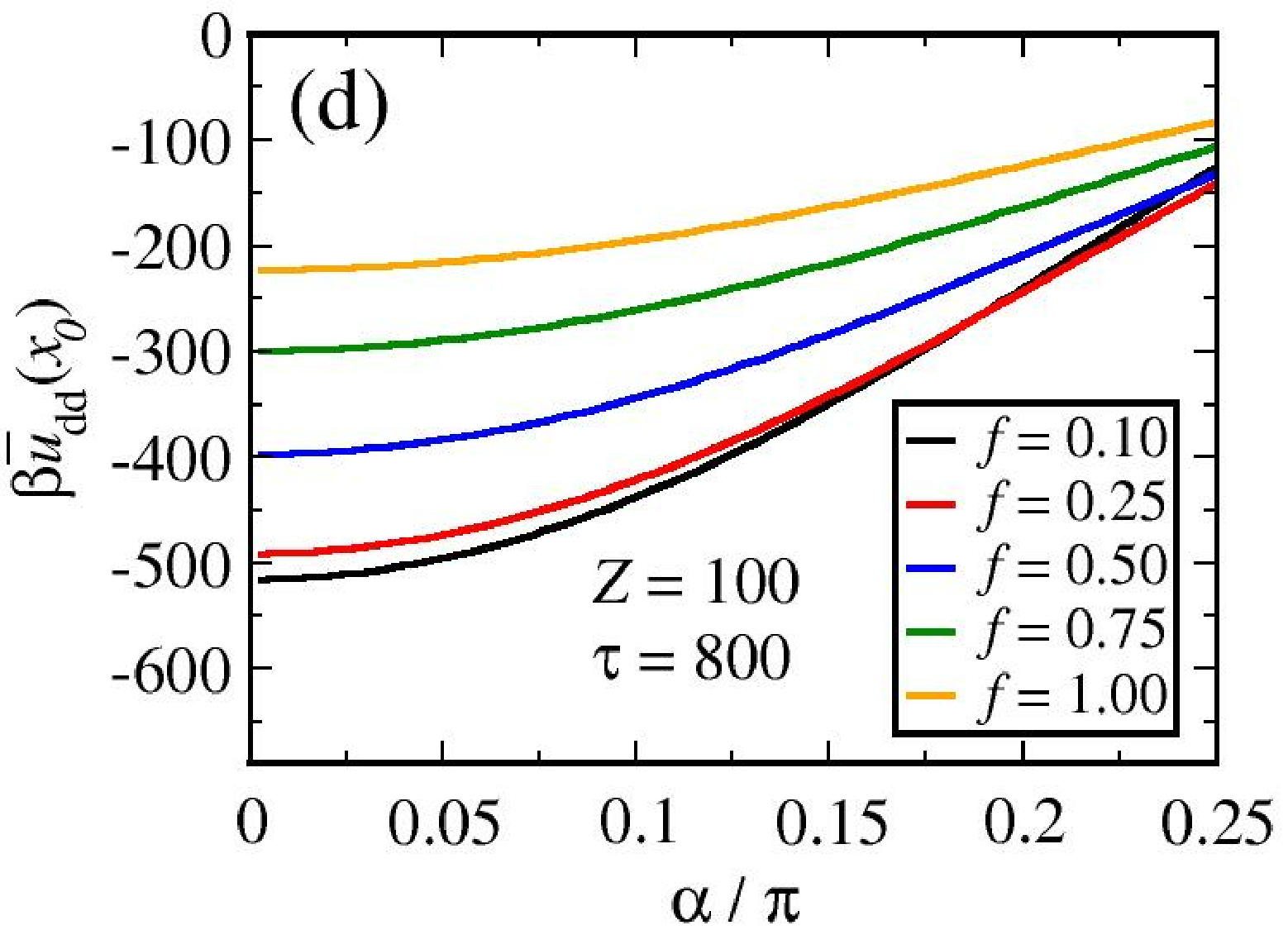}
    \caption{The $x$-positions at which the total interaction potential 
    $u(\bm{r}) = u_M(r) + \bar u_{dd}(\bm{r}) + u_{\mathrm H}(r)$ attains its minimum [panels (a) and (c)] and the corresponding values of that minimum [panels (b) and (d)] for different combinations of monopole- and dipole-strengths, expressed
    by the values of $Z$ and $\tau$, respectively.
    Results are shown for different choices of the ratio of the
    monopole- to dipole screening length, $f = \kappa_d/\kappa$.}
    \label{fig:fig6}
\end{figure}

We now proceed by investigating the interplay between isotropic repulsions and anisotropic attractions along the $x$- and $y$-directions resulting from the induced dipoles of different strengths $\tau$ and angles $\alpha$. Figure \ref{fig:fig5} shows the total pair interactions, including monopole and short-range repulsions, for the case os microgels with total charges of $Z=50$ and polarizations $\tau=400$ (top panels), as well as $Z=100$ and $\tau=800$ (lower panels), for the three representative values of $\alpha$. Here, both dipole and monopole charges are equally screened, $f=1$. Very similar behaviors are observed in both cases. Although the ratio between monopole and dipole charges is the same in both situations, the soft-particles with a large polarization display much more pronounced potential wells, in spite of their stronger monopole repulsion. The same behavior holds when the dipole and monopole interactions are not equally screened, as can be observed in Fig. \ref{fig:fig6}. This figure shows the positions (left panels) and strengths (right panels) of the potential wells along the $x$ direction for the two representative systems under consideration, at different fractions $f$ between monopole and dipole screening lengths. The overall qualitative behavior follows the same general trends as in the case of pure dipole interaction shown in Fig.~\ref{fig:fig4}. A clear distinction between different $f$ values is observed is observed in the case of small $\alpha$ (linear polarizations), while for circular polarizations ($\alpha\lesssim \pi/4$) the potential depths reduce in magnitude, and become closer to one another at different screenings. It is important to note, however, that in all cases the concomitant increase of monopole and dipole charges (keeping its proportion) leads to stronger field-induced attractions.   

\begin{figure}[h!]
    \centering
    \subfigure[]{\includegraphics[width = 3.6 cm, height = 5cm]{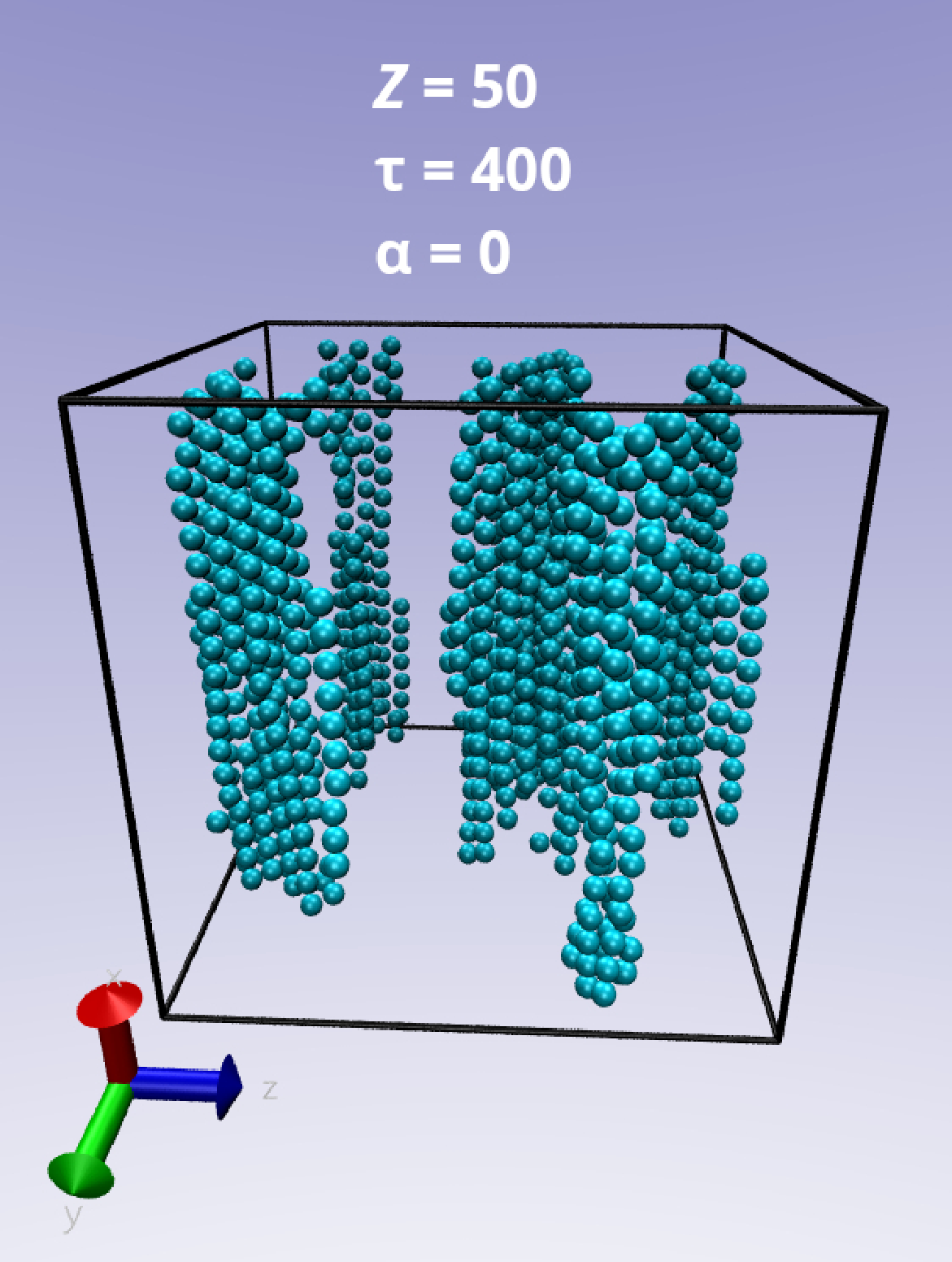}}  
    \subfigure[]{\includegraphics[width = 3.6 cm, height = 5cm]{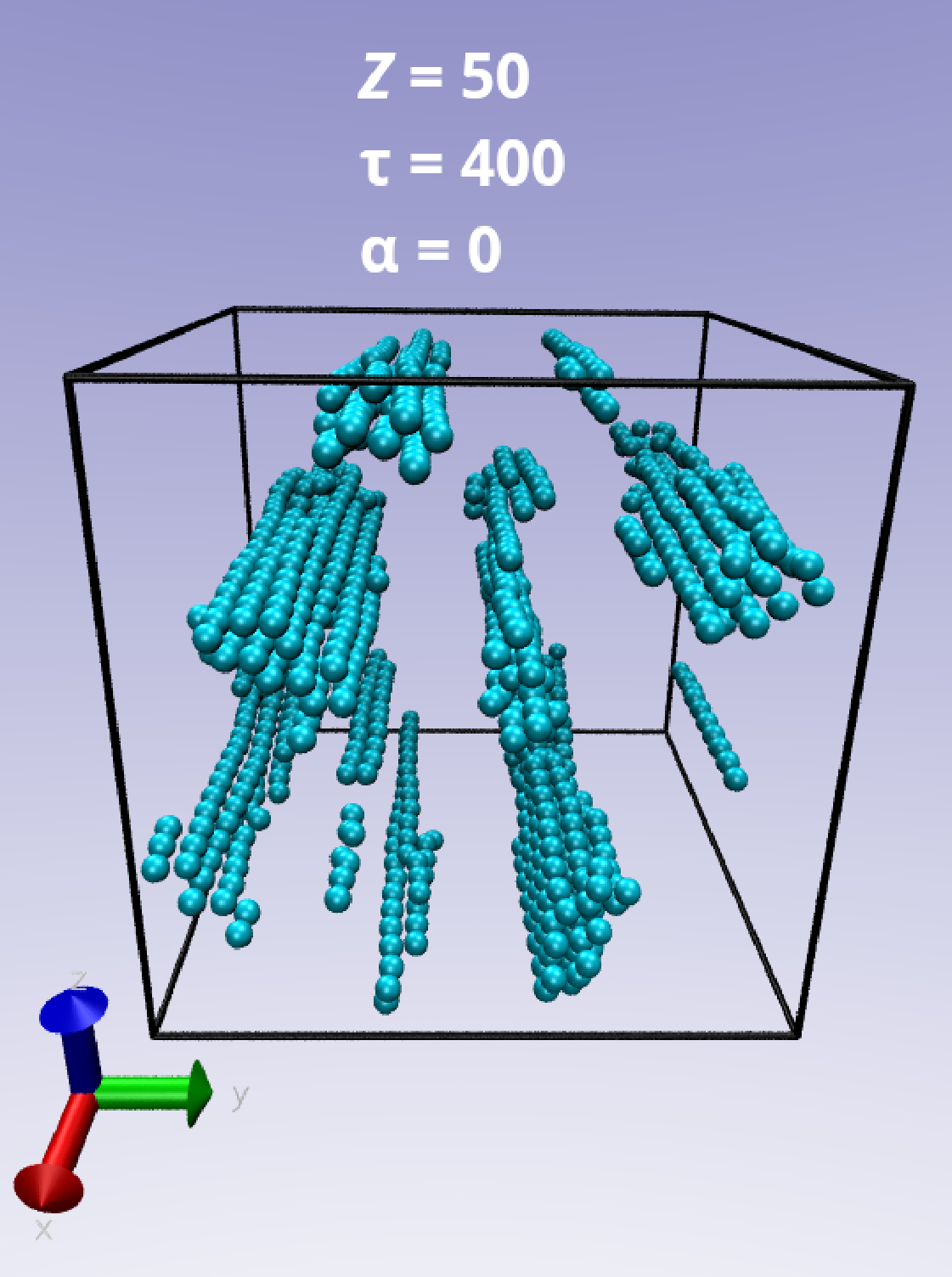}}  
    \subfigure[]{\includegraphics[width = 3.6 cm, height = 5cm]{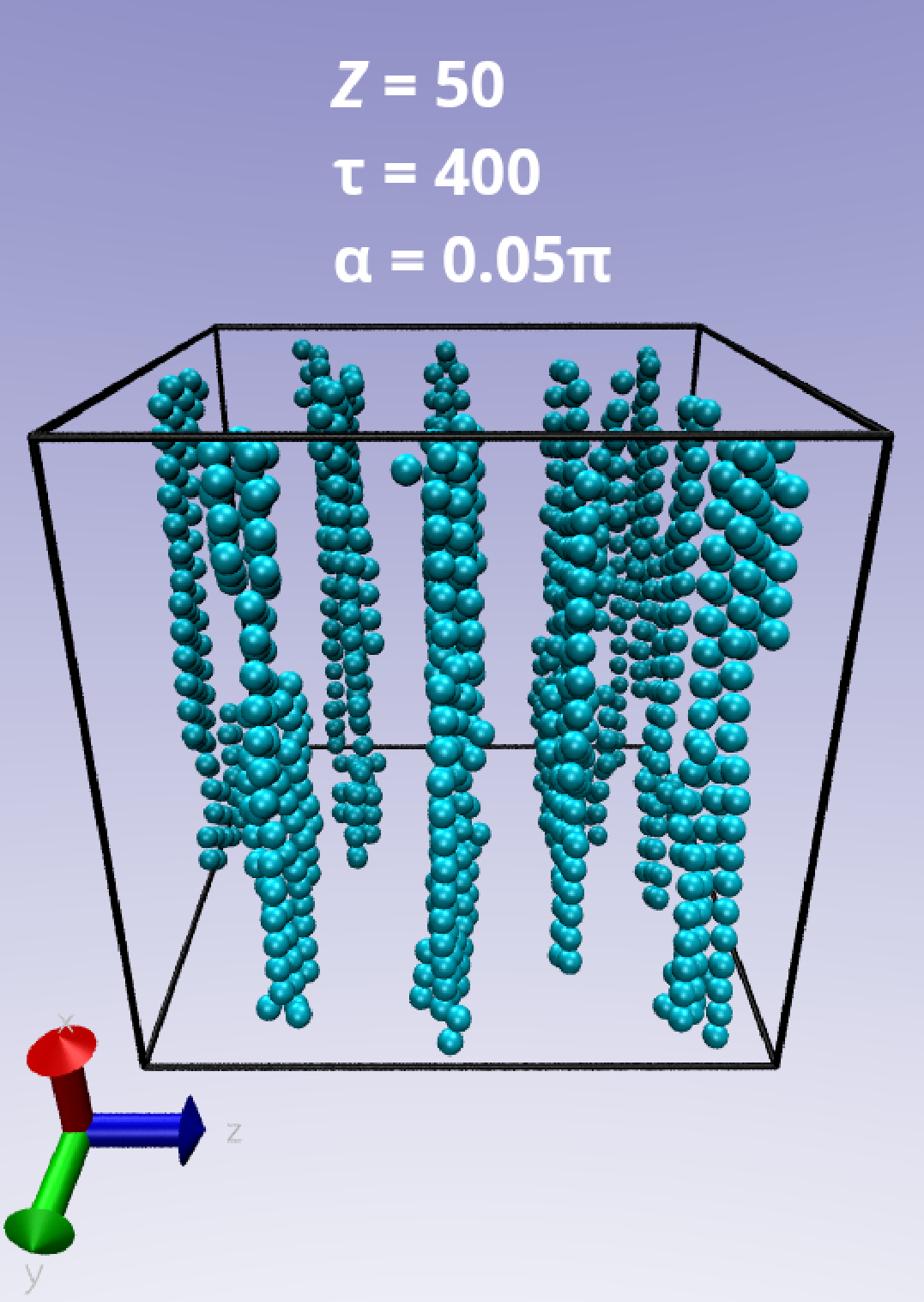}} 
    \subfigure[]{\includegraphics[width = 3.6 cm, height = 5cm]{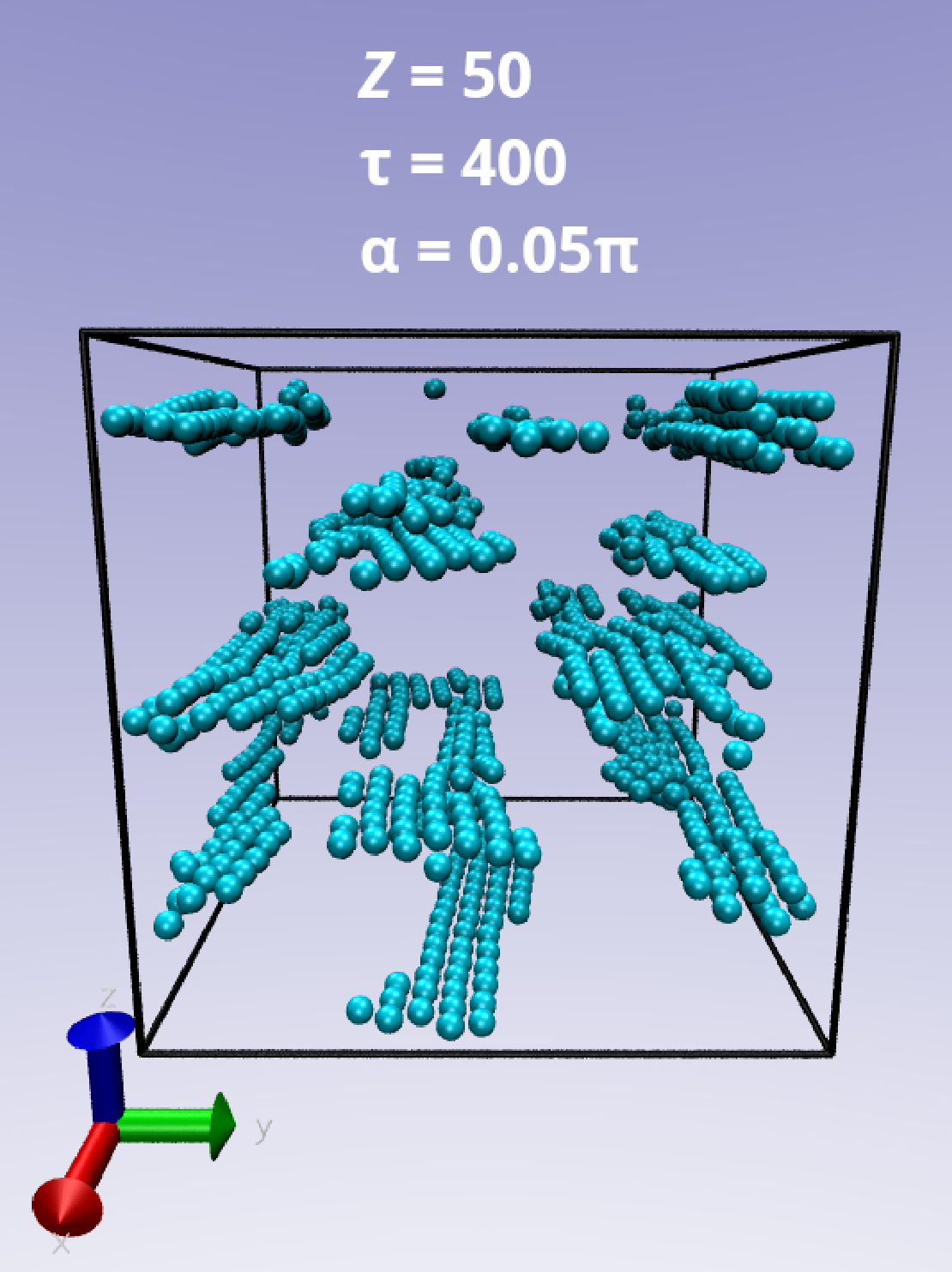}} \\
    \subfigure[]{\includegraphics[width = 3.6 cm, height = 5cm]{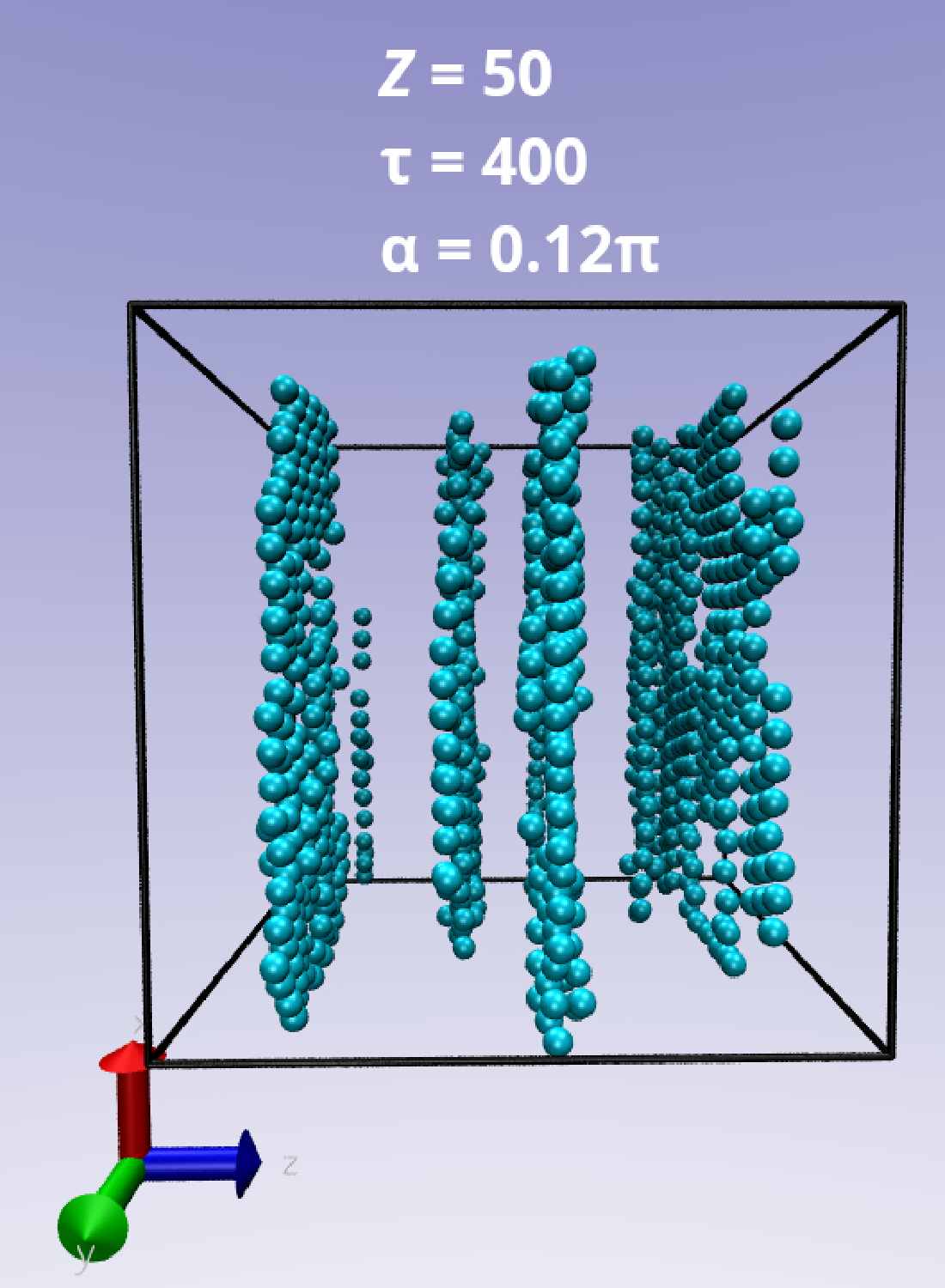}}  
    \subfigure[]{\includegraphics[width = 3.6 cm, height = 5cm]{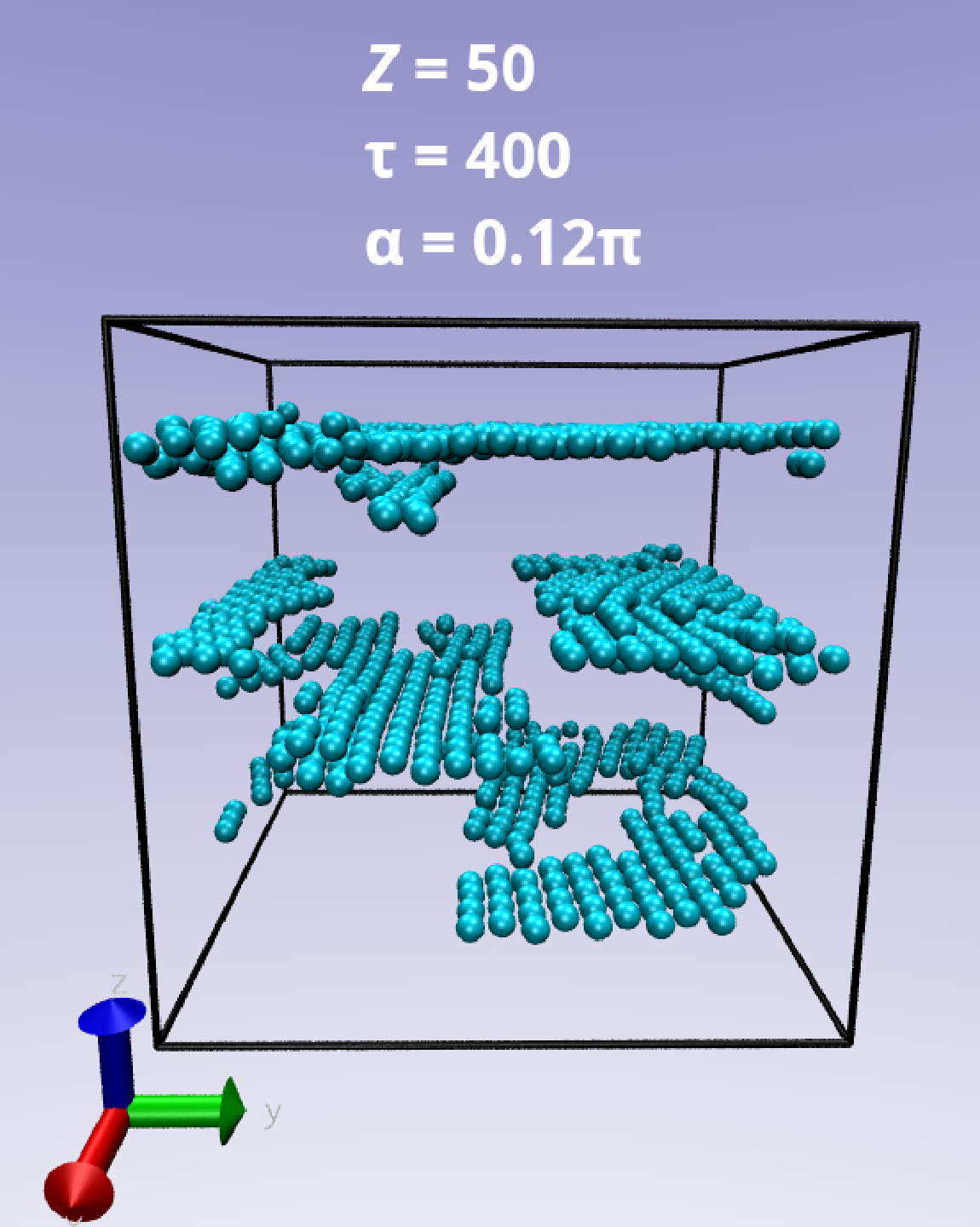}}  
    \subfigure[]{\includegraphics[width = 3.6 cm, height = 5cm]{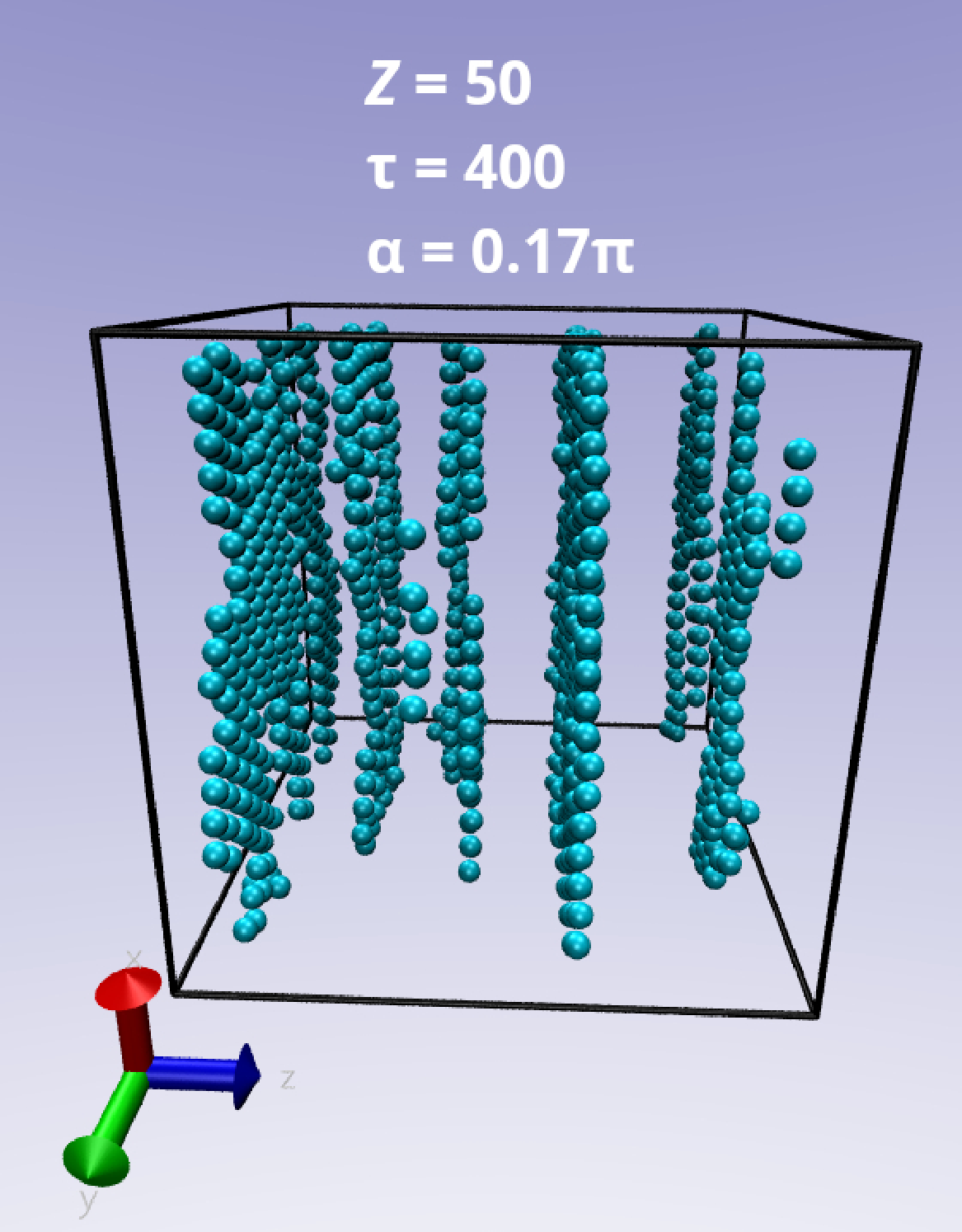}}  
    \subfigure[]{\includegraphics[width = 3.6 cm, height = 5cm]{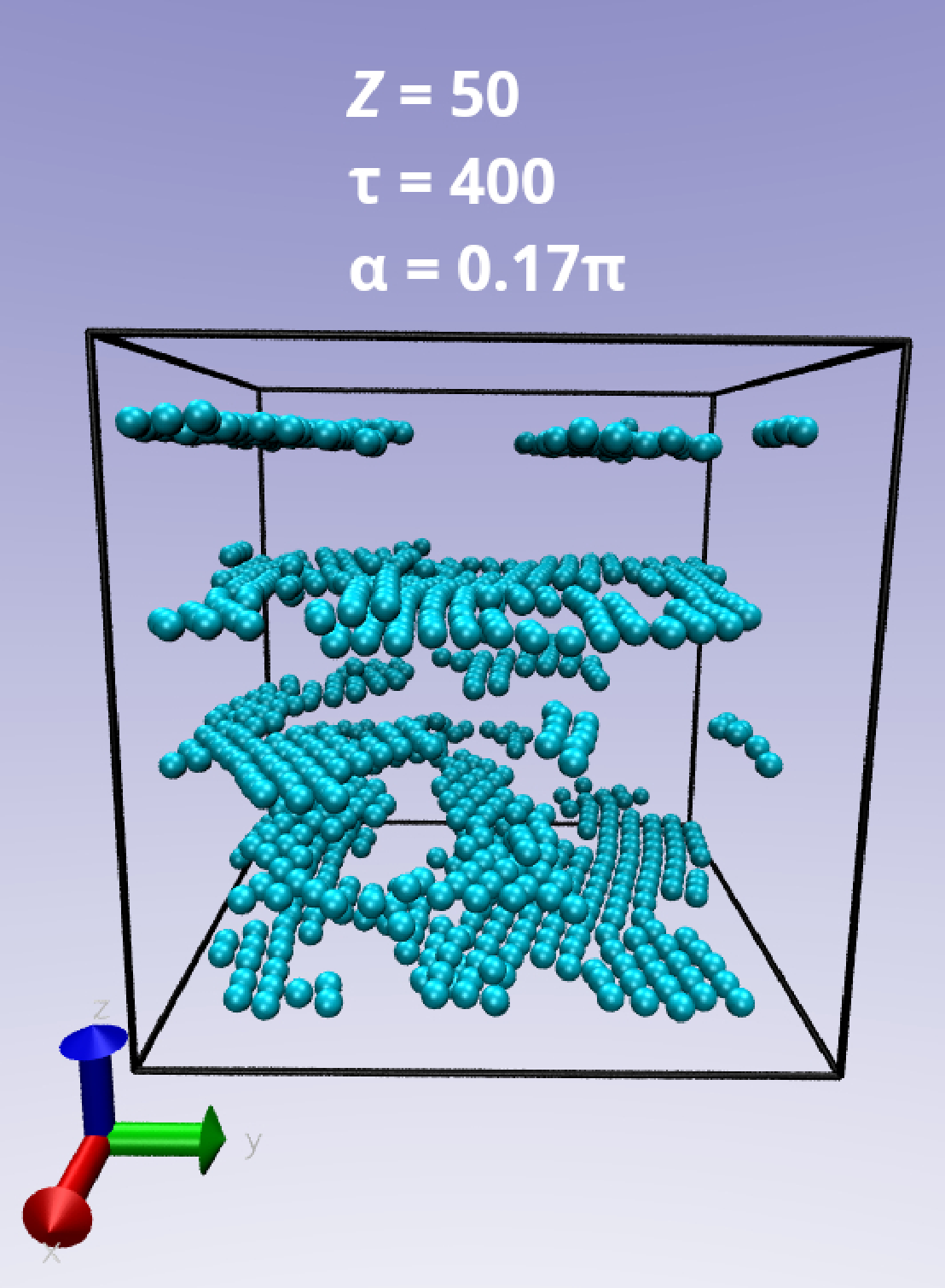}}\\  
    \subfigure[]{\includegraphics[width = 3.6 cm, height = 5cm]{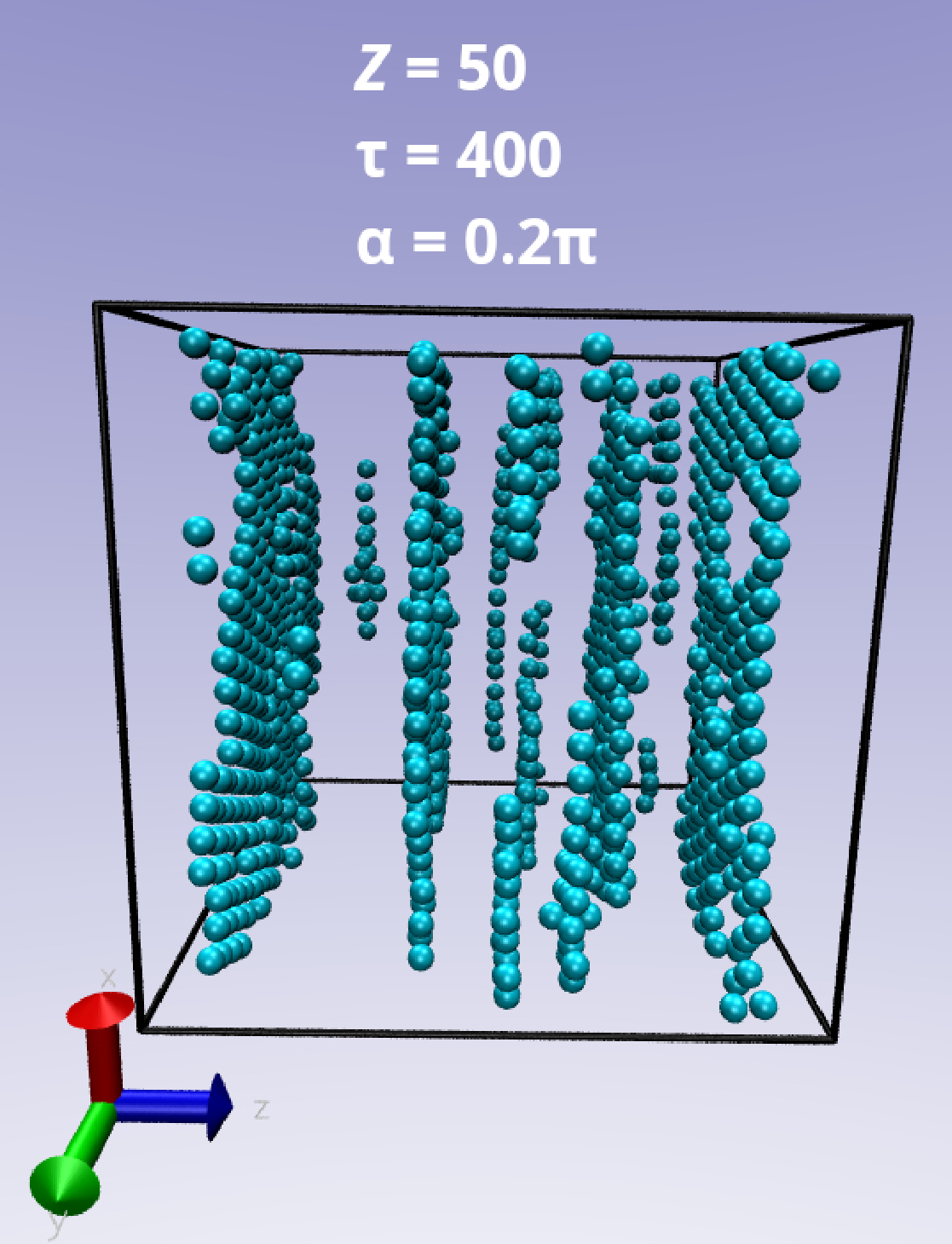}}  
    \subfigure[]{\includegraphics[width = 3.6 cm, height = 5cm]{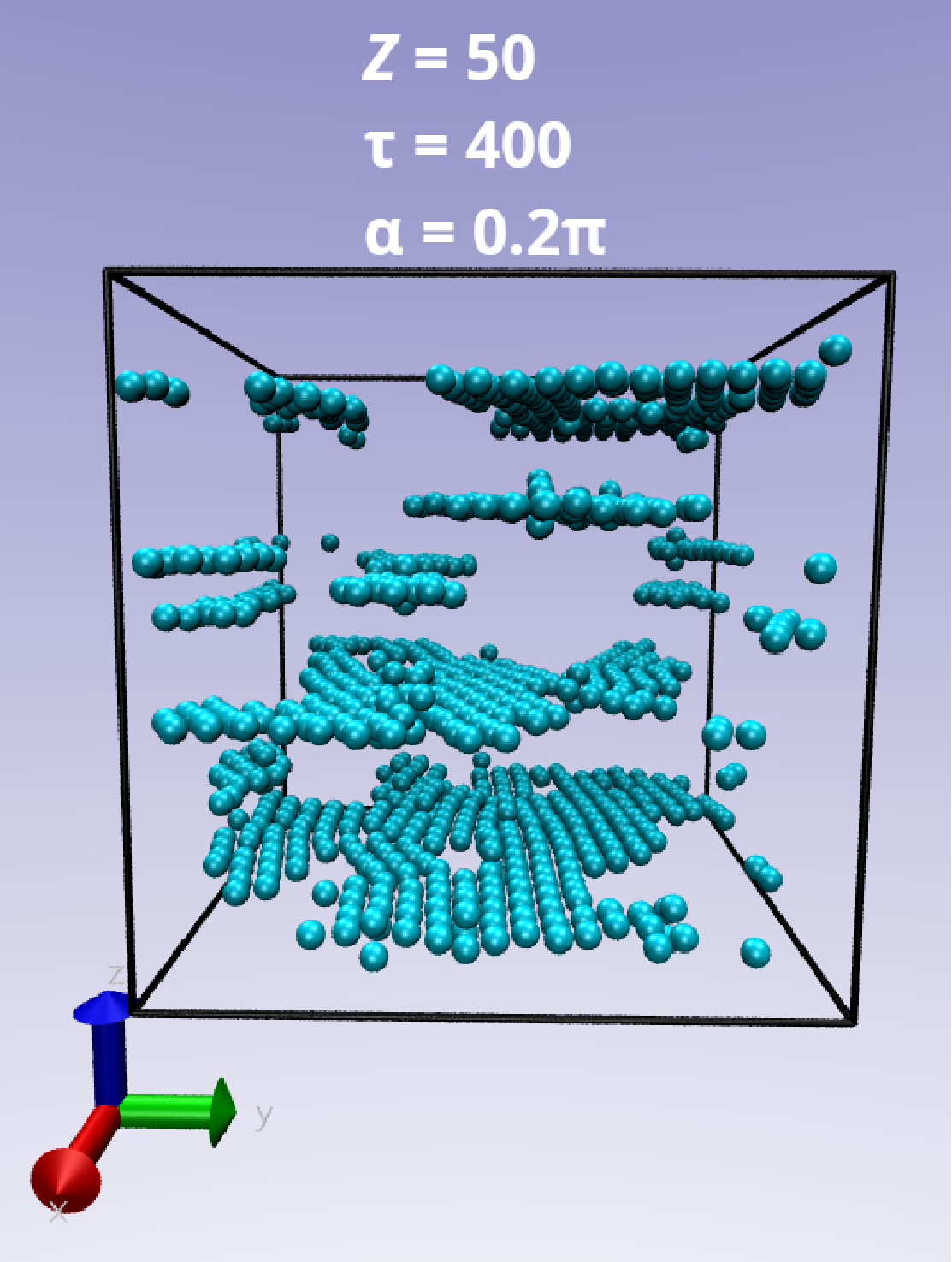}}  
    \subfigure[]{\includegraphics[width = 3.6 cm, height = 5cm]{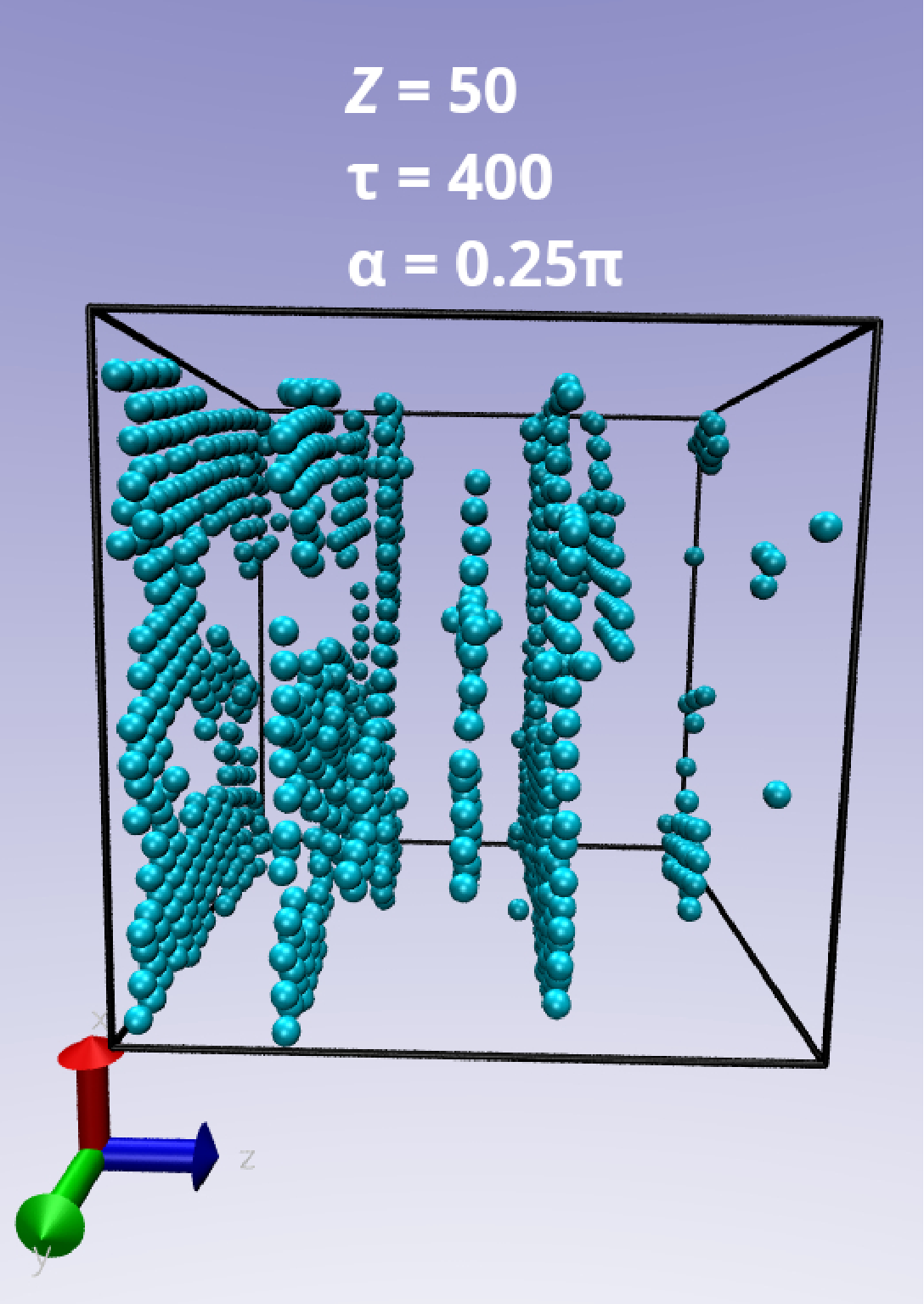}}  
     \subfigure[]{\includegraphics[width = 3.6 cm, height = 5cm]{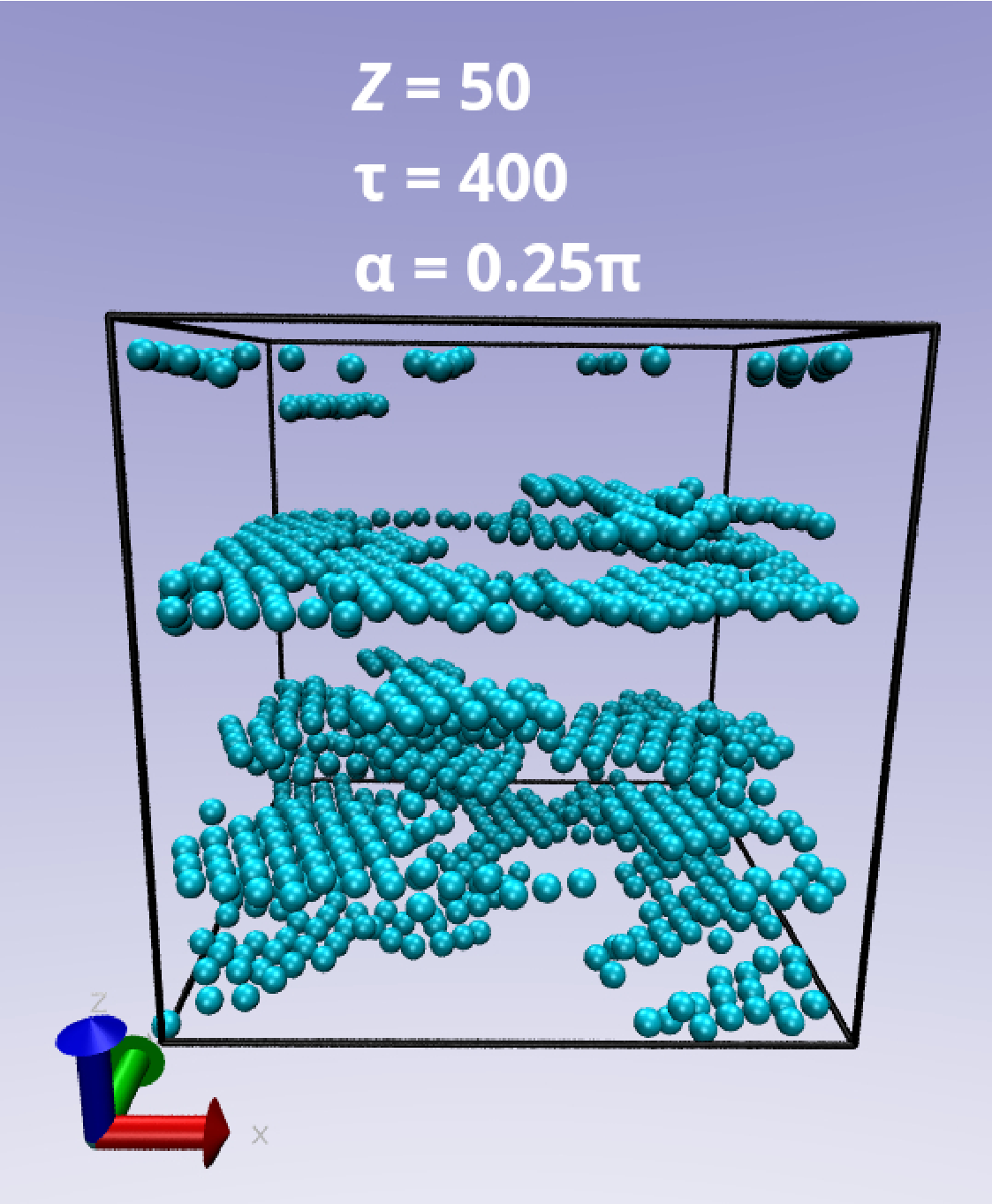}}  
    \caption{Simulation snapshots of the structural formations of polarized microgel particles subjected to fields of different polarization angles $\alpha$. The microgels have a bare monopole charge of $Z=50$, whereas their polarizations are fixed at $\tau=400$. The three arrows of the accompanying tripod at the 
    bottom left corner of each panel are colored red-green-blue and
    aligned along the $x$-, $y$-, and $z$-axes, respectively.}
    \label{fig:fig7}
\end{figure}

\begin{figure}[h!]
     \centering
     \subfigure[]{\includegraphics[width = 3.6 cm, height = 5cm]{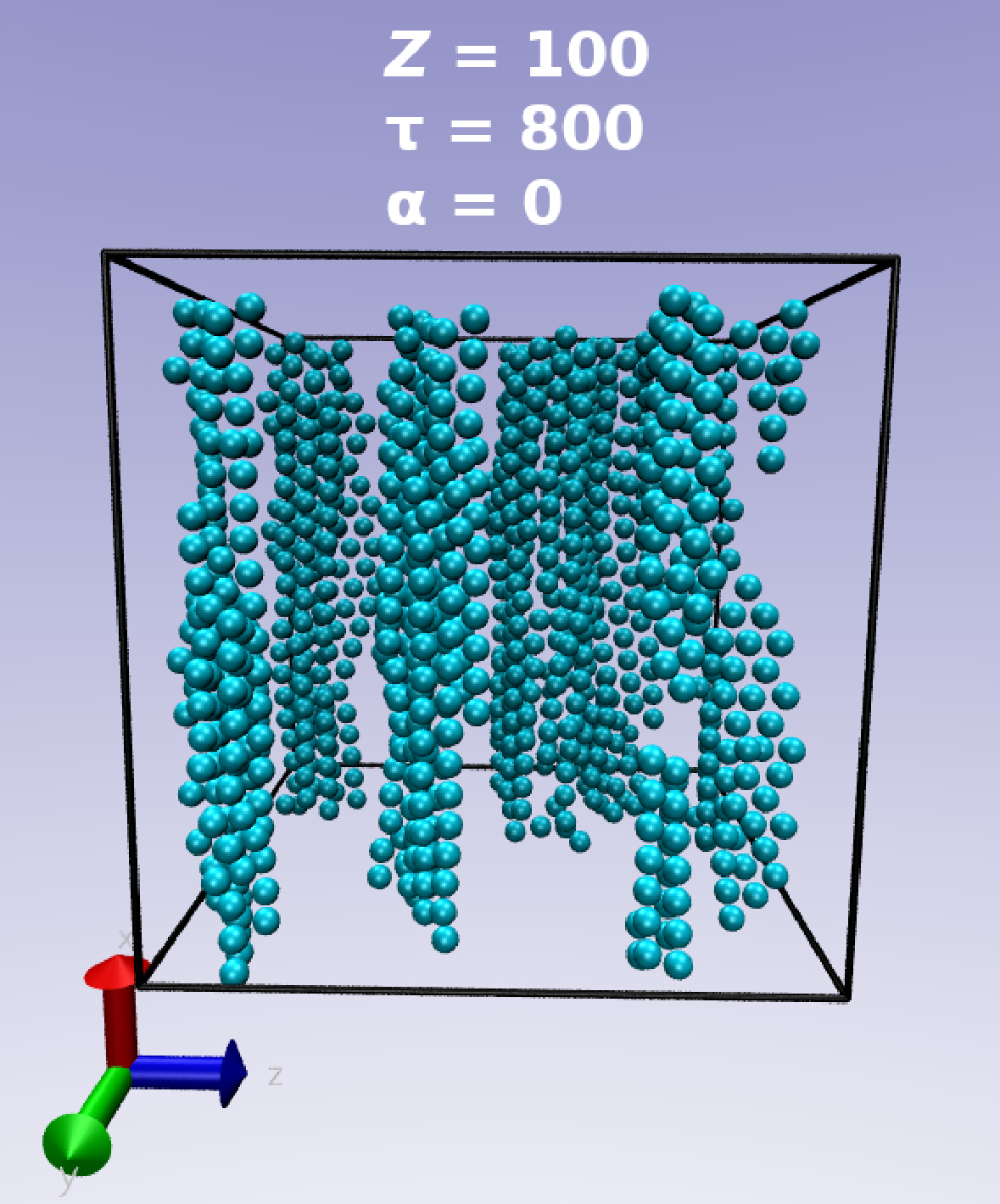}} 
     \subfigure[]{\includegraphics[width = 3.6 cm, height = 5cm]{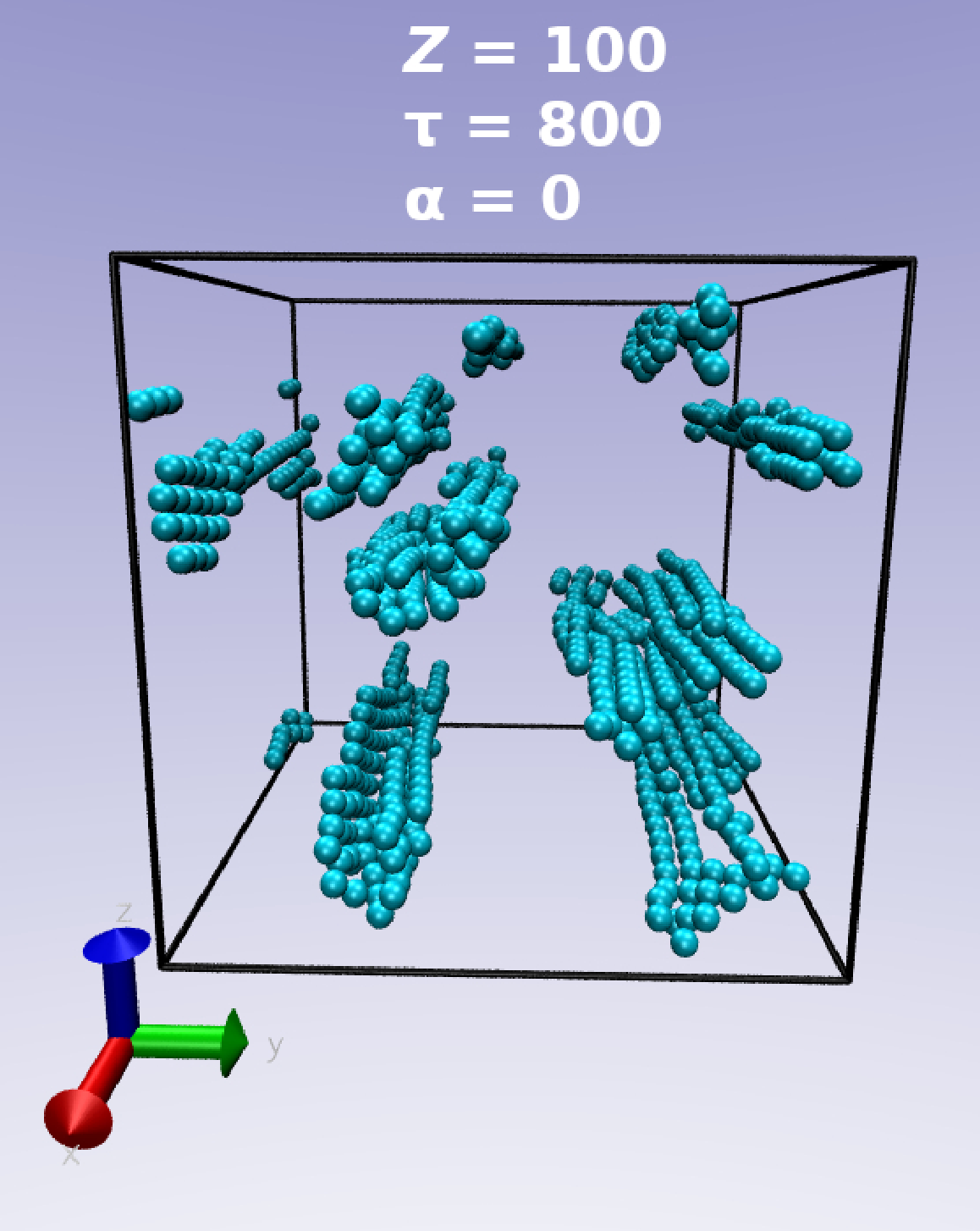}} 
     \subfigure[]{\includegraphics[width = 3.6 cm, height = 5cm]{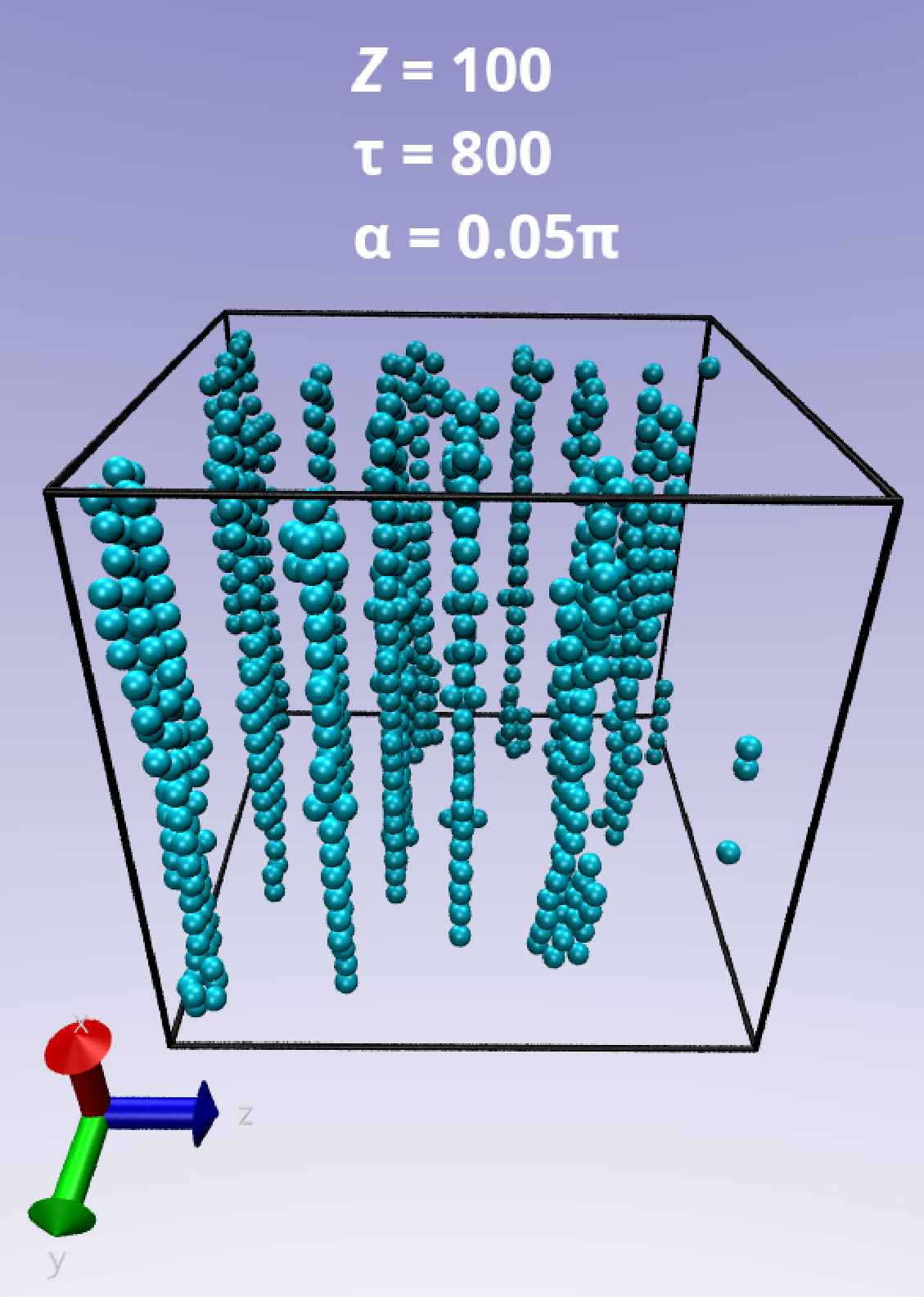}}
     \subfigure[]{\includegraphics[width = 3.6 cm, height = 5cm]{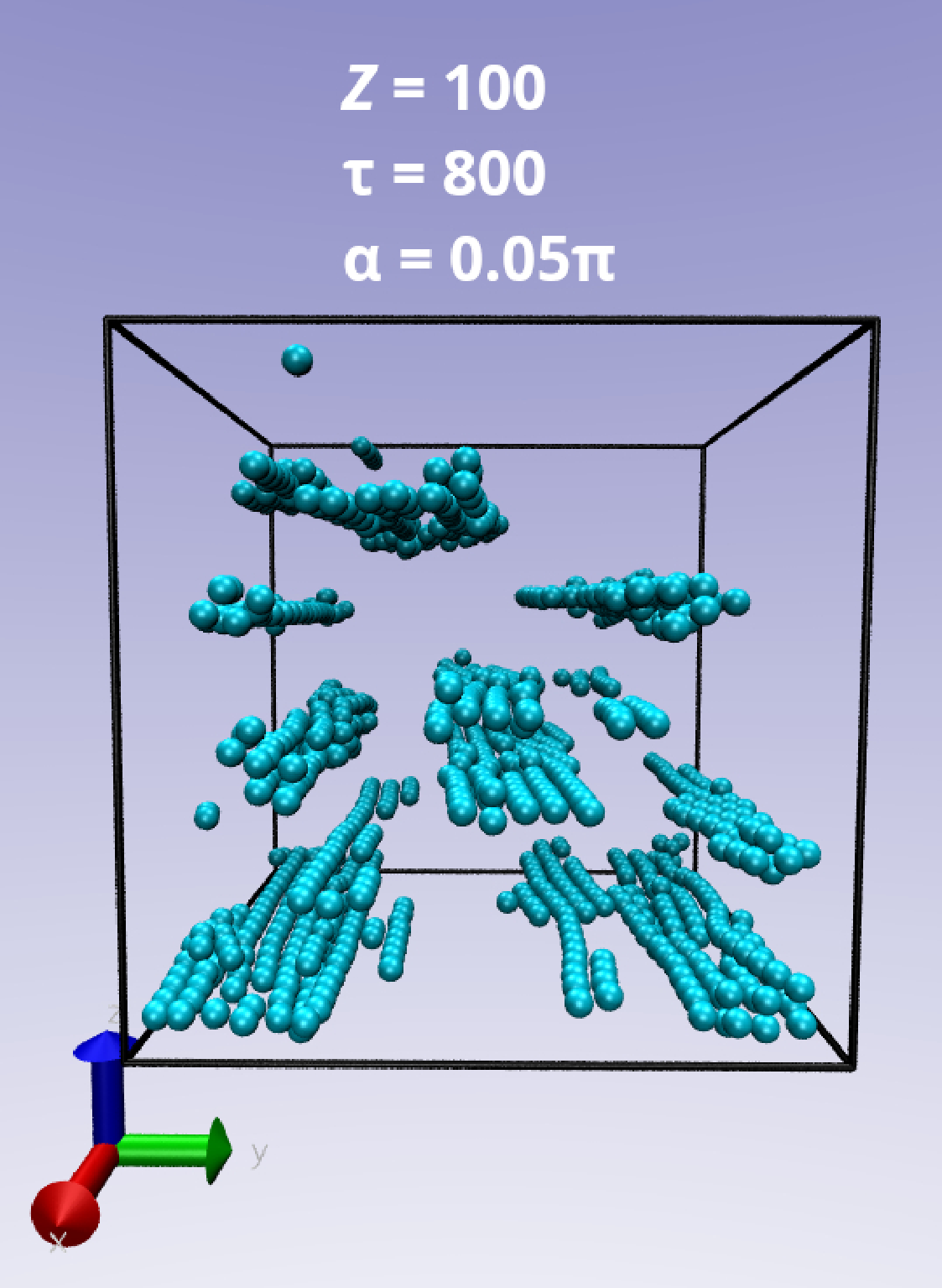}}\\
       \subfigure[]{\includegraphics[width = 3.6 cm, height = 5cm]{Z50_P400_a0.12_x.eps}} 
     \subfigure[]{\includegraphics[width = 3.6 cm, height = 5cm]{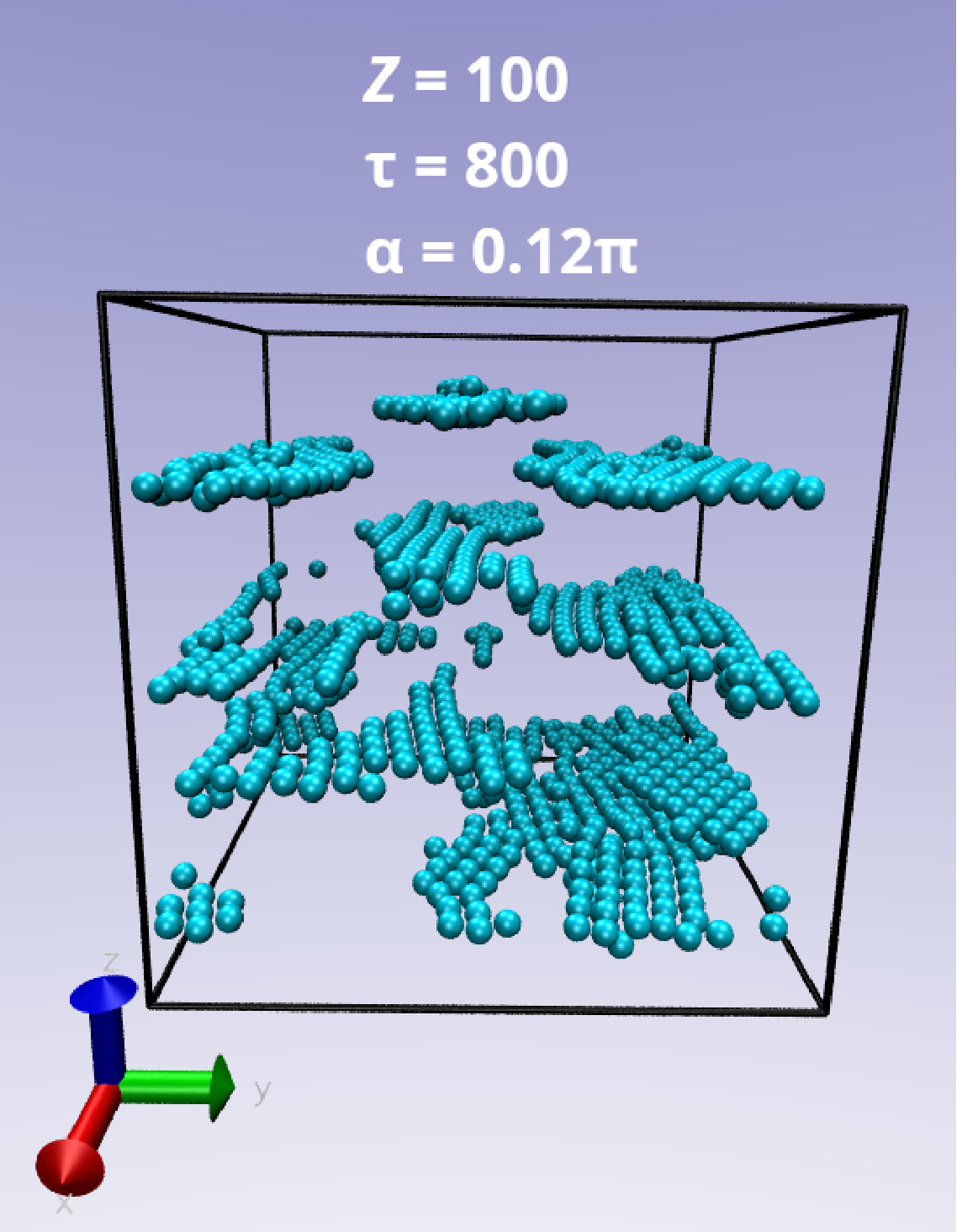}} 
     \subfigure[]{\includegraphics[width = 3.6 cm, height = 5cm]{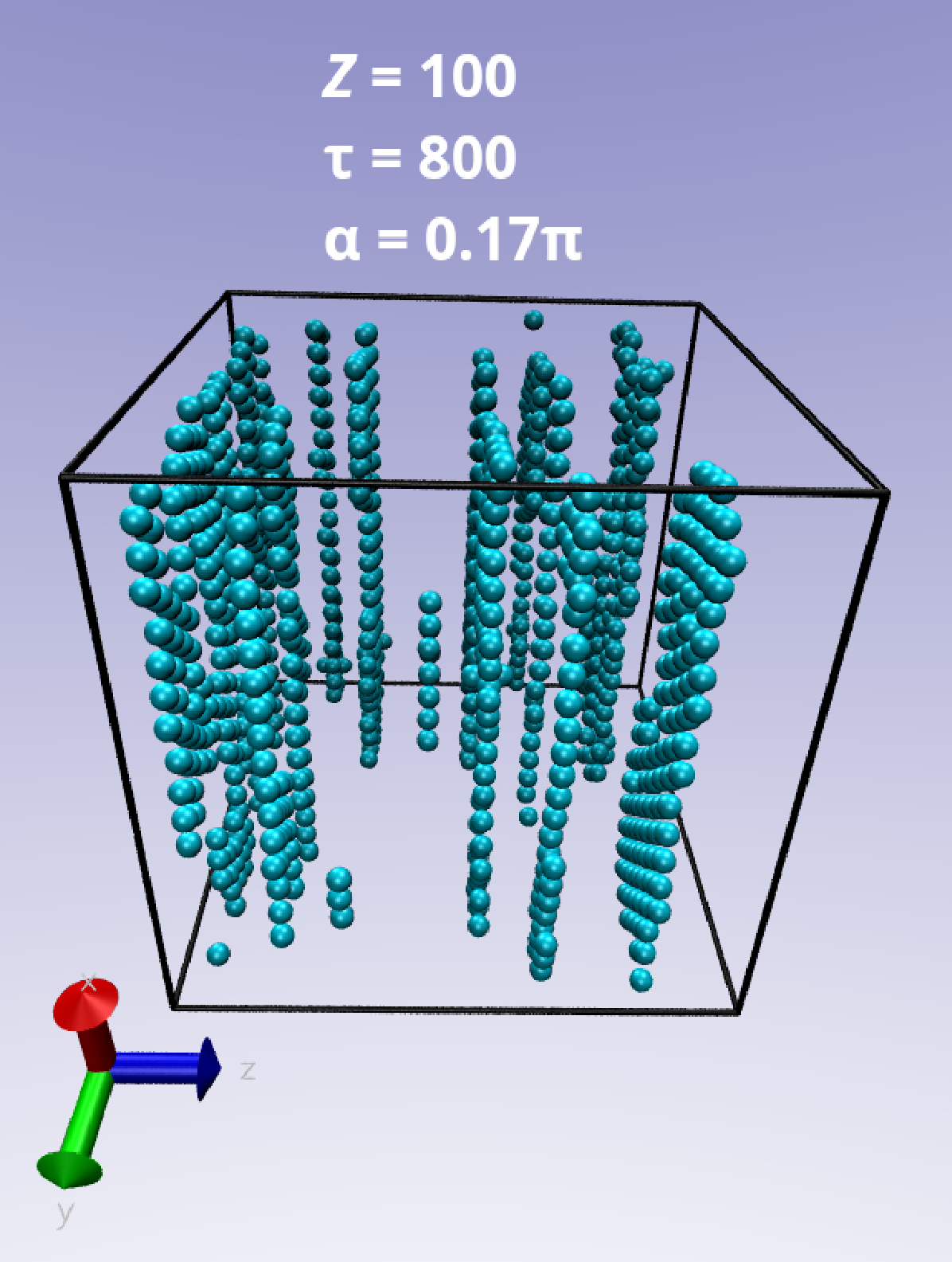}}
     \subfigure[]{\includegraphics[width = 3.6 cm, height = 5cm]{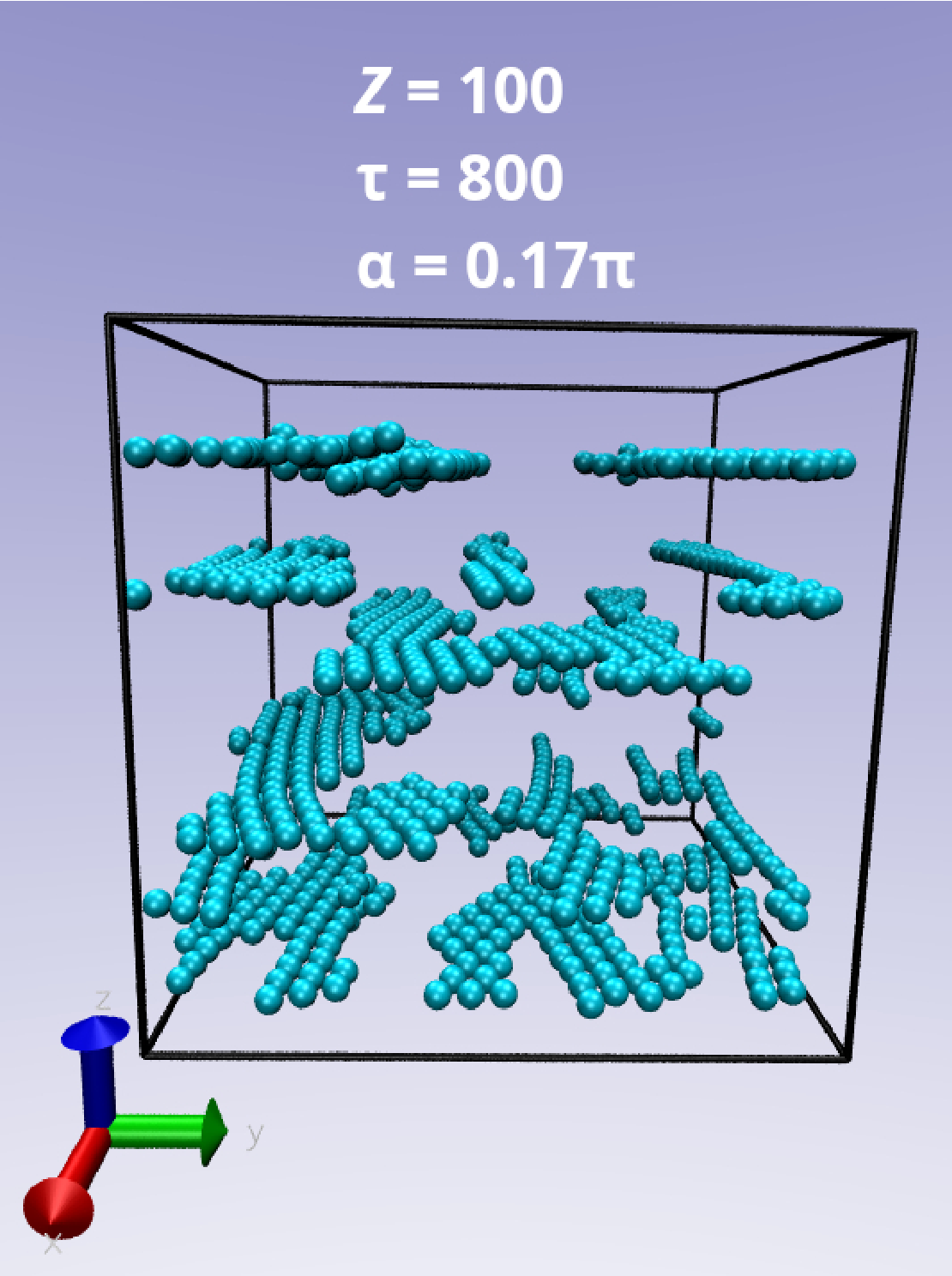}}\\
   \subfigure[]{\includegraphics[width = 3.6 cm, height = 5cm]{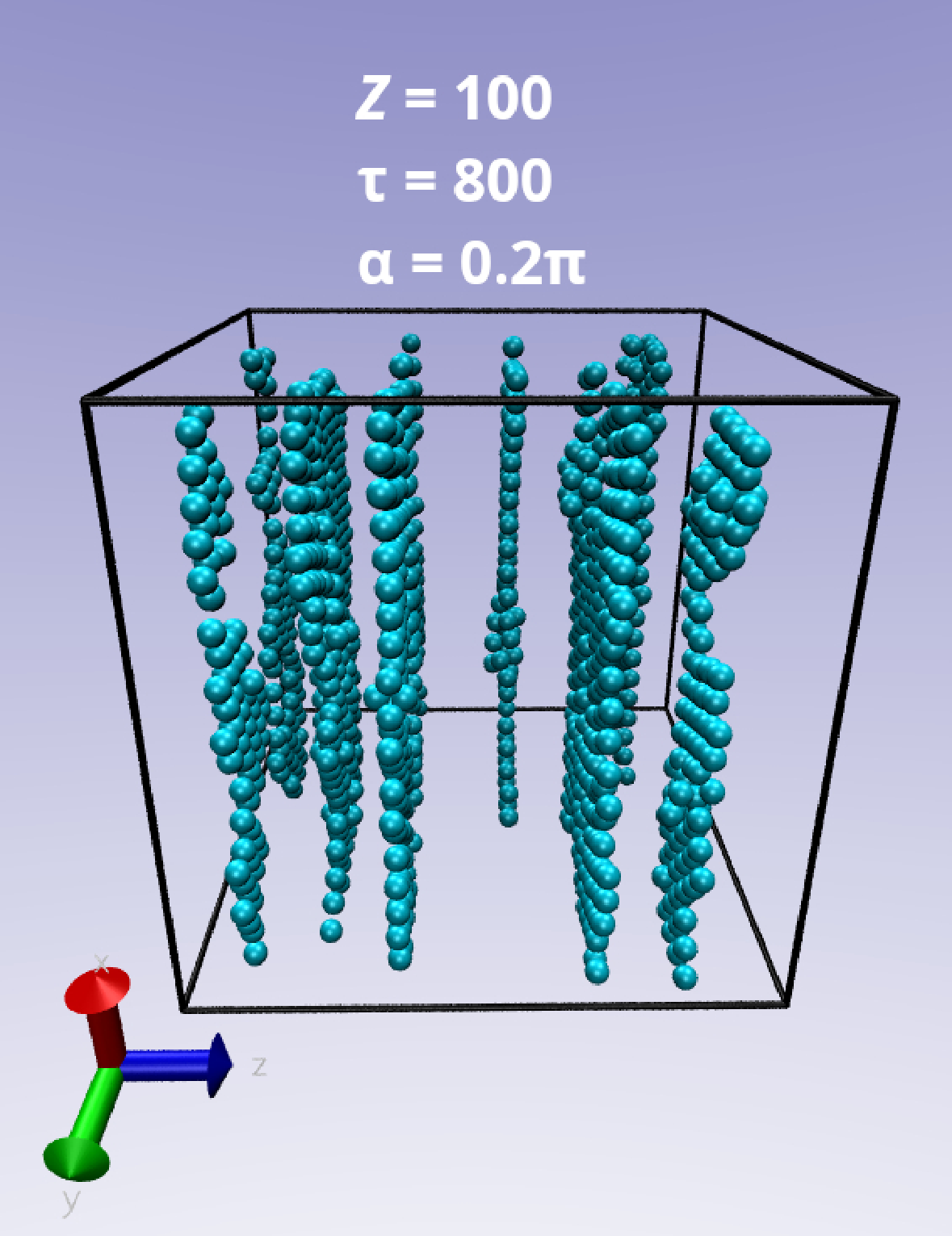}} 
     \subfigure[]{\includegraphics[width = 3.6 cm, height = 5cm]{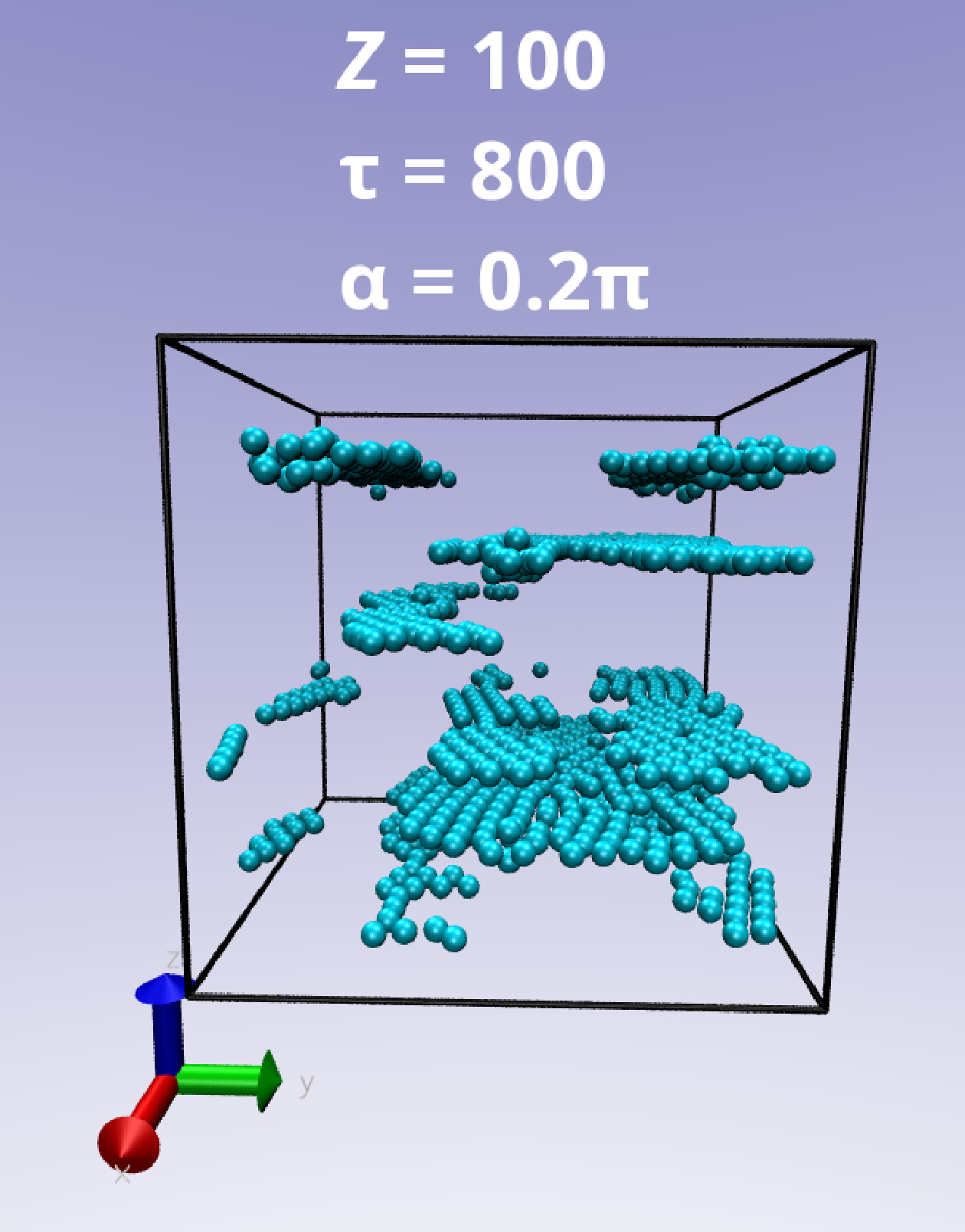}} 
     \subfigure[]{\includegraphics[width = 3.6 cm, height = 5cm]{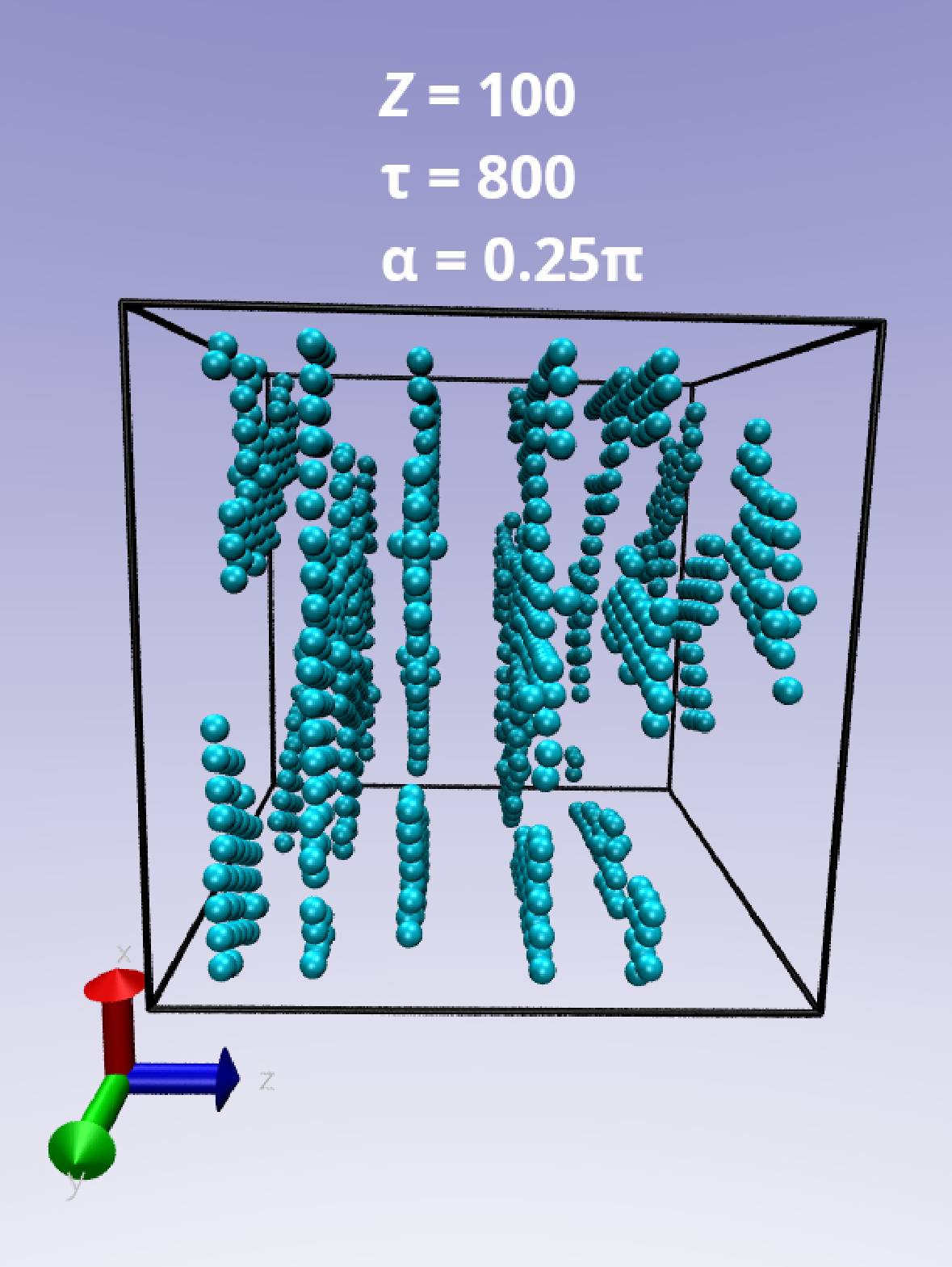}}
     \subfigure[]{\includegraphics[width = 3.6 cm, height = 5cm]{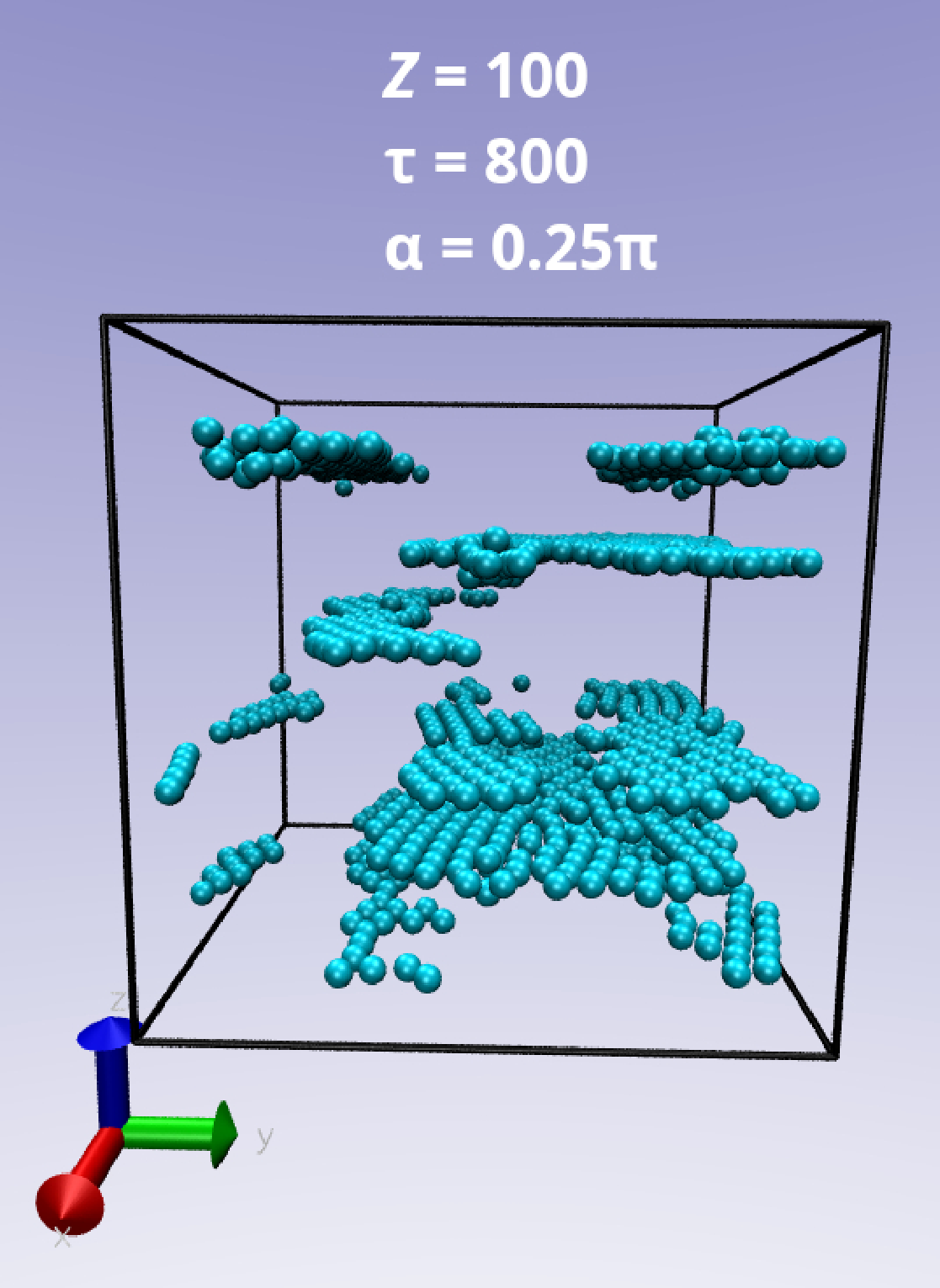}}
     \caption{Same as in Fig.~\ref{fig:fig7}, but now considering a second sample of microgels with monopole and dipole charges $Z=100$ and $\tau=800$, respectively. The three arrows of the accompanying tripod at the 
    bottom left corner of each panel are colored red-green-blue and
    aligned along the $x$-, $y$-, and $z$-axes, respectively.}
    \label{fig:fig8}
 \end{figure}

 \begin{figure}[h!]
     \centering
     \subfigure[]{\includegraphics[width = 4 cm, height = 4cm]{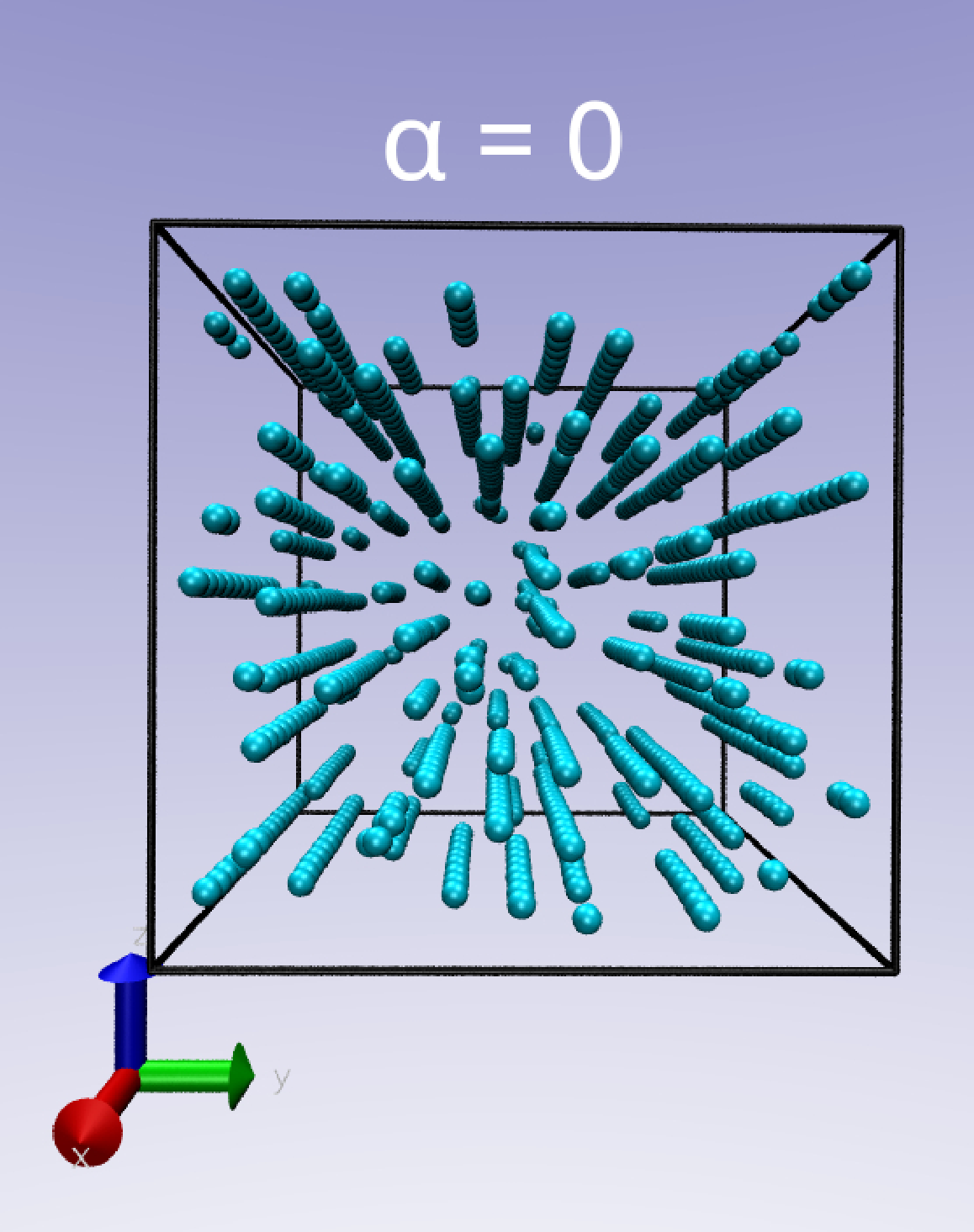}}
     \subfigure[]{\includegraphics[width = 4 cm, height = 4cm]{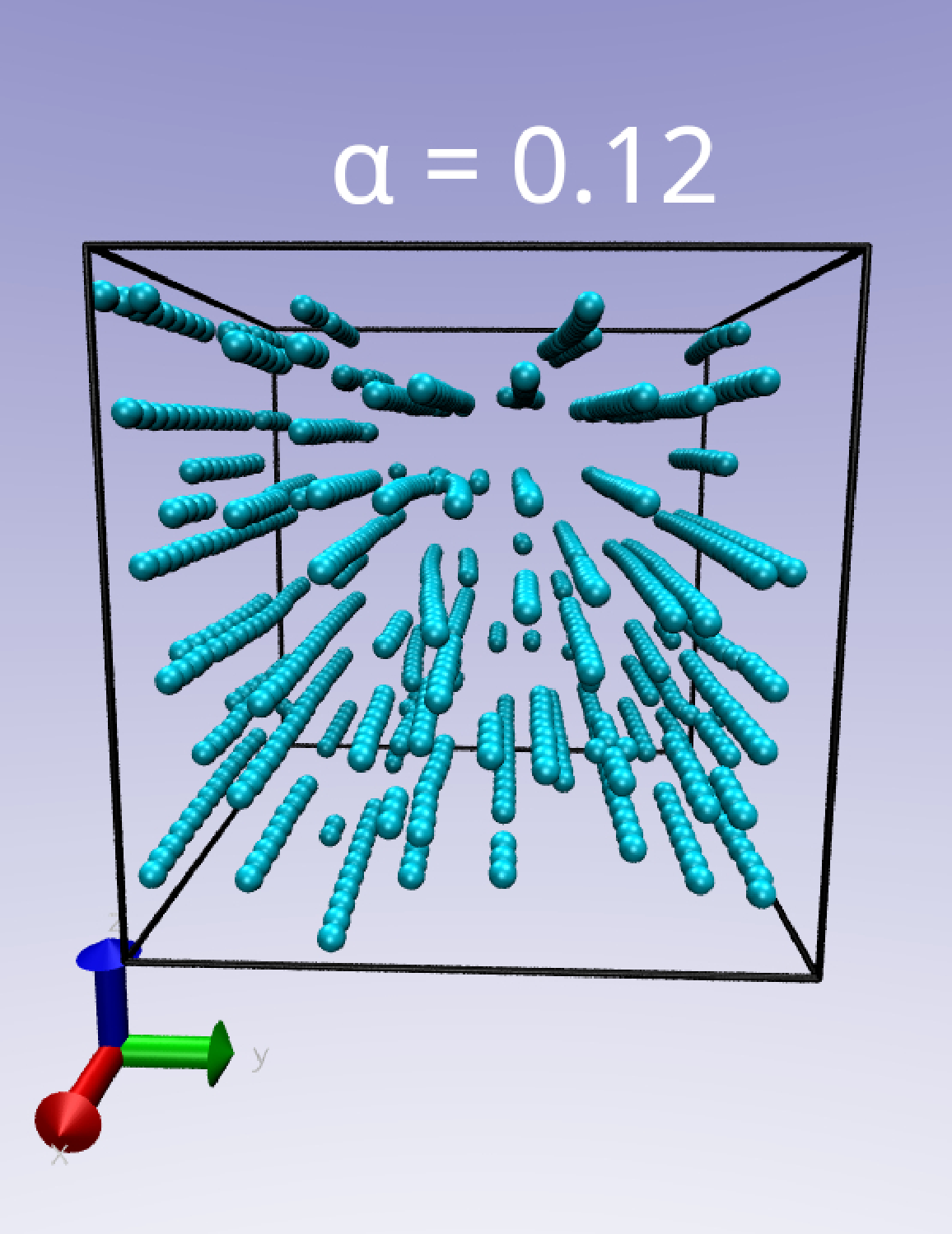}} 
     \subfigure[]{\includegraphics[width = 4 cm, height = 4cm]{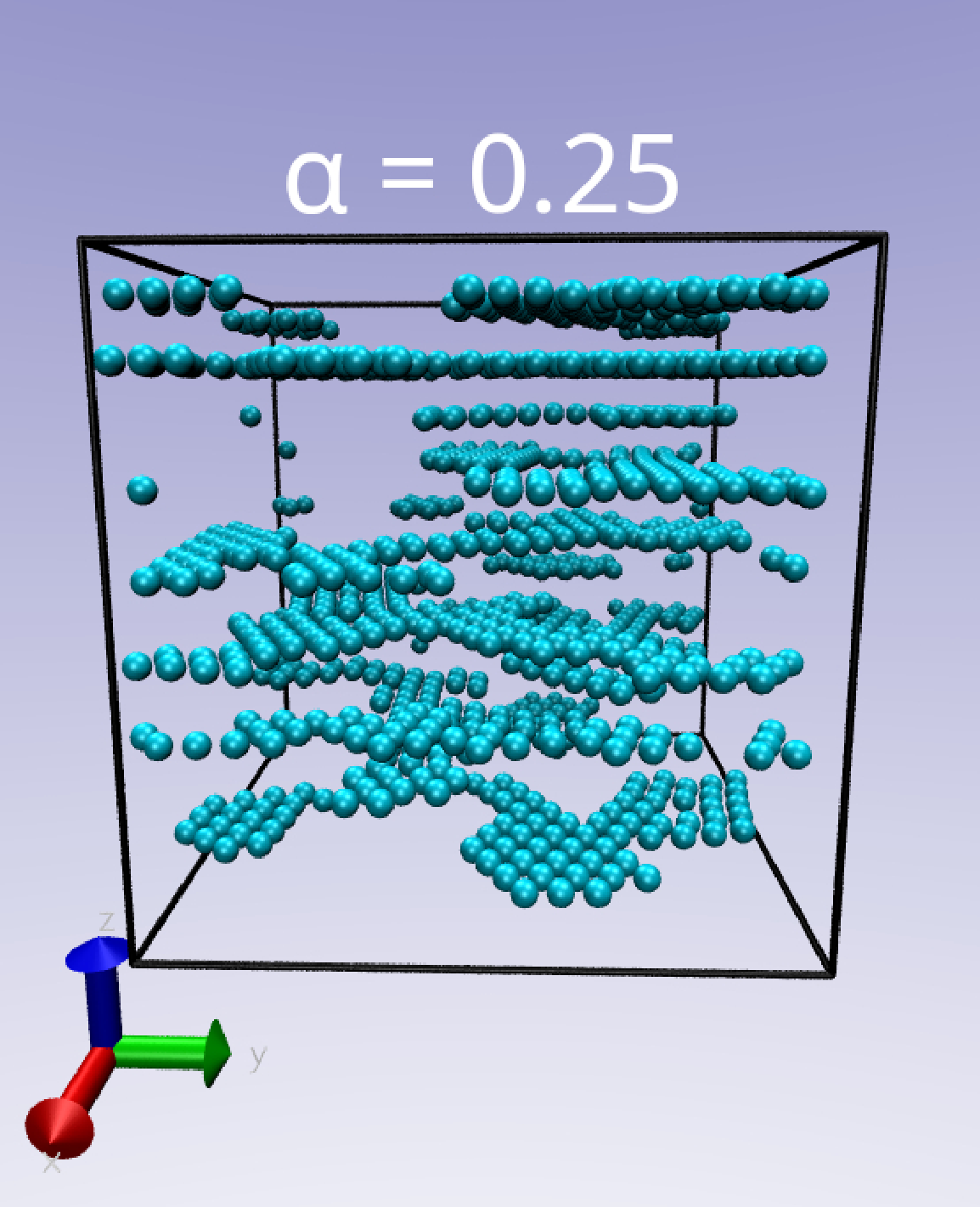}}\\
      \subfigure[]{\includegraphics[width = 4 cm, height = 4cm]{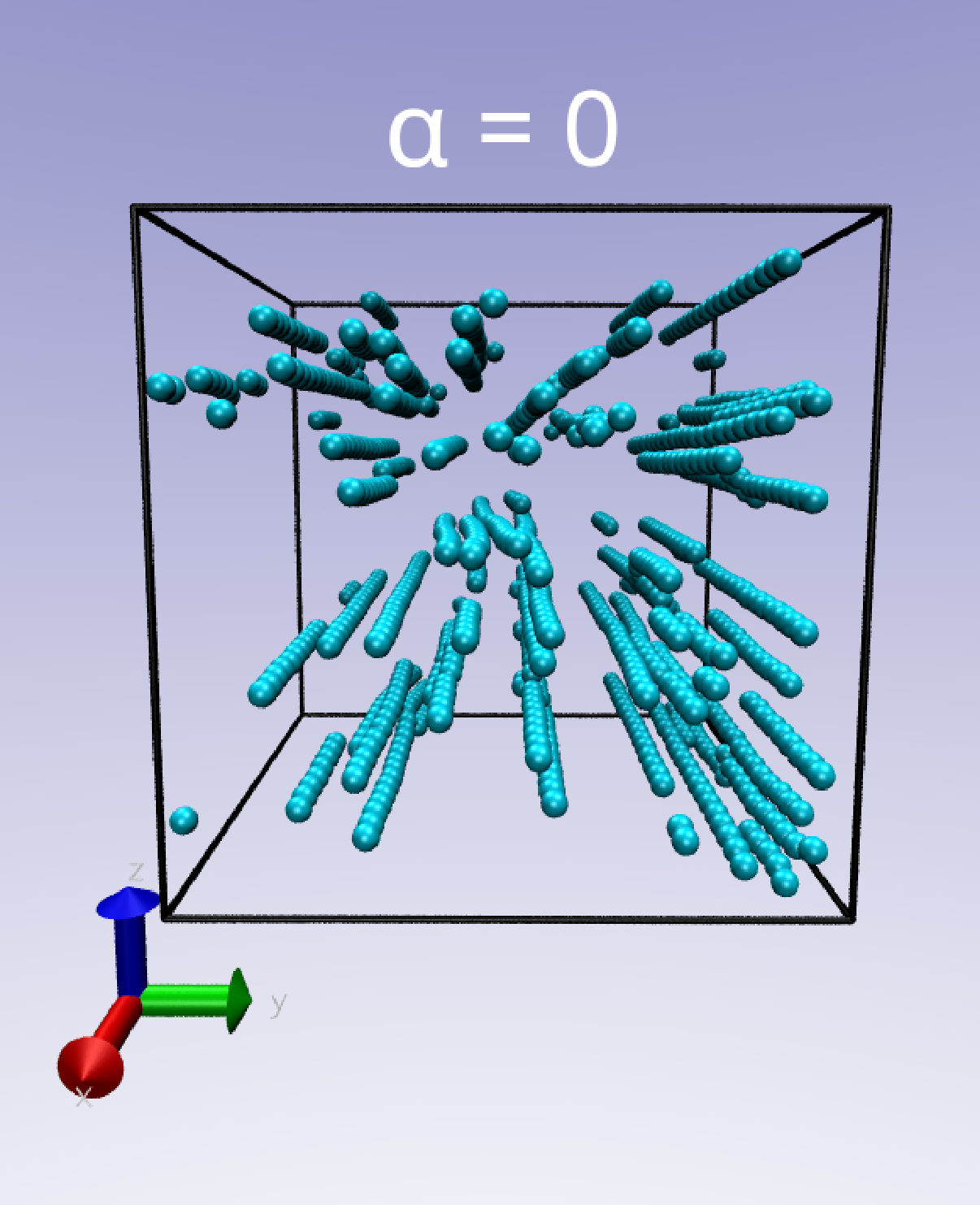}} 
     \subfigure[]{\includegraphics[width = 4 cm, height = 4cm]{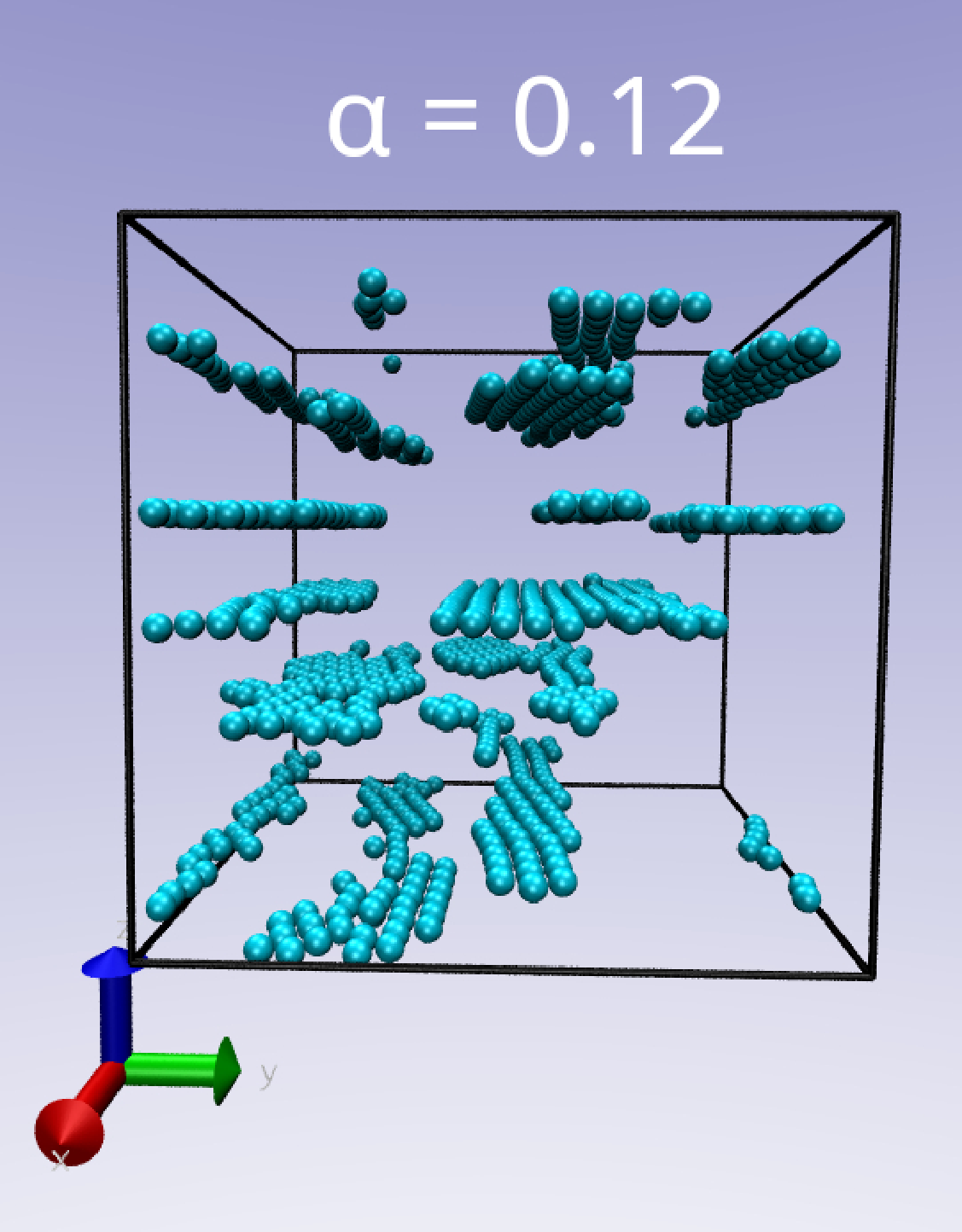}} 
     \subfigure[]{\includegraphics[width = 4 cm, height = 4cm]{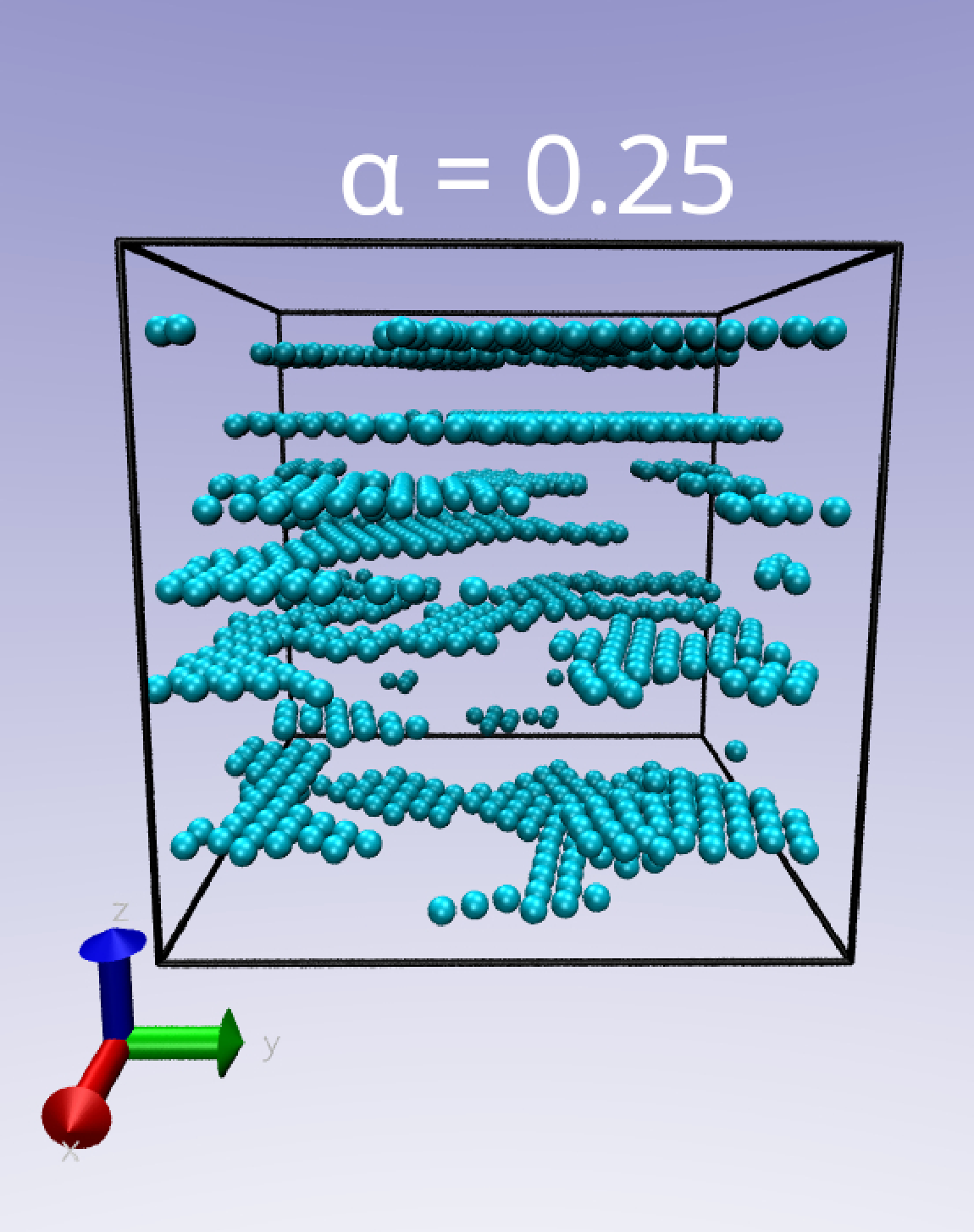}}\\
       \subfigure[]{\includegraphics[width = 4 cm, height = 4cm]{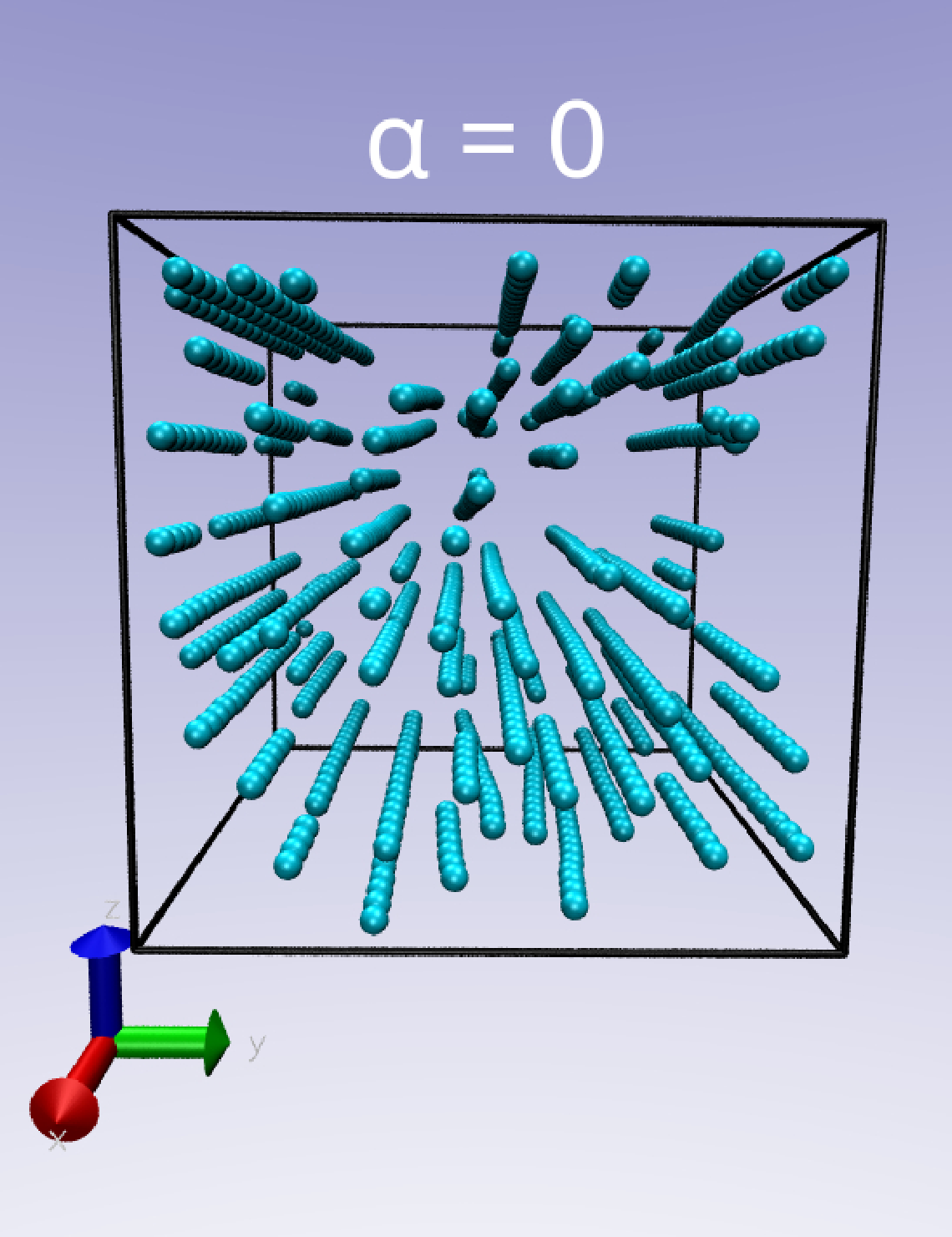}} 
     \subfigure[]{\includegraphics[width = 4 cm, height = 4cm]{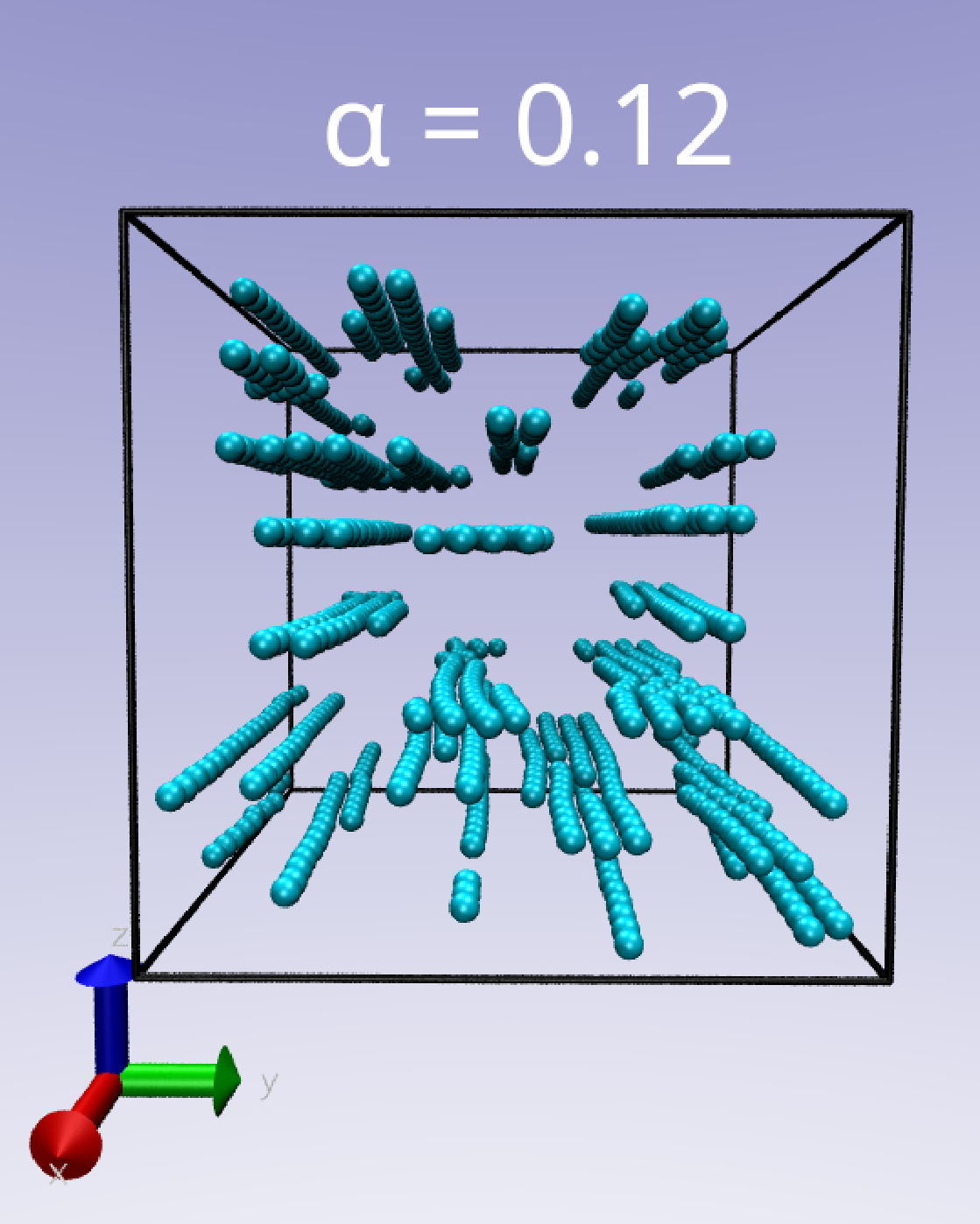}} 
     \subfigure[]{\includegraphics[width = 4 cm, height = 4cm]{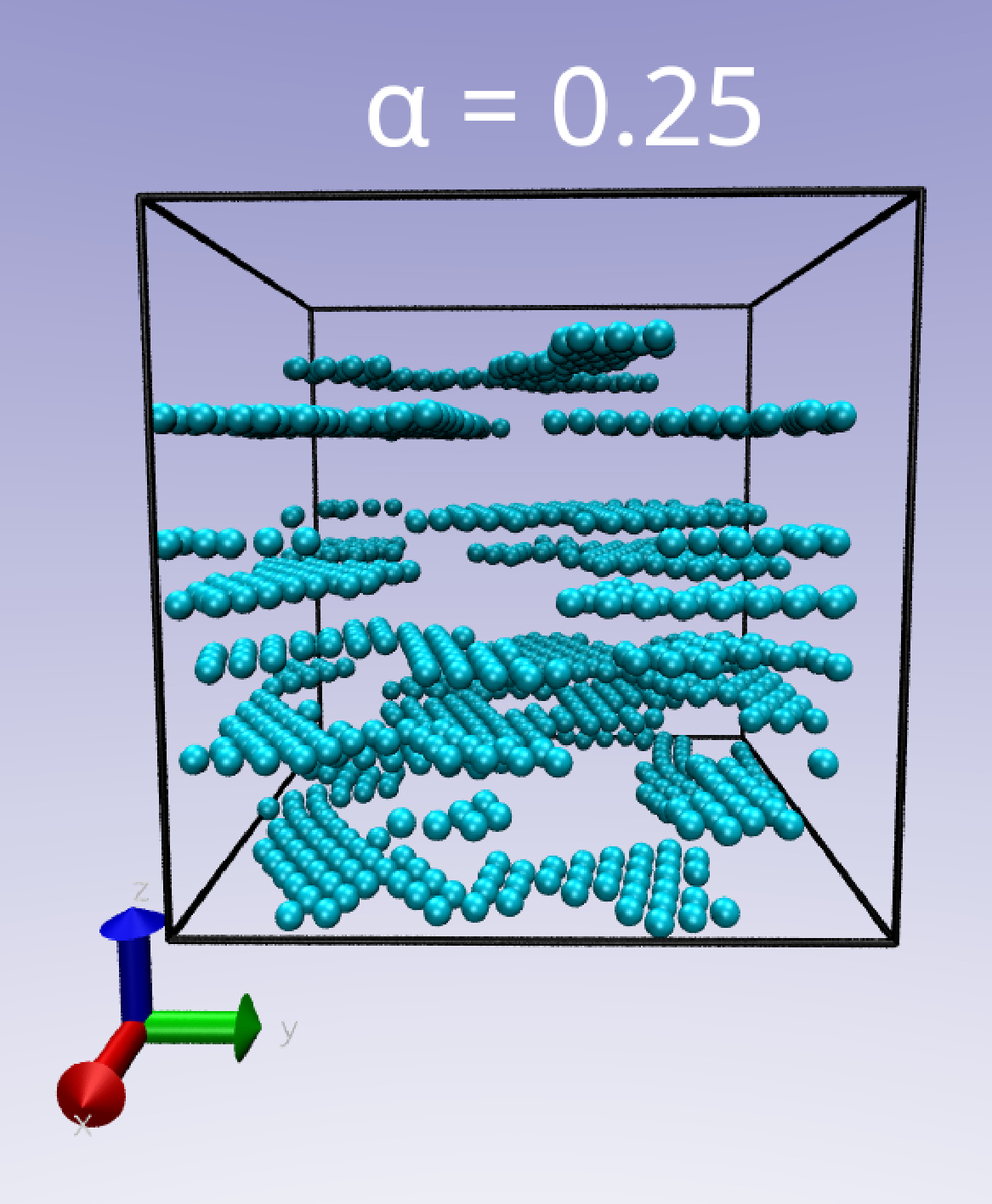}}\\
      \subfigure[]{\includegraphics[width = 4 cm, height = 4cm]{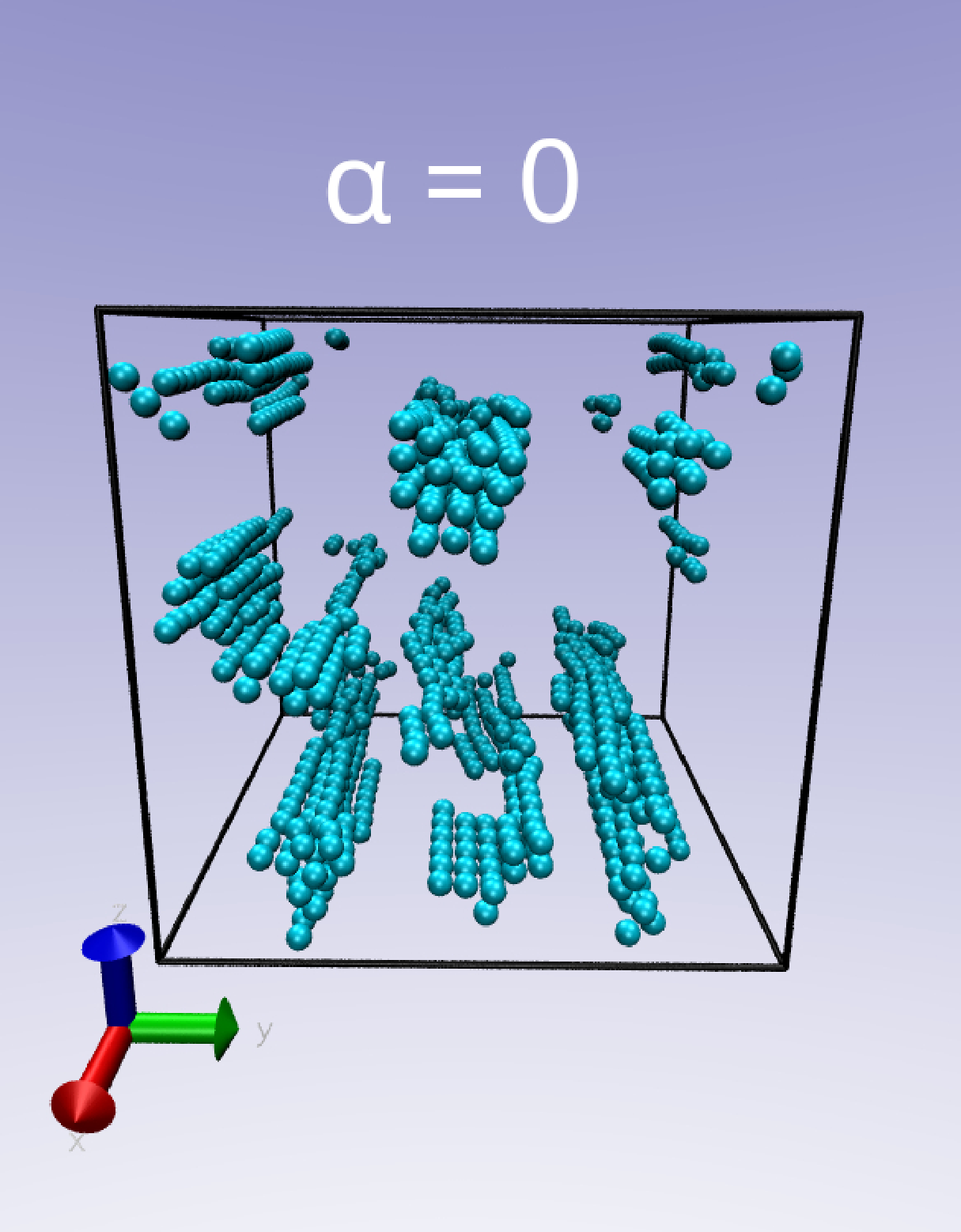}} 
     \subfigure[]{\includegraphics[width = 4 cm, height = 4cm]{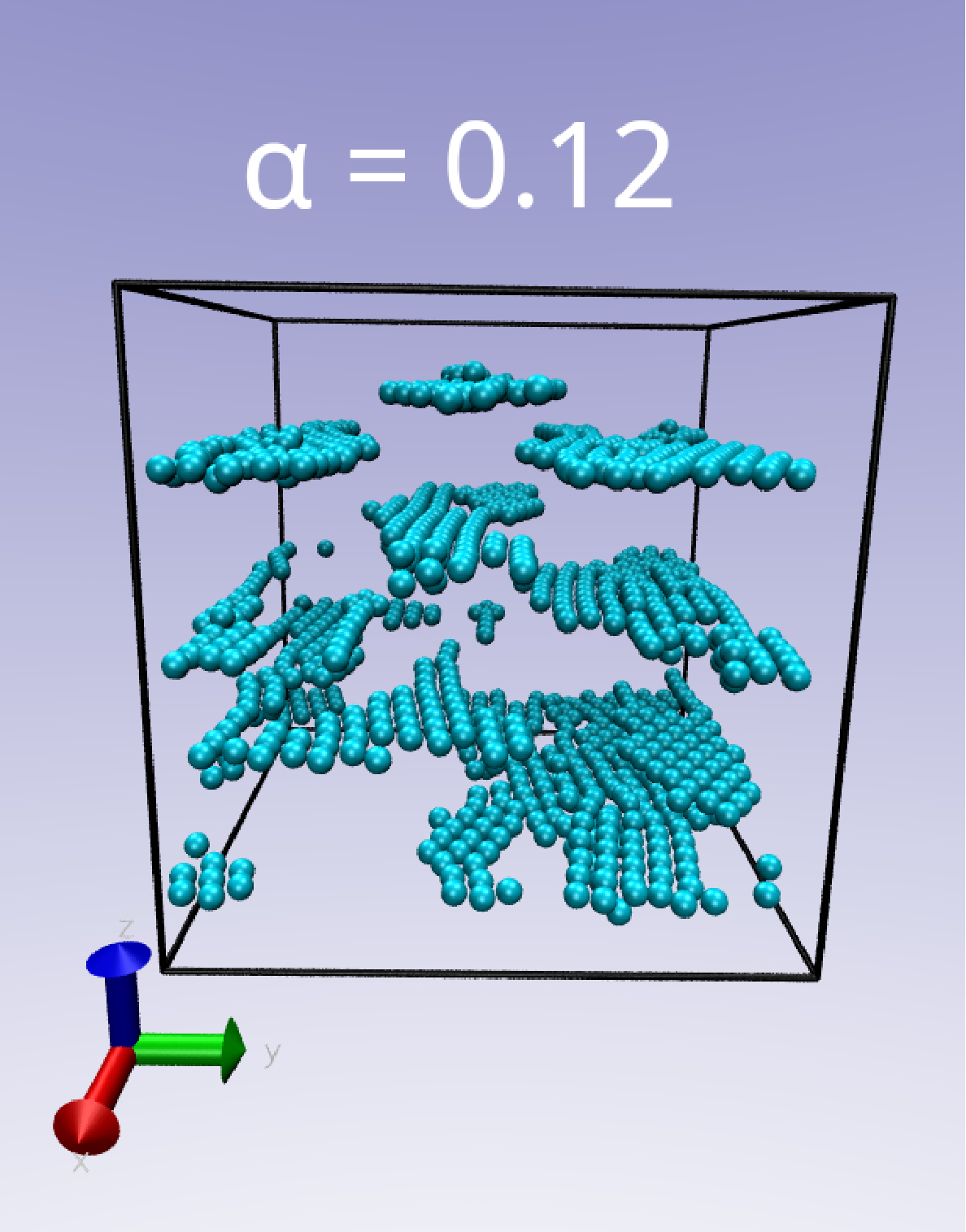}} 
     \subfigure[]{\includegraphics[width = 4 cm, height = 4cm]{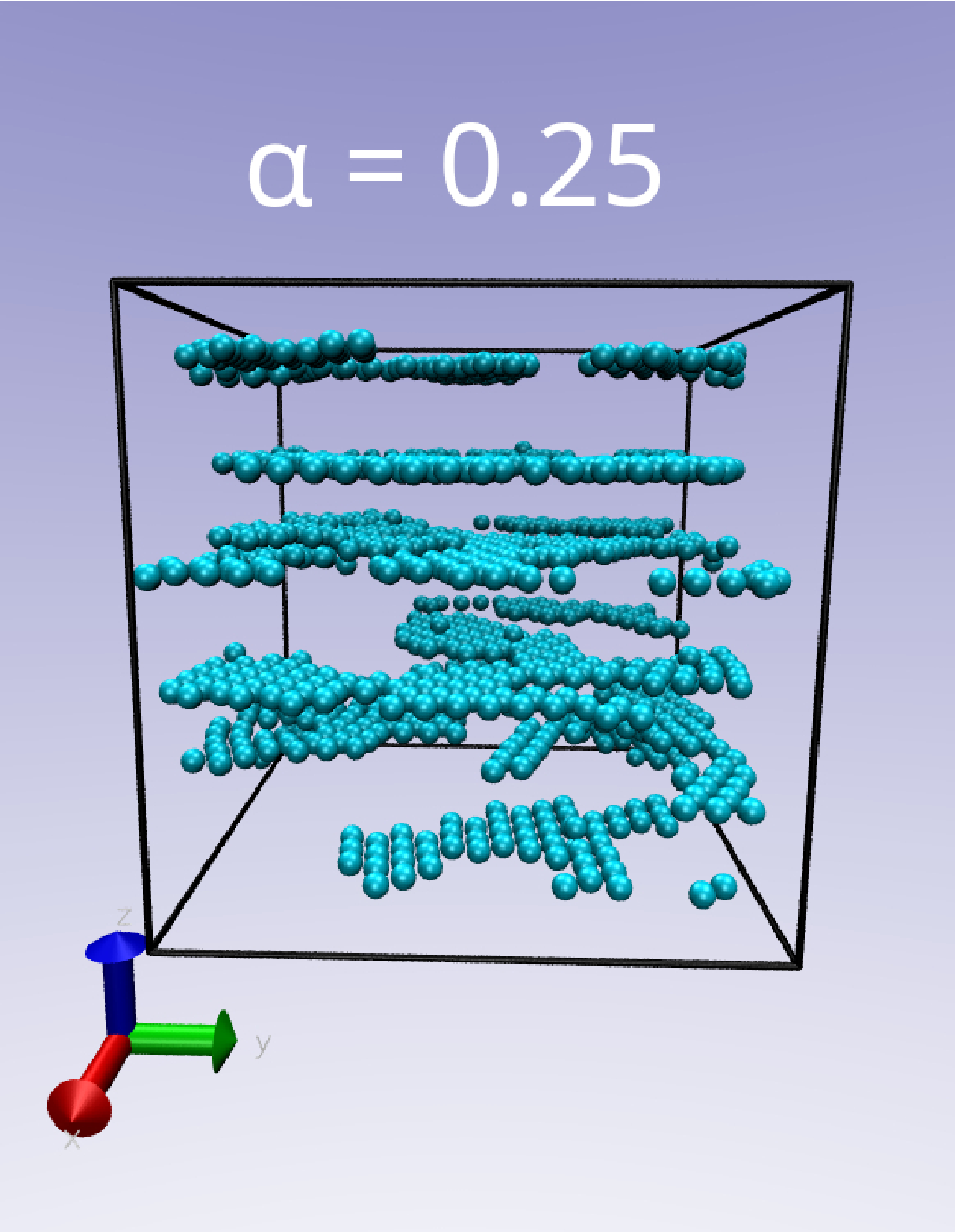}}\\
     \caption{Simulation snapshots considering varying the range of the dipole interactions,
     for parameters $Z = 100$ and $\tau = 800$. The values of $f=\kappa_d/\kappa$ are, from top to bottom, given by: $f=0.1$ [panels (a), (b), (c)], 
     $f=0.5$ [panels (d), (e), (f)], 
     $f=0.75$ [panels (g), (h), (i)], and 
     $f=1$ [panels (j), (k), (l)]. 
     Vertically organized panels correspond to linear polarizations
     [$\alpha = 0$, panels (a), (d), (g), (j)], 
     elliptical ones [$\alpha=\pi/8$, panels (b), (e), (h), (k)] 
     and circular ones [$\alpha = \pi/4$, panels (c), (f), (i), (l)]. The three arrows of the accompanying tripod at the 
    bottom left corner of each panel are colored red-green-blue and
    aligned along the $x$-, $y$-, and $z$-axes, respectively.}
     \label{fig:fig9}
 \end{figure}

\subsection{Structure}

We are now going to analyze the self-assembly structures for the two representative systems considered above (soft-particles with $Z=50$ and $\tau=400$, and a second sample with $Z=100$ and $\tau=800$). Figure \ref{fig:fig7} shows a series of simulation snapshots for the first system, at different values of anisotropy $\alpha$ of the underlying elliptically polarized field. In these figures, the polarization angle is progressively increased from $\alpha=0$ (linear) to $\alpha=\pi/4$ (circular). For each given configuration, snapshots are shown for two distinct configurations, with the vertical axis lying along the $x$- and $z$-directions. Fig.~\ref{fig:fig7} clearly shows how the configurations are gradually transforming from linear chains to planar sheets as $\alpha$ grows from $0$ to $\pi/4$. 
At $\alpha=0$, several branches of linear chains are formed, which are combined without any clear ordering along the perpendicular direction. As $\alpha$ increases, these chains start to re-organize in a way that they become more aligned along the $xy$-plane. Initially, the amorphous chain aggregates undergo an internal reconfiguration, in which the chains start to stick next to each other over the $y$-direction, see Fig.~\ref{fig:fig7}(d). As $\alpha$ increases further, the small aggregates of laterally arranged chains start to merge into larger planar aggregates, Figs.~\ref{fig:fig7}(e) - \ref{fig:fig7}(h). This clearly results from the onset of stronger attractions along the $y$ direction by increasing $\alpha$, seen in Fig.~\ref{fig:fig4}(c). Upon a close inspection at the merging chain structures at intermediate values of $\alpha$, it is possible to observe that the chains start to tilt along the $y$ direction, another result from the combination of weakened/stronger attractions along the $x$/$y$ directions. As some of the planar structures start to arrange into larger aggregates, several small particle agglomerates start to emerge, which are unable to fit into the larger ones. This situation persists until the interactions become fully isotropic along the $xy$ plane (circular polarization). In this case, it is no longer possible to distinguish chain formations along the planar structures. Instead, the aggregates start featuring hexagonal, crystal-like local arrangements in virtue of isotropic interactions over the $xy$-plane. At the same time, the crystalline
structure on the $xy$-plane is coexisting with large voids, 
reminiscent of a crystal-gas coexistence caused by the strong
attractive forces between microgels lying on this plane. 

The outlined pathway from linear chains to parallel, isotropic aggregates as the elliptical eccentricity is reduced from unity (linear chains) to zero (circular polarization) becomes also clear in Fig.~\ref{fig:fig8}, which shows snapshots from simulations of microgels with monopole charges $Z=100$ and polarizations $\tau=800$.  Despite the similar patterns, it is important to note that in this case the initial chains are combined into a larger number of smaller, planar aggregates. This is due to a stronger attraction along the $x$ direction, which makes it difficult to break up the chains in order to arrange them into larger aggregates. As a result, it is possible to observe the formation of large chains which remain bounded together along the lateral $y$ direction, even at larger $\alpha$. The self-assembly of smaller aggregates also results from enhanced repulsions along the $z$ direction, which avoids close contact between different planar structures in order to merge into larger agglomerates.

It is also interesting to investigate the effects of decreasing the range of dipole interactions on the underlying self-assembly structures. As shown in Figs.~\ref{fig:fig4} and \ref{fig:fig6}, increasing the ratio $f$ between dipole and monopole inverse screening lengths have drastic effects in the isotropic particle interactions. In particular, decreasing $f$ leads to much more pronounced dipole attractions along the $x$ direction, together with stronger repulsion across the $y$ axis. The effects of such stronger interactions on particle aggregation are summarized in Fig.~\ref{fig:fig9}, in which simulation snapshots for three representative values of $\alpha$
 are shown. From top to bottom, results are shown for $f=0.1$, $f=0.5$, $f=0.75$, and $f=1$. In all cases, the particle charges and polarizations are fixed at $Z=100$ and $\tau=800$, respectively. A strong dependence on dipole screening is observed, particularly at linear (left panels) and intermediate (middle panels) polarizations. Due to the stronger in-plane repulsions, the forming chains at the linear polarization regime no longer merge together when the dipole interactions are long-ranged. The chains are instead well separated due to strong lateral repulsion. A similar scenario has been observed in previous works, in accordance with experimental results in regimes of very high driven frequencies, in which case the dipole interactions are expected to be almost unscreened. As the parameter $f$ is increased, the reduced in-plane repulsions result in chains coming closer together, despite no clear aggregation between different chain taking place. The situation changes drastically at fully screened dipole interactions, where chains form small clusters at the in-plane directions. Quite different behaviors can also be seen at intermediate polarizations ($\alpha=\pi/8$, middle panels), whereby the structures quickly change from fully segregate chains (upper panels, $f=0.1$), to plane-chain structures along the in-plane directions. This behavior clearly results from a rapid onset of chain-chain attraction over the in-plane, as shown in Figs.~\ref{fig:fig4}c and \ref{fig:fig4}d). Again, a sharp difference is observed in the structure of chain-like aggregation as $f$ is increased from $f=0.75$ to $f=1$. In the latter case, much larger aggregates are formed, which also seem to be much more isotropic along the in-plane directions. Finally, the isotropic aggregations along the $yz$ plane formed in the case of circularly polarized light ($\alpha=\pi/4$, right-handed panels) seem to be much less sensitive to the range of dipole-dipole interactions. This is again in accordance with the general picture provided by Figs.~\ref{fig:fig4} and \ref{fig:fig6}, which show that the effects from varying $f$ are suppressed as the polarizations are close to the circular case. Still, a marked difference can be observed as $f$ increases for $f=0.75$ to $f=1$. Here we see that a number of in-plane structures are merged together, well separated from each other across the $x$-direction. This situation strongly contrasts with the cases of smaller $f$, where a large number of smaller, well-separated clusters along at the in-plane directions is observed.

\section{Conclusions}\label{sec:conclusions}

A theoretical approach has been designed to analyze the self-assembly of soft particles induced by planar, elliptical electric fields of different strengths and frequencies. The developed coarse-graining model is a direct extension of previous works, which have addressed the aggregation properties of ionic microgels in presence of both linear and circularly polarized fields, following earlier experimental works on these systems. Such an extension allowed one to fill the gap between structural properties interpolating between linear and circular cases, where the stronger anisotropy leads to structures made up of linear chains that start to arrange into planes as the elliptical eccentricity approaches one. Apart from a rich variety of intermediate structures, it was also found that the self-assembly formations are rather sensitive to the range of dipole interactions -- which can generally lie from unscreened to fully screened interactions -- a property which has not been looked at in detail in previous works. In particular, chain formation is found to be strongly dependent on the range of dipole attractions. Whereas in cases of long-range dipole interactions (i.e., weak ionic screenings) there is a predominant formation of large chains well segregated from each other, the situation of short-range dipoles (i.e., strong ionic screenings) favors the formation of aggregating chains. 

The theoretical framework developed herein is quite general, aimed to describe spherical particles of arbitrary polarization responses in the presence of elliptically polarized fields. Very similar approaches can be used to work out the situations of magnetic particles in the presence of oscillating magnetic fields, under few modifications in the general formalism. It is therefore important to note that the model can be further extended in a straightforward way to investigate different systems with varying polarization responses under different dynamic coupling to the external fields. Examples range from dielectric spheres under AC fields to soft particles of different sizes and internal conformations, such as microgels with different radial charge distributions. We thus expect that the present work can provide a solid basis for further investigations of the self-assembly of different particles induced by external fields.

\begin{acknowledgments}

A CC-BY public copyright license has been applied by the authors to the present document and will be applied to all subsequent versions up to the Author Accepted Manuscript (alternatively final peer-reviewed manuscript accepted for publication) arising from this submission, in accordance with the grant’s open access conditions.
\end{acknowledgments}

\section*{Author Declaration}
\subsection*{Conflict of Interest}
The authors have no conflicts to disclose.
\subsection*{Author Contributions}
\textbf{Carlos Eduardo Estanislau:} Performing the calculations (lead); Data curation (lead); formal analysis (lead); methodology (supporting); software (lead); visualization (lead); writing – original draft (supporting); writing – review and editing (supporting). \textbf{Thiago Colla:} Conceptualization (lead); formal analysis (lead); funding acquisition (lead); methodology (equal); project administration (equal); resources (lead); supervision (lead); validation (equal); writing – original draft (lead); writing – review and editing (equal). \textbf{Christos N.~Likos:} Conceptualization (lead); formal analysis (supporting); methodology (equal); project administration (equal); resources (supporting); supervision (supporting); validation (equal); writing – original draft (supporting); writing – review and editing (equal).

\section*{Data Availability}
The data that support the findings of this study are available from the corresponding author upon reasonable request.

\end{document}